\documentclass{JFM-FLM_Au}

\usepackage{pdflscape}
\usepackage{adjustbox}
\usepackage{makecell}
\usepackage{arydshln}
\usepackage{float}
\usepackage{subcaption}
\usepackage{multirow}
\usepackage{upgreek}
\usepackage{color}

\lefttitle{F. Guan, M. Liu, Y. Hasegawa}
\righttitle{Journal of Fluid Mechanics}

\title{Dissimilar heat transfer enhancement in spatially developing flow between parallel perforated plates by inducing a streamwise travelling-wave disturbance}

\author{Fengbo Guan\aff{1}, Ming Liu\aff{1} \and Yosuke Hasegawa\aff{1}}

\affiliation{\aff{1} Center for Research on Innovative Simulation Software, Institute of Industrial Science, The University of Tokyo, 4-6-1 Komaba, Meguro-ku, Tokyo 153-8505, Japan}

\corresau{Yosuke Hasegawa, \email{ysk@iis.u-tokyo.ac.jp}}

\begin{document}
\maketitle

\begin{abstract}
Travelling-wave-like wall blowing and suction is an effective approach for enhancing heat transfer with a minimal pressure drag penalty. However, achieving such a dissimilar heat transfer enhancement effect in a passive manner remains a challenge, and it remains unclear whether travelling-wave disturbances can be sustained particularly at low Reynolds numbers. In the present study, we propose introducing parallel perforated plates to induce travelling-wave-like disturbances passively. Pore-resolving simulations of a spatially developing laminar flow between parallel perforated plates are performed across a wide range of Reynolds numbers of $Re = 500-1500$ and pore-to-solid length ratios of $L_\mathrm{p}/L_\mathrm{s} = 0-10$. The flow between the perforated plates remains steady at $Re = 500$ with the heat transfer and pressure drop characteristics agreeing well with those of impermeable solid plates. At higher Reynolds numbers, increasing $L_\mathrm{p}/L_\mathrm{s}$ induces a travelling wave perturbation. Accordingly, dissimilar heat transfer enhancement is confirmed for $9 \leq L_\mathrm{p}/L_\mathrm{s} \leq 10$ at $Re = 1000$ and $4 \leq L_\mathrm{p}/L_\mathrm{s} \leq 7$ at $Re = 1500$. The highest analogy factor, i.e., the ratio of the Stanton number to the friction coefficient is obtained at $Re = 1500$ and $L_\mathrm{p}/L_\mathrm{s} = 6$, yielding an increase of more than $30$\% compared to that of an impermeable solid plate. Analysis of the fluctuating fields shows that, in the travelling-wave flow regime, a pressure-induced wall-normal velocity fluctuation transports temperature fluctuations away from the perforated plate, while breaking the correlation between the streamwise and wall-normal velocity fluctuations. This enhances the turbulent heat flux relative to the Reynolds shear stress near the perforated plate. When the Reynolds number and $L_\mathrm{p}/L_\mathrm{s}$ are further increased, the travelling wave disturbance breaks down into chaotic motions, amplifying the streamwise velocity fluctuation and increasing the Reynolds shear stress, which reduces the analogy factor. The present results indicate that introducing a perforated plate with a suitable porosity induces travelling-wave velocity disturbances and also achieves a considerable dissimilar heat transfer effect even at low Reynolds numbers where a standard impermeable flat wall yields a steady laminar flow.
\end{abstract}

\begin{keywords}
mixing enhancement, porous media
\end{keywords}

\section{Introduction}
\label{sec:introduction}
In response to global challenges such as the energy crisis and environmental degradation, highly efficient thermo-fluids systems are increasingly needed. For the efficient use of heat, heat exchangers play a key role in various applications such as air conditioning systems, chemical and power plants, and waste-heat recovery systems, to name a few. One of the common challenges in the design of heat exchangers is to achieve heat transfer enhancement with a minimal pressure-drop penalty. However, owing to the strong similarity between momentum and heat transport, which is widely known as the Reynolds analogy \citep{Reynolds1874}, realizing such dissimilar heat transfer enhancement is quite challenging \citep{SuzukiSuzukiSato1988, KasagiHasegawaFukagataIwamoto2012, MotokiKawaharaShimizu2018}.

Over the past decades, various heat transfer surfaces such as dimpled fins \citep{ElyyanRozatiTafti2008}, offset fins \citep{ManglikBergles1995}, wavy fins \citep{IsmailVelraj2009}, louvered fins \citep{DongChenChenZhangZhou2007}, riblets \citep{NaganoHattoriHoura2004}, vortex generators \citep{EiamsaArdPromvonge2011}, and rough surfaces \citep{ForooghiWeidenlenerMagagnatoBohmKubachKochFrohnapfel2018} have been proposed and widely applied to improve thermal performance. Although heat transfer can be greatly enhanced by modifying the surface geometry, such modifications typically increase the pressure drop simultaneously \citep{AlamKim2018}. As a result, the ratio of the Stanton number $St$ to the friction coefficient $C_f$ is often reduced compared to that of a flat surface.

There have also been attempts to achieve dissimilar heat transfer enhancement by actively manipulating the flow field. Typical active control strategies include the introduction of spanwise anti-cyclonic vortices \citep{MotokiTsugawaShimizuKawahara2022, KuboKawaharaShimizu2021}, anti-cyclonic rotation \citep{Brethouwer2023}, and wall blowing and suction \citep{HigashiMamoriFukagata2011, HasegawaKasagi2011}. \citet{HasegawaKasagi2011} introduced a suboptimal control theory \citep{LeeKimChoi1998} to determine the spatio-temporal distribution of wall blowing and suction for achieving dissimilar heat transfer enhancement in a fully developed turbulent channel flow. Interestingly, the resulting control input is characterized by coherent streamwise travelling wave-like wall blowing and suction. Later, \citet{YamamotoHasegawaKasagi2013} applied optimal control theory to this problem and revealed the optimal wavelength and phase speed of the travelling wave. A systematic parametric survey was also conducted by \citet{KaithakkalKametaniHasegawa2020} to clarify the effects of the wavelength and the phase velocity on heat transfer and pressure-drop characteristics in turbulent flow configurations. In addition, \citet{HigashiMamoriFukagata2011} and \citet{KaithakkalKametaniHasegawa2021} demonstrated that such a control strategy is also effective in achieving dissimilar heat transfer enhancement in laminar channel flows at low Reynolds numbers. Although the effectiveness of travelling wave-like wall blowing and suction has been validated conceptually, a practical strategy to generate such a control input in real heat exchangers still requires further research. In particular, because active control requires actuators and additional power input to drive them, it is desirable to achieve similar effects in a passive manner.

It is known that certain wall roughnesses, such as riblets, induce streamwise travelling-wave-like disturbances in turbulent flow and consequently achieve the breakdown of the Reynolds analogy \citep{RezaeiJafariHoangArjomandi2025}. For example, \citet{GarciaMayoralJimenez2011} showed that spanwise rollers appear near a riblet surface due to the Kelvin-Helmholtz instability, which has also been confirmed through recent experiments \citep{AbuRowinDeshpandeWangKozulChungSandbergHutchins2025}. This is a primary reason for the breakdown of the drag-reduction effect of riblets when their spanwise spacing becomes large. Similar travelling wave-like disturbances have also been reported near porous walls. Owing to the high computational cost of resolving flows around complex porous structures, homogenized models have often been adopted to model flows over porous structures. For a turbulent channel flow between porous walls at a bulk Reynolds number of $Re_b = 5660$, \citet{JimenezUhlmannPinelliKawahara2001} employed a typical Darcy model to represent a porous medium by determining the wall-normal velocity based on the local pressure difference across the porous wall. The wall-normal fluid motion through the porous wall results in spanwise rollers and thereby a 40\% increase in the overall drag compared with a smooth wall. Such a homogenized model was also employed by \citet{MotokiTsugawaShimizuKawahara2022}, and the existence of spanwise rollers was attributed to Kelvin-Helmholtz instability across the porous wall. In addition, the spanwise rolls with blowing and suction are shown to be effective in breaking the similarity between momentum and heat transfer for bulk Reynolds numbers exceeding $10^4$.

More recently, there have been attempts to resolve flows around porous structures so as to clarify the underlying microscopic mechanisms of heat and momentum transport inside them. \citet{ChandesrisDHueppeMathieuJametGoyeau2013}, \citet{HabibiKhorasani2024} and \citet{Brethouwer2026} performed direct numerical simulations for turbulent flow over a cube-type porous medium, while \citet{KuwataSuga2016, KuwataSuga2017, KuwataSuga2024} adopted a lattice Boltzmann method to resolve turbulent flow over similar cube-type porous structures and more complex Kelvin-type porous structures. Turbulent heat transfer over a packed bed of spheres and rectangular rods is investigated by \citet{Jadidi2022} through pore-scale large eddy simulations. Their simulations also confirmed the presence of spanwise rollers travelling downstream, which intensify turbulent mixing near the porous wall and therefore enhance momentum and heat transfer.

The existing studies mentioned above commonly consider fully developed turbulent flows at relatively high Reynolds numbers above 2000. Meanwhile, in many engineering applications, such as air conditioning systems and cooling devices, the surface density of heat transfer surfaces has rapidly increased to dissipate heat more effectively in a smaller volume. Consequently, the Reynolds number becomes lower, so that the flow often remains laminar. In addition, with the miniaturization of modern heat exchangers, internal flows are often streamwise developing. It remains unclear whether travelling-wave-like disturbances can still be induced in such practical scenarios, where viscous effects become more dominant. To this end, we conduct pore-resolving simulations of a developing flow between parallel perforated plates to investigate the possibility of inducing travelling-wave-like disturbances at Reynolds numbers below 2000 and to clarify their impacts on heat transfer and pressure-drop characteristics, with particular focus on the dissimilarity between momentum and heat transfer.

The paper is organized as follows. The numerical method, configurations, and performance indices considered in the present study are introduced in \S\ref{sec:numerical_methods}. The global and local performance indices are summarized and discussed in \S\ref{sec:global_local_performances}. The effects of the travelling-wave characteristics on the performance indices are discussed in \S\ref{sec:flow_characteristics}. Finally, the conclusions of the present study are summarized in \S\ref{sec:conclusions}.

\section{Numerical methods and configurations}
\label{sec:numerical_methods}
\subsection{Problem description}
\label{subsec::problem}

\begin{figure*}[htbp]
\centering

\begin{subfigure}[t]{0.48\linewidth}
    \centering
    \caption{}
    \vspace{2mm}
    \includegraphics[width=\linewidth]{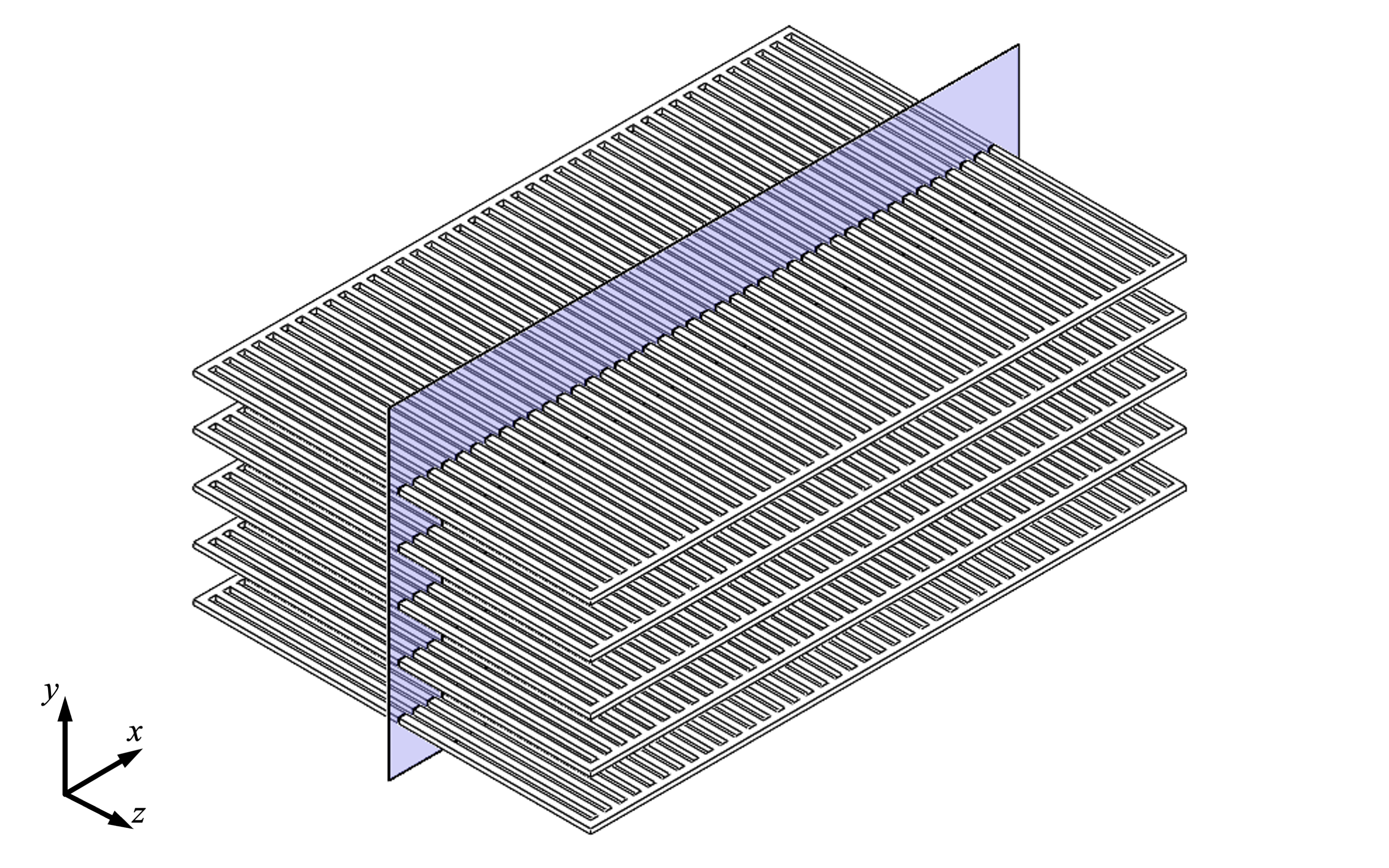}
\end{subfigure}
\hfill
\begin{subfigure}[t]{0.48\linewidth}
    \centering
    \caption{}
    \vspace{2mm}
    \includegraphics[width=\linewidth]{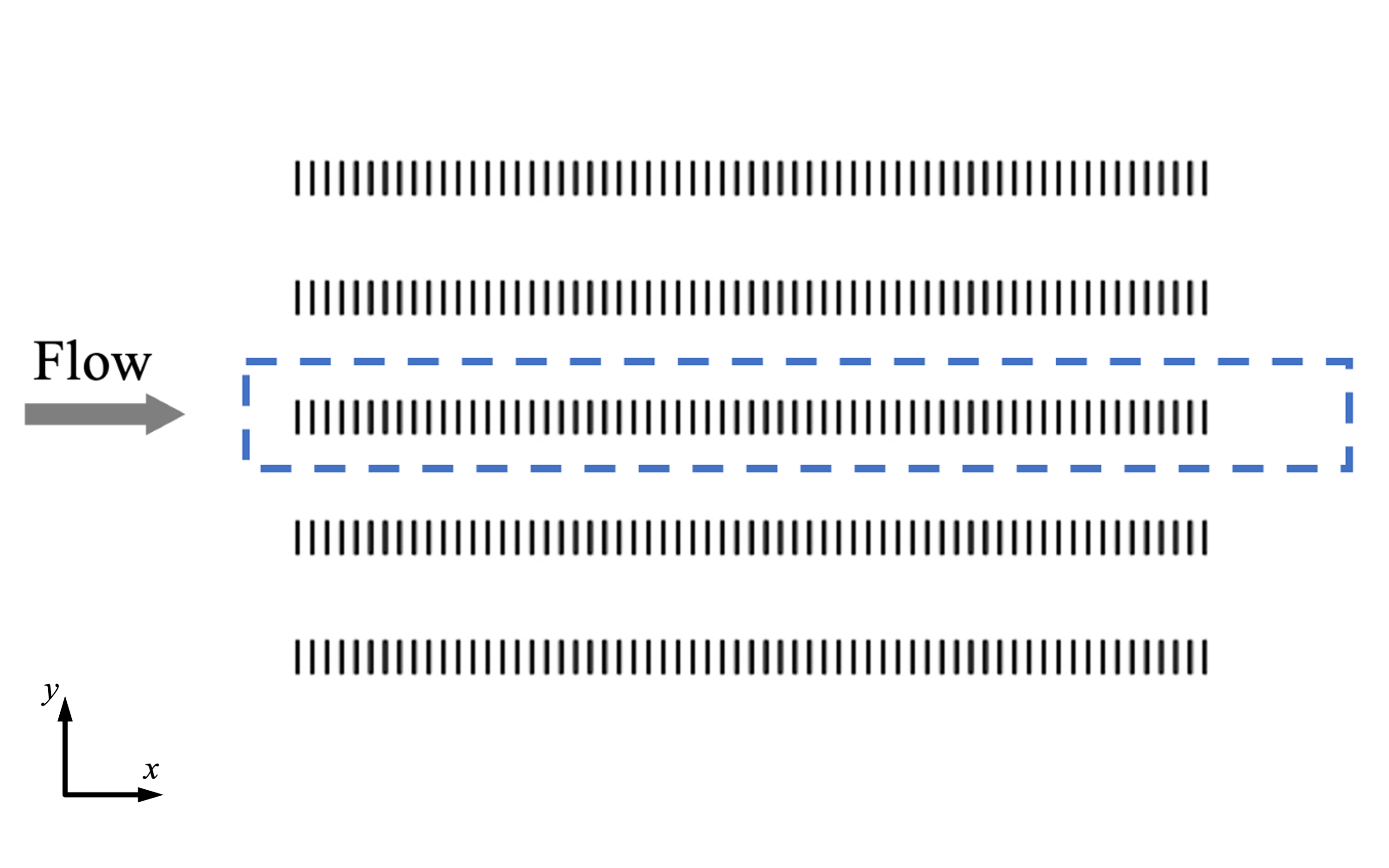}
\end{subfigure}

\vspace{4mm}

\begin{subfigure}[t]{0.98\linewidth}
    \centering
    \caption{}
    \vspace{2mm}
    \includegraphics[width=\linewidth]{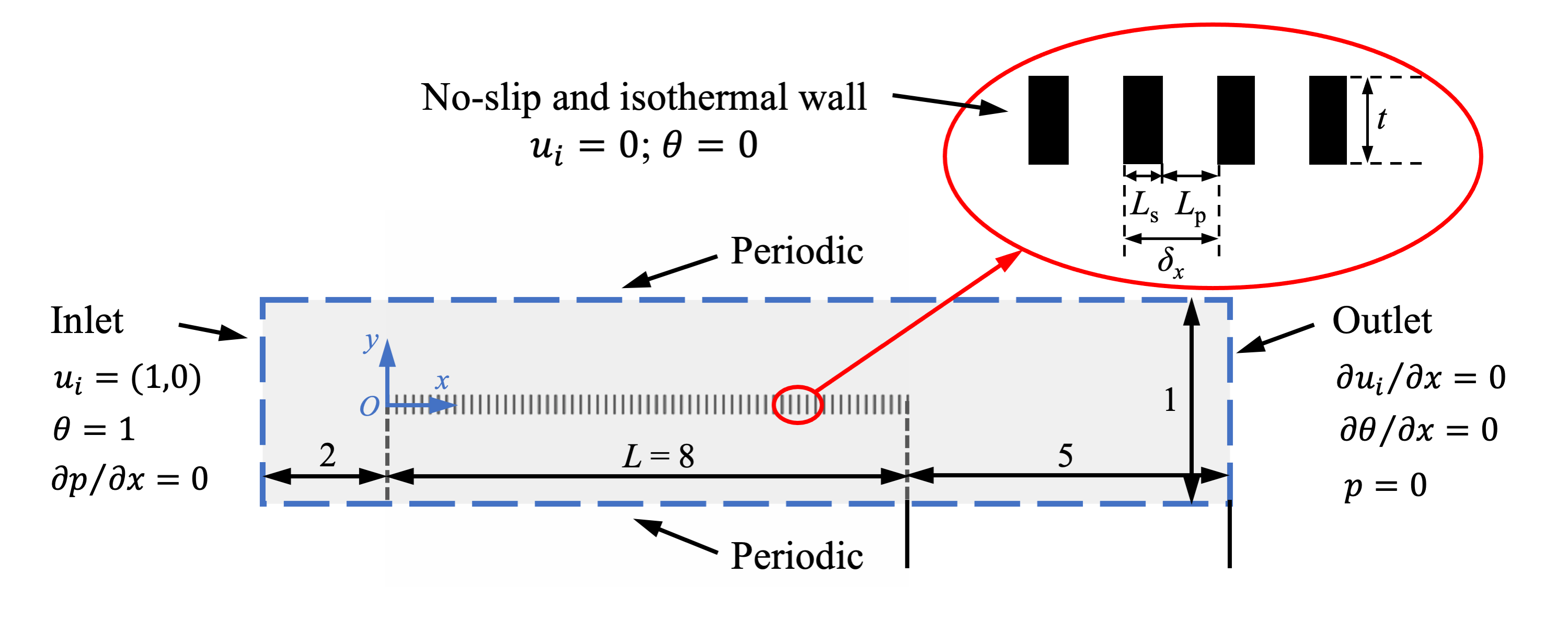}
\end{subfigure}

\caption{Schematic diagram of computational domain with the perforated plate: (a) Schematic of multiple parallel perforated plates, (b) one periodic unit, and (c) 2D computational domain and the coordinate system considered in the present study.}
\label{fig1}
\end{figure*}

The perforated plates that have been adopted in \citet{Shahzad2022} and \citet{Hoang2024} to modulate near-wall turbulence, is introduced in the present study. We consider a fluid flow along multiple parallel perforated plates with a finite length $L$, as shown in Fig.~\ref{fig1}(a). To simplify the numerical simulations, we extract a single unit of the two-dimensional computational domain with periodic conditions at the top and bottom boundaries, as shown in Figs.~\ref{fig1}(b) and \ref{fig1}(c). The streamwise and wall-normal coordinates are denoted as $x$ and $y$, respectively. Throughout this study, dimensional quantities are denoted with a superscript *, whereas dimensionless quantities are expressed without a superscript. All dimensionless quantities are normalized by the vertical spacing $H^*$ of two adjacent plates, the inlet streamwise velocity $u_\mathrm{in}^*$, and the temperature difference between the inflow and the wall $\theta_{\mathrm{in}}^*=T_{\mathrm{in}}^*-T_{\mathrm{w}}^*$. The solid temperature $T_w^*$ is assumed to be isothermal by neglecting thermal conduction inside the solid. This ideal boundary condition is chosen so that the boundary conditions for the velocity and temperature fields become similar, which is useful for understanding their intrinsic differences \citep{HasegawaKasagi2011}.
The streamwise and wall-normal dimensions of the present computational domain are $[-2, 13]\times[-0.5, 0.5]$, where the origin of the coordinate system is located at the middle of the leading edge of the perforated plate, as shown in Fig.~\ref{fig1}(c). The outlet is set at 5 unit lengths downstream of the trailing edge of the perforated plates to eliminate the influence of the outlet boundary condition. In the present study, the total length and the thickness of the perforated plate are fixed and set as $L = 8$ and $t = 0.1$, respectively. Uniform inflow velocity and temperature of $\boldsymbol{u}_\mathrm{in}^* = (u_\mathrm{in}^*, v_\mathrm{in}^*) = (1.0, 0.0)$ and $\theta_{\mathrm{in}}^* = 1.0$ are given at the inlet, and zero-gradient conditions are given for the velocity and temperature at the outlet. As for pressure, the reference zero value is fixed at the outlet, and the zero-gradient condition is given at the inlet. We assume that the flow is two-dimensional and incompressible and that all physical properties are independent of temperature, so that the temperature is treated as a passive scalar. The resulting governing equations of the velocity and thermal fields are the following momentum, continuity, and energy transport equations:
\begin{equation}
\frac{\partial u_i}{\partial t}
+
\frac{\partial (u_j u_i)}{\partial x_j}
=
-\frac{\partial p}{\partial x_i}
+
\frac{1}{Re}
\frac{\partial^2 u_i}{\partial x_j^2},
\end{equation}

\begin{equation}
\frac{\partial u_i}{\partial x_i}
=
0,
\end{equation}

\begin{equation}
\frac{\partial \theta}{\partial t}
+
\frac{\partial (u_j \theta)}{\partial x_j}
=
\frac{1}{Re\,Pr}
\frac{\partial^2 \theta}{\partial x_j^2},
\end{equation}
where $u_i$ represents the velocity component in the $i$-th direction, whilst $t$, $p$ and $\theta$ denote time, pressure, and temperature, respectively. The Reynolds number and the Prandtl number are respectively defined as $Re = u_\mathrm{in}^* H^*/ \nu ^*$ and $Pr = \nu ^*/\alpha ^*$, where $\nu^*$ and $\alpha^*$ represent the kinematic viscosity and thermal diffusivity of the fluid. We set $Pr = 0.7$, assuming that the working fluid is air under standard conditions. The Reynolds number is varied over $Re = 500$, 1000 and 1500.

In the present simulations, the flow field around the microscopic structures of the perforated plate is directly resolved without introducing a homogenized model. We arrange solid and pore regions periodically along the $x$ direction. 
Within each case, the horizontal lengths of the pores and solid blocks are uniform and denoted by $L_\mathrm{p}$ and $L_\mathrm{s}$, respectively, as illustrated in the inset of Fig. 1(c). Throughout the present study, the value of $L_\mathrm{s}$ is fixed as 0.01, while the values of $L_\mathrm{p}$ are systematically varied from 0 to 0.10, corresponding to the porosity $L_\mathrm{p}/(L_\mathrm{s} + L_\mathrm{p})$ ranging from 0 to 0.91. Note that the case with $L_\mathrm{p}/L_\mathrm{s} = 0$ corresponds to the reference case of a solid flat plate with no pores, i.e., zero permeability. The geometries of all the cases are summarized in Table~\ref{tab1}, where $S/S_0$ represents the ratio of the total surface area of the perforated plate to that of the zero-permeability solid plate. In the present study, each case is named by the ratio $L_\mathrm{p}/L_\mathrm{s}$ and the Reynolds number. For instance, Case r7Re1000 represents the case with $L_\mathrm{p}/L_\mathrm{s} = 7$ at a Reynolds number of 1000. In total, 33 simulation cases corresponding to 11 geometries at three Reynolds numbers, i.e., Cases r0Re500--r10Re500, r0Re1000--r10Re1000, and r0Re1500--r10Re1500, are considered in the present study.

\begin{table}
\caption{Geometries of perforated plates.}
\label{tab1}
\begin{tabular}{cccccccccccc}
\textbf{Case}
& \textbf{r0} & \textbf{r1} & \textbf{r2} & \textbf{r3} & \textbf{r4}
& \textbf{r5} & \textbf{r6} & \textbf{r7} & \textbf{r8} & \textbf{r9} & \textbf{r10} \\
$L_\mathrm{p}^{*}$
& 0 & 0.01 & 0.02 & 0.03 & 0.04 & 0.05 & 0.06 & 0.07 & 0.08 & 0.09 & 0.10 \\

$L_\mathrm{s}^{*}$
& 0.01 & 0.01 & 0.01 & 0.01 & 0.01 & 0.01 & 0.01 & 0.01 & 0.01 & 0.01 & 0.01 \\

$L_\mathrm{p}/L_\mathrm{s}$
& solid & 1 & 2 & 3 & 4 & 5 & 6 & 7 & 8 & 9 & 10 \\

$S/S_0$
& 1.00 & 5.50 & 3.66 & 2.75 & 2.20 & 1.84 & 1.57 & 1.38 & 1.22 & 1.10 & 1.00 \\
\end{tabular}
\centering
\end{table}

\subsection{Level-set functions for representing perforated plates}
\label{subsec::level_set_function}

\begin{figure*}[htbp]
  \centerline{\includegraphics[width=3in]{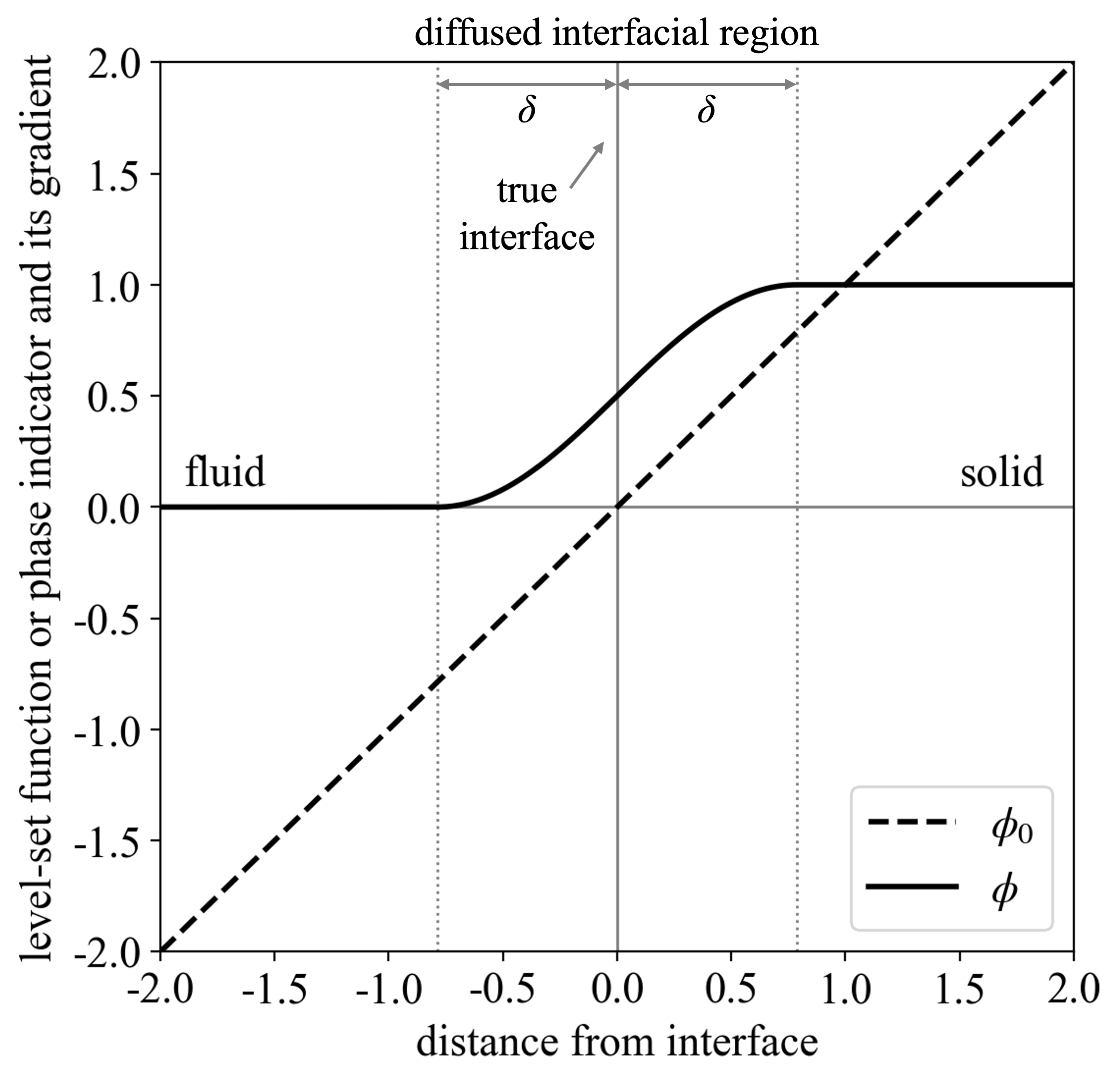}}%
  \caption{Profiles of the level-set function $\phi_0$ and the phase indicator $\phi$ near the solid surface.}
\label{fig2}
\end{figure*}

To represent the perforated plates in the computational domain, we introduce a level-set function $\phi_0$, which allows the solid surface to be embedded in a Cartesian grid system \citep{LiuHasegawa2023, ChenKumarKametaniHasegawa2024}. The level-set function is a signed distance function from the fluid-solid interface. As shown in Fig.~\ref{fig2}, the value of the level-set function is defined to be negative in the fluid region, zero on the interface, and positive in the solid region. Then, the level-set function can be transformed into the phase indicator function $\phi$ ($\phi = 0$ for fluid and $\phi = 1$ for solid) as follows:

\begin{equation}
\phi =
\begin{cases}
0,
& \phi_0 < -\delta
\\[6pt]
\dfrac{1}{2}
\left[
\sin\left(
\dfrac{\pi}{2}
\dfrac{\phi_0}{\delta}
\right)
+ 1
\right],
& -\delta \leq \phi_0 \leq \delta
\\[6pt]
1,
& \phi_0 > \delta
\end{cases},
\label{level_set_function}
\end{equation}
where $\delta$ is the half-length of an interfacial region ($0 < \phi < 1$) between the fluid and solid regions. Specifically, the fluid-solid interface is represented as a diffuse interface with a thickness of $2\delta$ in the present study. As indicated in Eq.~(\ref{level_set_function}) and Fig.~\ref{fig2}, the phase indicator $\phi$ is zero in the fluid region and unity in the solid region, whereas it varies smoothly from zero to one within the interfacial region defined above. 

\subsection{Volume penalization method and governing equations}
\label{subsec::volume_penalization_method}

In recent years, immersed boundary methods (IBMs) have attracted much attention for solving flow and heat transfer problems in complex geometries because they allow arbitrary geometries to be embedded in a Cartesian grid system \citep{Peskin2002, ChoiOberoiEdwardsRosati2007}. This unique feature of IBMs avoids the tedious task of body-fitted grid generation. In the present study, the volume penalization method (VPM) \citep{KadochKolomenskiyAngotSchneider2012, LiuHasegawa2023}, one of the diffuse-interface IBMs, is introduced to perform two-dimensional simulations for the flow through multiple parallel perforated plates. Based on the VPM, the boundary conditions on the complex fluid-solid interfaces are realized by artificial source terms so as to solve the velocity and temperature fields inside the fluid and solid phases in a unified manner. Accordingly, the incompressible Navier-Stokes, continuity, and energy equations solved in the VPM are given as

\begin{equation}
\frac{\partial u_i}{\partial t}
+
\frac{\partial (u_j u_i)}{\partial x_j}
=
-\frac{\partial p}{\partial x_i}
+
\frac{1}{Re}
\frac{\partial^2 u_i}{\partial x_j^2}
-
\eta \phi u_i,
\label{VPM_momentum}
\end{equation}

\begin{equation}
\frac{\partial u_i}{\partial x_i}
=
0,
\end{equation}

\begin{equation}
\frac{\partial \theta}{\partial t}
+
\frac{\partial (u_j \theta)}{\partial x_j}
=
\frac{1}{Re\,Pr}
\frac{\partial^2 \theta}{\partial x_j^2}
-
\eta \phi \theta,
\label{VPM_energy}
\end{equation}

The final terms on the RHS of Eqs.~(\ref{VPM_momentum}) and (\ref{VPM_energy}) are artificial body force and heat sink terms that impose the no-slip and isothermal boundary conditions on the surface of each solid region in the plate \citep{ChenKumarKametaniHasegawa2024}. These two terms are zero in the fluid region where $\phi = 0$, and act only to make the velocity and thermal fields zero inside the interfacial and solid regions where the phase indicator $\phi$ is non-zero. By assigning a sufficiently large value to the penalization coefficients $\eta$, i.e., $\eta = 10^6$, the no-slip and isothermal boundary conditions can be satisfied properly \citep{ChenKumarKametaniHasegawa2024}. As presented in \citet{KadochKolomenskiyAngotSchneider2012}, the present VPM scheme has first-order accuracy with respect to the grid size. As will be discussed in \S\ref{subsec::grid_test}, the grid convergence test was conducted to confirm that further grid refinement does not affect the results and conclusions drawn from the present simulation.
In this study, the simulation is conducted using the open-source CFD toolbox \texttt{OpenFOAM} (version 8). For solving Eqs.~(\ref{VPM_momentum}) to (\ref{VPM_energy}), we apply the finite volume method, where the transient term is discretized by the first-order Euler implicit scheme, whereas the convection and diffusion terms are discretized by a first-order Gauss upwind scheme and a second-order Gauss linear scheme, respectively.

\subsection{Indices of thermal-hydraulic performance}
\label{subsec::indices}

In the VPM simulations, the conservation of the total momentum over the computational domain allows the total drag force acting in the streamwise direction on the solid surface, $F_D$, to be calculated by the volume integral of the artificial body force in the streamwise direction as

\begin{equation}
F_D
=
\frac{1}{T}
\int_{0}^{T}
\int_{V}
\eta \phi u_x
\, dV \, dt.
\end{equation}
Similarly, considering the conservation of the total heat within the computational domain, the total heat flux from the fluid to the solid, $Q$, can be calculated by the volume integral of the artificial heat sink term in Eq.~(\ref{VPM_energy}) as

\begin{equation}
Q
=
\frac{1}{T}
\int_{0}^{T}
\int_{V}
\eta \phi \theta
\, dV \, dt.
\end{equation}

All the time-averaged quantities are integrated after the flow field reaches an equilibrium state over a sufficiently long period $T$ to obtain converged statistics. More specifically, $T$ is set to 30 in the present study, which roughly corresponds to four times the flow-through time for the total plate length $L$. Then, the total area on both sides of a zero-permeability solid plate for a unit spanwise length perpendicular to the $x-y$ plane, $2L$, is adopted as a reference wetted area to compare the total drag and heat transfer among perforated plates with different porosities. Thus, the difference in the surface area is included in the evaluations of the drag coefficients and the heat transfer rate. As shown in Table~\ref{tab1}, the actual surface area is larger than that of the zero-permeability solid plate in the present configuration. More specifically, as $L_\mathrm{p}/L_\mathrm{s}$ (or the porosity) decreases, the actual surface area increases. The corresponding drag force $f_D$ and heat flux $q$ per unit reference wetted area are defined as

\begin{equation}
f_D
=
\frac{F_D}{2L}
=
\frac{1}{2LT}
\int_{0}^{T}
\int_{V}
\eta \phi u_x
\, dV \, dt,
\end{equation}

\begin{equation}
q
=
\frac{Q}{2L}
=
\frac{1}{2LT}
\int_{0}^{T}
\int_{V}
\eta \phi \theta
\, dV \, dt,
\end{equation}

Accordingly, for the evaluation of the global heat transfer and pressure-drop characteristics, as well as their ratio, the Stanton number $St^G$, the friction factor $C_f^G$, and the analogy factor $A^G$ are calculated as 

\begin{equation}
St^{G}
=
\frac{h^{*}}
{\rho^{*} c_p^{*} u_{\mathrm{in}}^{*}}
=
\frac{q}
{u_{\mathrm{in}} \theta_{\mathrm{LMTD}}},
\label{St_G}
\end{equation}

\begin{equation}
C_f^{G}
=
\frac{
2F_D^{*}
}{
\rho^{*}
{u_{\mathrm{in}}^{*}}^{2}
S_{r0}^{*}
}
=
\frac{2f_D}{u_{\mathrm{in}}^{2}},
\label{Cf_G}
\end{equation}

\begin{equation}
A^{G}
=
\frac{St^{G}}{C_f^{G}}.
\label{A_G}
\end{equation}
Here, $h^*$, $\rho^*$, $c_p^*$, $S_{r0}^*$, and $F_D^*$ denote the heat transfer coefficient, the density of the fluid, the specific heat capacity of the fluid, the surface area of a zero-permeability solid plate, and drag force acting on the plate, respectively. The logarithmic mean temperature difference (LMTD) $\theta_\mathrm{LMTD}$ is defined as follows,

\begin{equation}
\theta_{\mathrm{LMTD}}
=
\frac{
\theta_{\mathrm{bulk,in}}
-
\theta_{\mathrm{bulk,out}}
}{
\ln \theta_{\mathrm{bulk,in}}
-
\ln \theta_{\mathrm{bulk,out}}
},
\end{equation}

\begin{equation}
\theta_{\mathrm{bulk,in}}
=
\frac{
\dfrac{1}{T}
\int_{0}^{T}
\int_{-1/2}^{1/2}
u(x_{\mathrm{in}},y,t)
\theta(x_{\mathrm{in}},y,t)
\, dy \, dt
}{
\dfrac{1}{T}
\int_{0}^{T}
\int_{-1/2}^{1/2}
u(x_{\mathrm{in}},y,t)
\, dy \, dt
},
\end{equation}

\begin{equation}
\theta_{\mathrm{bulk,out}}
=
\frac{
\dfrac{1}{T}
\int_{0}^{T}
\int_{-1/2}^{1/2}
u(x_{\mathrm{out}},y,t)
\theta(x_{\mathrm{out}},y,t)
\, dy \, dt
}{
\dfrac{1}{T}
\int_{0}^{T}
\int_{-1/2}^{1/2}
u(x_{\mathrm{out}},y,t)
\, dy \, dt
},
\end{equation}
Here, $\theta_\mathrm{bulk,in}$ and $\theta_\mathrm{bulk,out}$ are the time-averaged bulk mean temperatures at the leading and trailing edges of the plate.
In addition to the global indices in Eqs.~(\ref{St_G})--(\ref{A_G}), we also introduce the local bulk mean temperature $\theta_\mathrm{bulk}^L(x)$, heat flux $q^L(x)$, and streamwise drag $f_D^L(x)$ \citep{IncroperaDeWitt2002, KaysCrawfordWeigand2010} for local analysis,

\begin{equation}
\theta_{\mathrm{bulk}}^{L}(x)
=
\frac{1}{\delta x \, T}
\int_{0}^{T}
\int_{x-\delta x/2}^{x+\delta x/2}
\int_{-1/2}^{1/2}
u(x,y,t)\theta(x,y,t)
\, dy \, dx \, dt
\end{equation}

\begin{equation}
u_{\mathrm{bulk}}^{L}(x)
=
\frac{1}{\delta x \, T}
\int_{0}^{T}
\int_{x-\delta x/2}^{x+\delta x/2}
\int_{-1/2}^{1/2}
u(x,y,t)
\, dy \, dx \, dt,
\end{equation}

\begin{equation}
q^{L}(x)
=
\frac{1}{2 \delta x \, T}
\int_{0}^{T}
\int_{x-\delta x/2}^{x+\delta x/2}
\int_{-1/2}^{1/2}
\eta \phi \theta(x,y)
\, dy \, dx \, dt,
\end{equation}

\begin{equation}
f_D^{L}(x)
=
\frac{1}{2 \delta x \, T}
\int_{0}^{T}
\int_{x-\delta x/2}^{x+\delta x/2}
\int_{-1/2}^{1/2}
\eta \phi u_x(x,y)
\, dy \, dx \, dt,
\end{equation}
where $\delta x = L_\mathrm{p} + L_\mathrm{s}$ is the streamwise length of one periodic unit cell of a porous structure (see, the inset of Fig.~\ref{fig1}(c)). Thus, local averaging is performed over one microscopic periodic unit of the porous structure. Accordingly, the local Stanton number $St^L(x)$, friction factor $C_f^L(x)$, and analogy factor $A^L(x)$ are defined to evaluate the local performance as follows:

\begin{equation}
St^{L}(x)
=
\frac{
q^{L}(x)
}{
u_{\mathrm{bulk}}^{L}(x)
\theta_{\mathrm{bulk}}^{L}(x)
},
\end{equation}

\begin{equation}
C_f^{L}(x)
=
\frac{
2f_D^{L}(x)
}{
\left(
u_{\mathrm{bulk}}^{L}(x)
\right)^2
},
\end{equation}

\begin{equation}
A^{L}(x)
=
\frac{
2St^{L}(x)
}{
C_f^{L}(x)
}.
\end{equation}

\subsection{Grid dependence test}
\label{subsec::grid_test}

\begin{figure*}[htbp]
  \centerline{\includegraphics[width=5in]{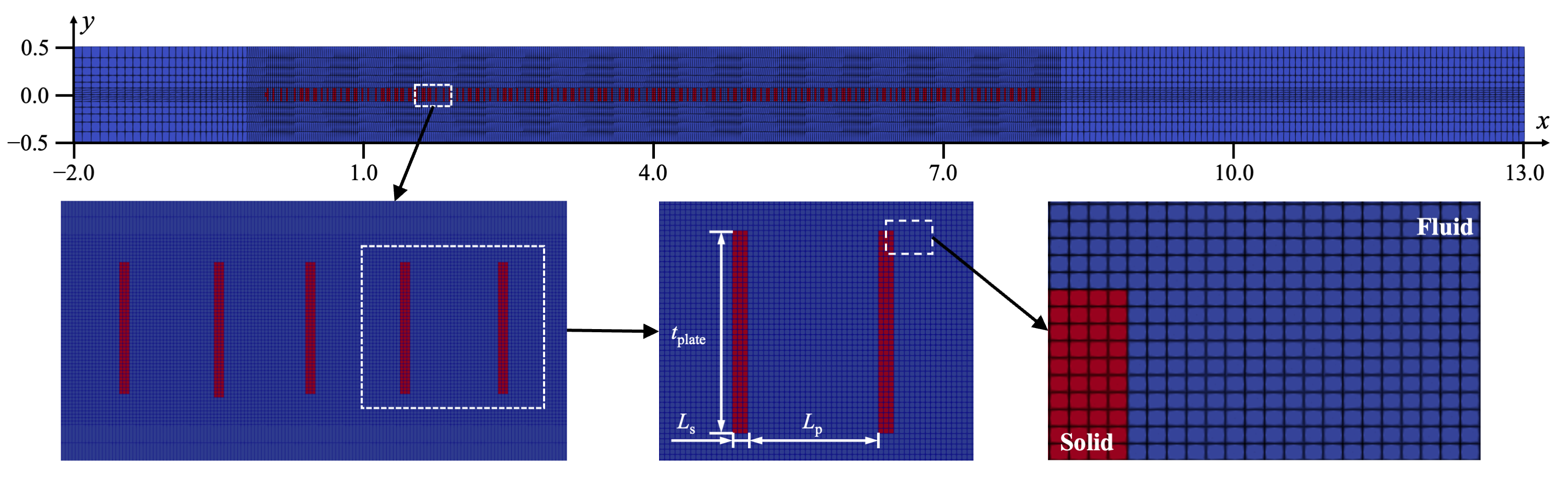}}%
  \caption{Computational domain and Cartesian grid system.}
\label{fig3}
\end{figure*}

We generate a Cartesian grid system throughout the computational domain, as shown in Fig.~\ref{fig3}. The grids around the perforated plate are refined to accurately capture the flow and thermal fields in the near-wall regions. Note that the actual grid resolution is two times finer than that shown in Fig.~\ref{fig3}, and the resolution is summarized in Table~\ref{tab2}.

\begin{table}
\caption{Computational mesh and the resultant heat transfer and pressure-drop characteristics in the grid convergence test for Case r6Re1500.}
\label{tab2}
\begin{tabular}{ccccc}
\textbf{Items}
& \makecell{\textbf{Number of mesh cells within $L_\mathrm{s}$ and $t$}}
& \makecell{\textbf{Total number of mesh cells}}
& $\boldsymbol{St^G}$
& $\boldsymbol{C_f^G}$ \\
Mesh-1 & $4 \times 40$   & $0.420 \times 10^6$ & 0.0404 & 0.0889 \\
Mesh-2 & $6 \times 60$   & $0.945 \times 10^6$ & 0.0307 & 0.0673 \\
Mesh-3 & $8 \times 80$   & $1.680 \times 10^6$ & 0.0272 & 0.0580 \\
Mesh-4 & $10 \times 100$ & $2.625 \times 10^6$ & 0.0264 & 0.0562 \\
Mesh-5 & $12 \times 120$ & $3.780 \times 10^6$ & 0.0262 & 0.0559 \\
\end{tabular}
\centering
\end{table}

To clarify the effects of the grid resolution on the numerical results, the grid resolution is systematically refined in Case r6Re1500. Specifically, the number of mesh cells across a single solid block ($L_\mathrm{s}$) in the $x$ direction and the thickness of the solid block in the $y$ direction are changed from 4 to 10 and from 40 to 100, respectively (see Fig.~\ref{fig3}). The obtained $St^G$ and $C_f^G$ using the different mesh resolutions are summarized in Table~\ref{tab2}. Once the total number of mesh cells reaches 1.680 $\times$ 10$^6$, which corresponds to 8 cells for a single solid block along the $x$ direction, further grid refinement has a small impact on both $St^G$ and $C_f^G$. As shown in Table~\ref{tab2}, further grid refinement to 2.625 $\times$ 10$^6$ mesh cells only results in the relative difference in $St^G$ and $C_f^G$ of 3.03\% and 3.28\%, respectively.
Thus, we employ the Cartesian mesh with 1.68 $\times$ 10$^6$ cells in the following simulations. Using a grid of the same size, the accuracy of the present VPM is also validated through a spatially developing channel flow under an isothermal wall condition. Its details are summarized in Appendix~\ref{app:verification}.

\section{Global and local performance}
\label{sec:global_local_performances}
\subsection{Global performance}
\label{subsec:global_performance}
\begin{figure*}[htbp]
\centering

\begin{subfigure}[t]{0.48\linewidth}
    \centering
    \caption{}
    \vspace{2mm}
    \includegraphics[width=\linewidth]{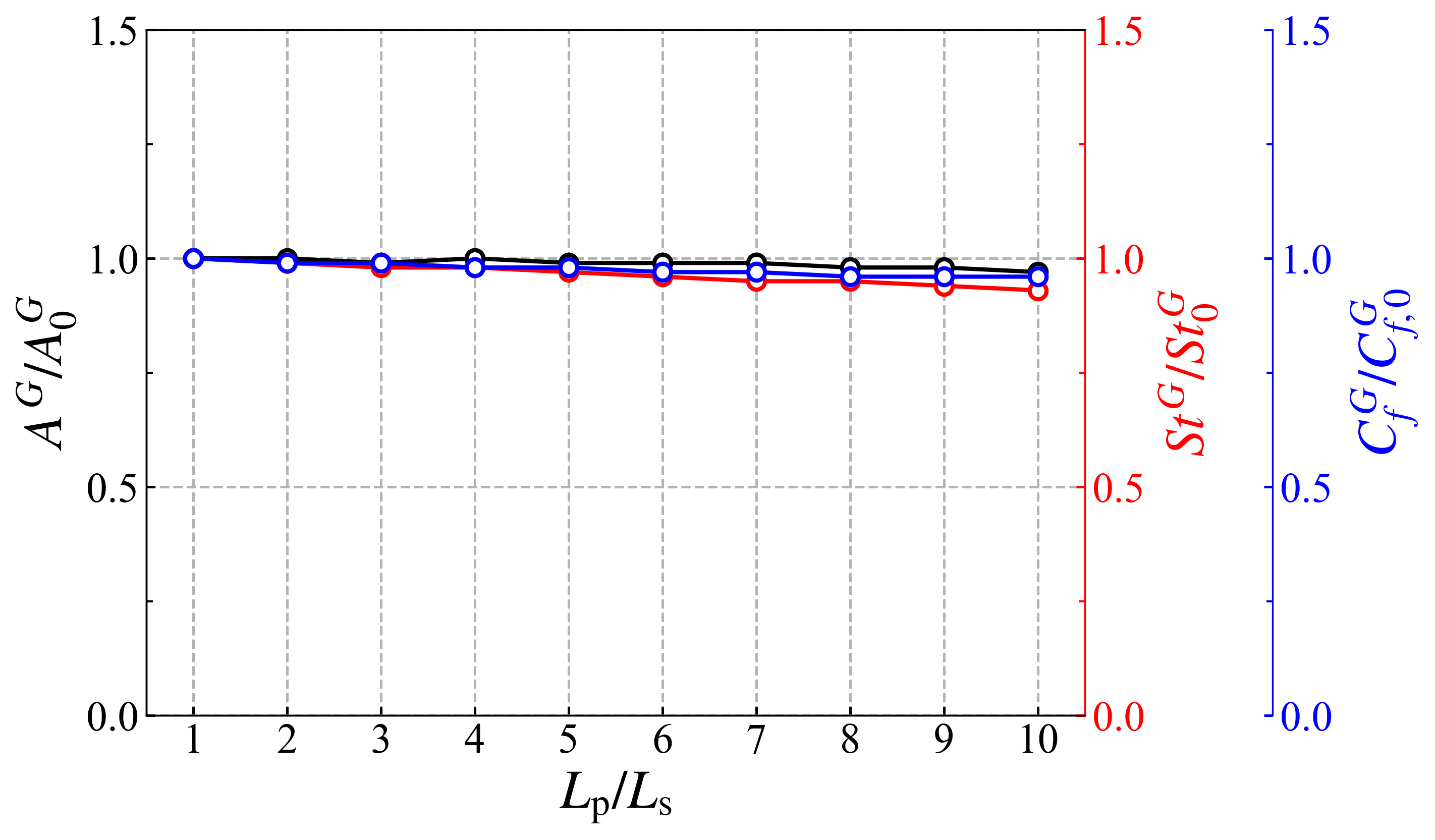}
\end{subfigure}
\hfill
\begin{subfigure}[t]{0.48\linewidth}
    \centering
    \caption{}
    \vspace{2mm}
    \includegraphics[width=\linewidth]{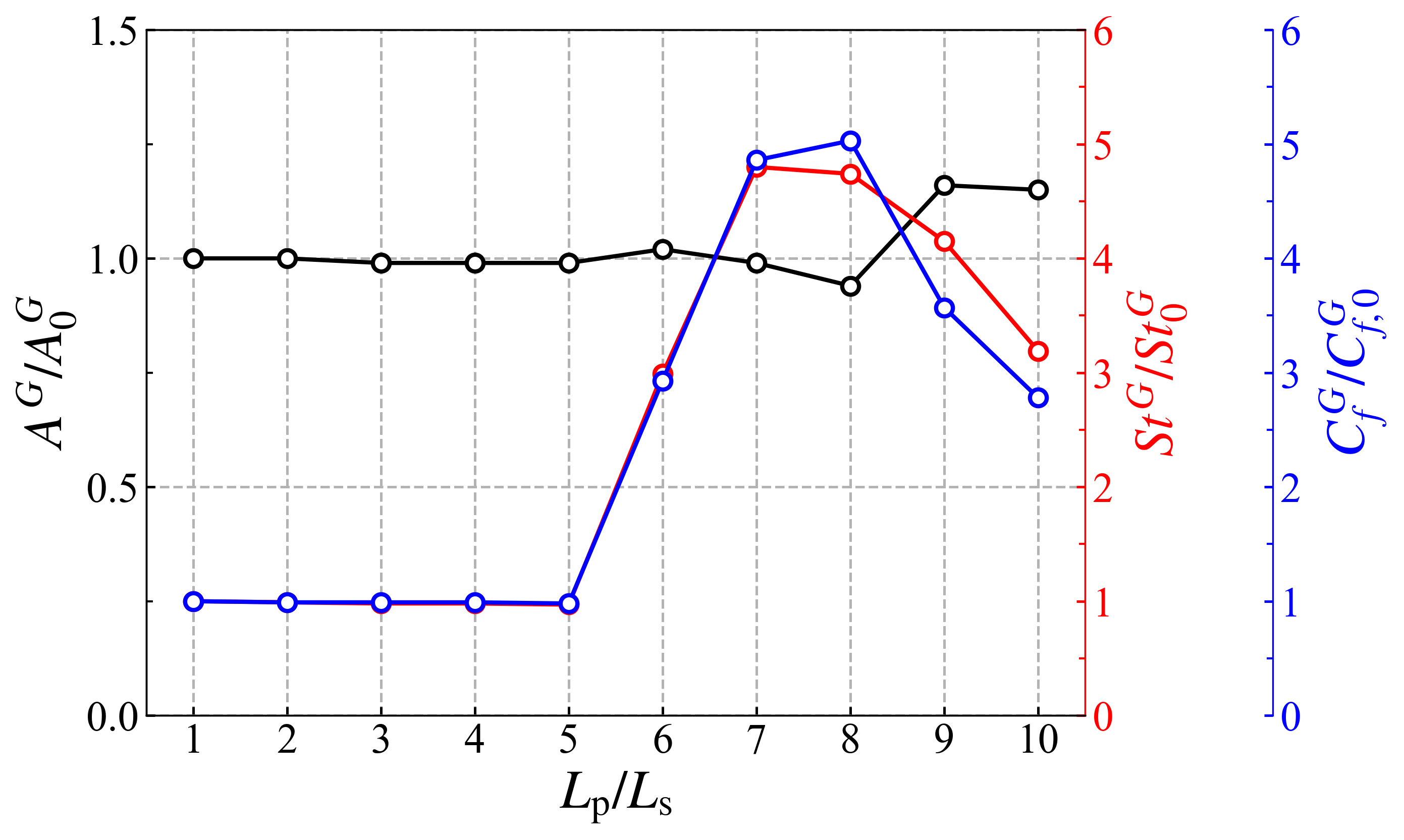}
\end{subfigure}

\vspace{4mm}

\begin{subfigure}[t]{0.48\linewidth}
    \centering
    \caption{}
    \vspace{2mm}
    \includegraphics[width=\linewidth]{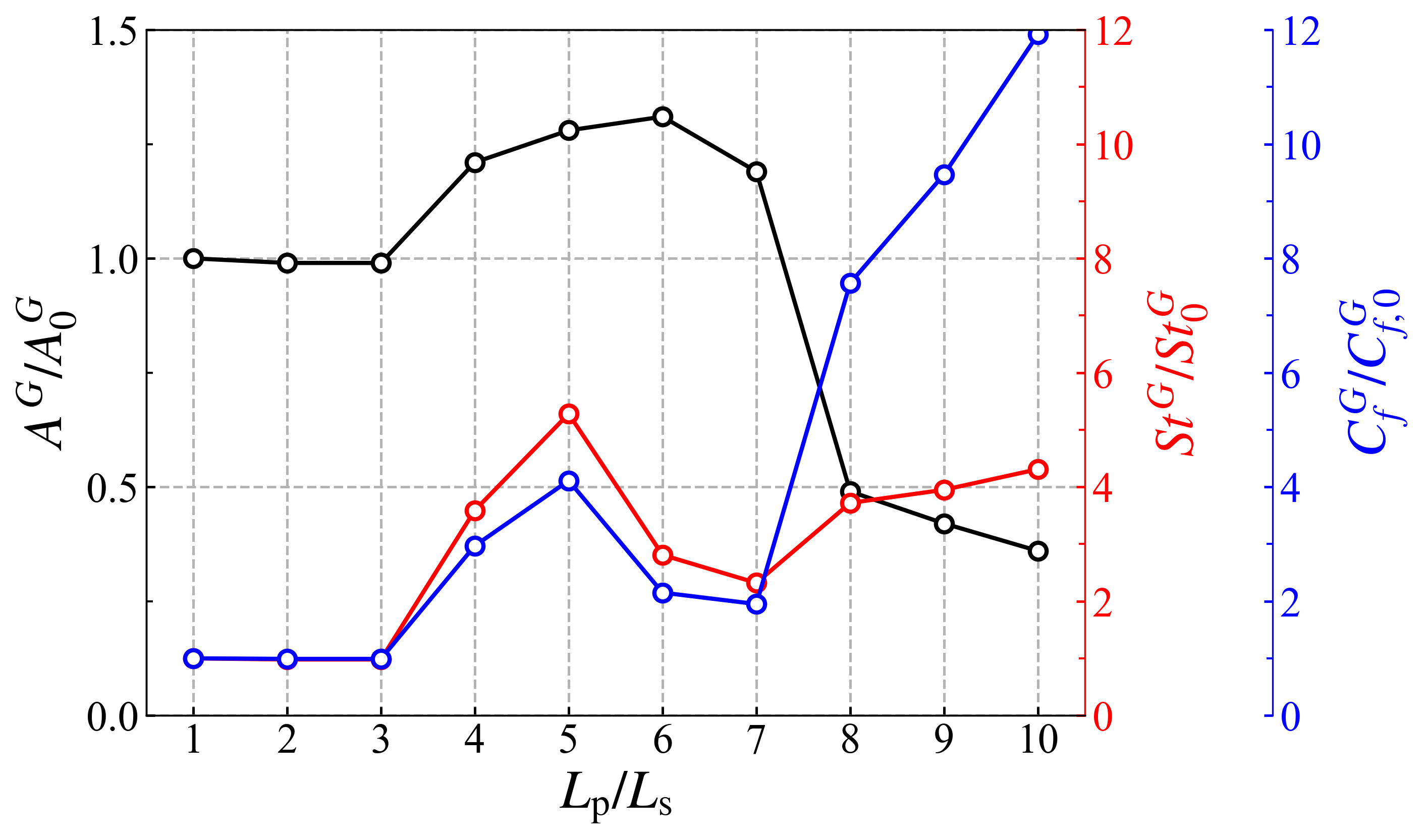}
\end{subfigure}

\caption{$A^G/A_0^G$, $St^G/St_0^G$, and $C_f^G/C_{f,0}^G$ against $L_\mathrm{p}/L_\mathrm{s}$ at (a) $Re = 500$, (b) $Re = 1000$, and (c) $Re = 1500$.}
\label{fig4}
\end{figure*}

The global Stanton number $St^G$, global friction factor $C_f^G$, and global analogy factor $A^G$ of the perforated plates with various geometries at Reynolds numbers of 500, 1000, and 1500 are summarized in Tables~\ref{tab3}--\ref{tab5}, respectively. 
The tables also present the global performance indices normalized by the zero-permeability solid plate values, i.e., $A^G/A_0^G$, $St^G/St_0^G$, and $C_f^G/C_{f,0}^G$, where subscript 0 denotes the reference case at the corresponding Reynolds number.
To illustrate the general trend observed in Tables~\ref{tab3}--\ref{tab5}, Figure~\ref{fig4} plots $A^G/A_0^G$, $St^G/St_0^G$, and $C_f^G/C_{f,0}^G$ against $L_\mathrm{p}/L_\mathrm{s}$ for different Reynolds numbers.

\begin{table}
\caption{Global performance indices for the perforated plate with different geometries $(L_\mathrm{p}/L_\mathrm{s})$ at $Re = 500$.}
\label{tab3}
\begin{adjustbox}{max width=\textwidth}
\begin{tabular}{cccccccccccc}
$\boldsymbol{L_\mathrm{p}/L_\mathrm{s}}$
& \textbf{0} & \textbf{1} & \textbf{2} & \textbf{3} & \textbf{4} & \textbf{5} & \textbf{6} & \textbf{7} & \textbf{8} & \textbf{9} & \textbf{10} \\
$St^G$
& 0.0195 & 0.0195 & 0.0193 & 0.0192 & 0.0191 & 0.0189 & 0.0187 & 0.0186 & 0.0185 & 0.0183 & 0.0181 \\

$St^G/St_0^G$
& 1.00 & 1.00 & 0.99 & 0.98 & 0.98 & 0.97 & 0.96 & 0.95 & 0.95 & 0.94 & 0.93 \\

$C_f^G$
& 0.0555 & 0.0555 & 0.0550 & 0.0549 & 0.0546 & 0.0542 & 0.0539 & 0.0537 & 0.0535 & 0.0532 & 0.0531 \\

$C_f^G/C_{f,0}^G$
& 1.00 & 1.00 & 0.99 & 0.99 & 0.98 & 0.98 & 0.97 & 0.97 & 0.96 & 0.96 & 0.96 \\

$A^G$
& 0.703 & 0.703 & 0.702 & 0.699 & 0.700 & 0.697 & 0.694 & 0.693 & 0.692 & 0.688 & 0.682 \\

$A^G/A_0^G$
& 1.00 & 1.00 & 1.00 & 0.99 & 1.00 & 0.99 & 0.99 & 0.99 & 0.98 & 0.98 & 0.97 \\
\end{tabular}
\end{adjustbox}
\centering
\end{table}

\begin{table}
\caption{Global performance indices for the perforated plate with different geometries $(L_\mathrm{p}/L_\mathrm{s})$ at $Re = 1000$.}
\label{tab4}
\begin{adjustbox}{max width=\textwidth}
\begin{tabular}{cccccccccccc}
$\boldsymbol{L_\mathrm{p}/L_\mathrm{s}}$
& \textbf{0} & \textbf{1} & \textbf{2} & \textbf{3} & \textbf{4} & \textbf{5} & \textbf{6} & \textbf{7} & \textbf{8} & \textbf{9} & \textbf{10} \\
$St^G$
& 0.0124 & 0.0124 & 0.0123 & 0.0122 & 0.0121 & 0.0120 & 0.0371 & 0.0595 & 0.0588 & 0.0514 & 0.0396 \\

$St^G/St_0^G$
& 1.00 & 1.00 & 0.99 & 0.98 & 0.98 & 0.97 & 2.99 & 4.80 & 4.74 & 4.15 & 3.19 \\

$C_f^G$
& 0.0348 & 0.0348 & 0.0345 & 0.0344 & 0.0343 & 0.0410 & 0.1021 & 0.1693 & 0.1752 & 0.1243 & 0.0967 \\

$C_f^G/C_{f,0}^G$
& 1.00 & 1.00 & 0.99 & 0.99 & 0.99 & 0.98 & 2.93 & 4.86 & 5.03 & 3.57 & 2.78 \\

$A^G$
& 0.713 & 0.713 & 0.713 & 0.709 & 0.706 & 0.704 & 0.727 & 0.703 & 0.671 & 0.827 & 0.819 \\

$A^G/A_0^G$
& 1.00 & 1.00 & 1.00 & 0.99 & 0.99 & 0.99 & 1.02 & 0.99 & 0.94 & 1.16 & 1.15 \\
\end{tabular}
\end{adjustbox}
\centering
\end{table}

\begin{table}
\caption{Global performance indices for the perforated plate with different geometries $(L_\mathrm{p}/L_\mathrm{s})$ at $Re = 1500$.}
\label{tab5}
\begin{adjustbox}{max width=\textwidth}
\begin{tabular}{cccccccccccc}
$\boldsymbol{L_\mathrm{p}/L_\mathrm{s}}$
& \textbf{0} & \textbf{1} & \textbf{2} & \textbf{3} & \textbf{4} & \textbf{5} & \textbf{6} & \textbf{7} & \textbf{8} & \textbf{9} & \textbf{10} \\
$St^G$
& 0.0097 & 0.0097 & 0.0095 & 0.0095 & 0.0348 & 0.0512 & 0.0273 & 0.0225 & 0.0361 & 0.0383 & 0.0418 \\

$St^G/St_0^G$
& 1.00 & 1.00 & 0.98 & 0.98 & 3.59 & 5.28 & 2.81 & 2.32 & 3.72 & 3.95 & 4.31 \\

$C_f^G$
& 0.0270 & 0.0270 & 0.0268 & 0.0268 & 0.0803 & 0.1109 & 0.0580 & 0.0526 & 0.2045 & 0.2557 & 0.3218 \\

$C_f^G/C_{f,0}^G$
& 1.00 & 1.00 & 0.99 & 0.99 & 2.97 & 4.11 & 2.15 & 1.95 & 7.57 & 9.47 & 11.92 \\

$A^G$
& 0.719 & 0.719 & 0.709 & 0.709 & 0.867 & 0.923 & 0.941 & 0.856 & 0.353 & 0.300 & 0.260 \\

$A^G/A_0^G$
& 1.00 & 1.00 & 0.99 & 0.99 & 1.21 & 1.28 & 1.31 & 1.19 & 0.49 & 0.42 & 0.36 \\
\end{tabular}
\end{adjustbox}
\centering
\end{table}

At $Re = 500$, $St^G$ and $C_f^G$ slightly decrease as $L_\mathrm{p}/L_\mathrm{s}$ increases as shown in Fig.~\ref{fig4}(a), possibly due to the reduction in the actual wetted area. Overall, the performance indices of all the perforated plates are almost the same as those of the solid plate, and $A^G/A_0^G$ remains close to unity. For $Re = 1000$, $St^G$ and $C_f^G$ remain close to the zero-permeability values for $L_\mathrm{p}/L_\mathrm{s} \leq 5$. When $L_\mathrm{p}/L_\mathrm{s}$ is increased to 6, both quantities abruptly rise, while $A^G/A_0^G$ remains close to unity. For $L_\mathrm{p}/L_\mathrm{s} = 9$ and 10, $St^G$ increases more rapidly than $C_f^G$, yielding dissimilar heat transfer enhancement with $A^G/A_0^G > 1$. At $Re = 1500$, $St^G$ and $C_f^G$ are nearly unchanged for $L_\mathrm{p}/L_\mathrm{s} \leq 3$. Increasing $L_\mathrm{p}/L_\mathrm{s}$ from 3 to 5 enhances both quantities, with a larger relative increase in $St^G$, leading to dissimilar heat transfer enhancement. This enhancement is maintained for $5 \leq L_\mathrm{p}/L_\mathrm{s} \leq 7$, although both $St^G$ and $C_f^G$ decrease gradually. For $L_\mathrm{p}/L_\mathrm{s} \geq 8$, $A^G/A_0^G$ decreases below unity because $C_f^G$ rises significantly, while the increase in $St^G$ is moderate. As will be discussed in \S\ref{sec:flow_characteristics}, the sudden increases of $St^G/St_0^G$ and $C_f^G/C_{f,0}^G$ from unity are caused by the growth of a streamwise travelling-wave-like disturbance around the perforated plate. The present results suggest that the optimal porous structure for dissimilar heat transfer enhancement strongly depends on the Reynolds number.

\subsection{Local performance}
\label{subsec:local_performance}

\begin{figure*}[htbp]
\centering

\begin{subfigure}[t]{0.48\linewidth}
    \centering
    \caption{}
    \vspace{2mm}
    \includegraphics[width=\linewidth]{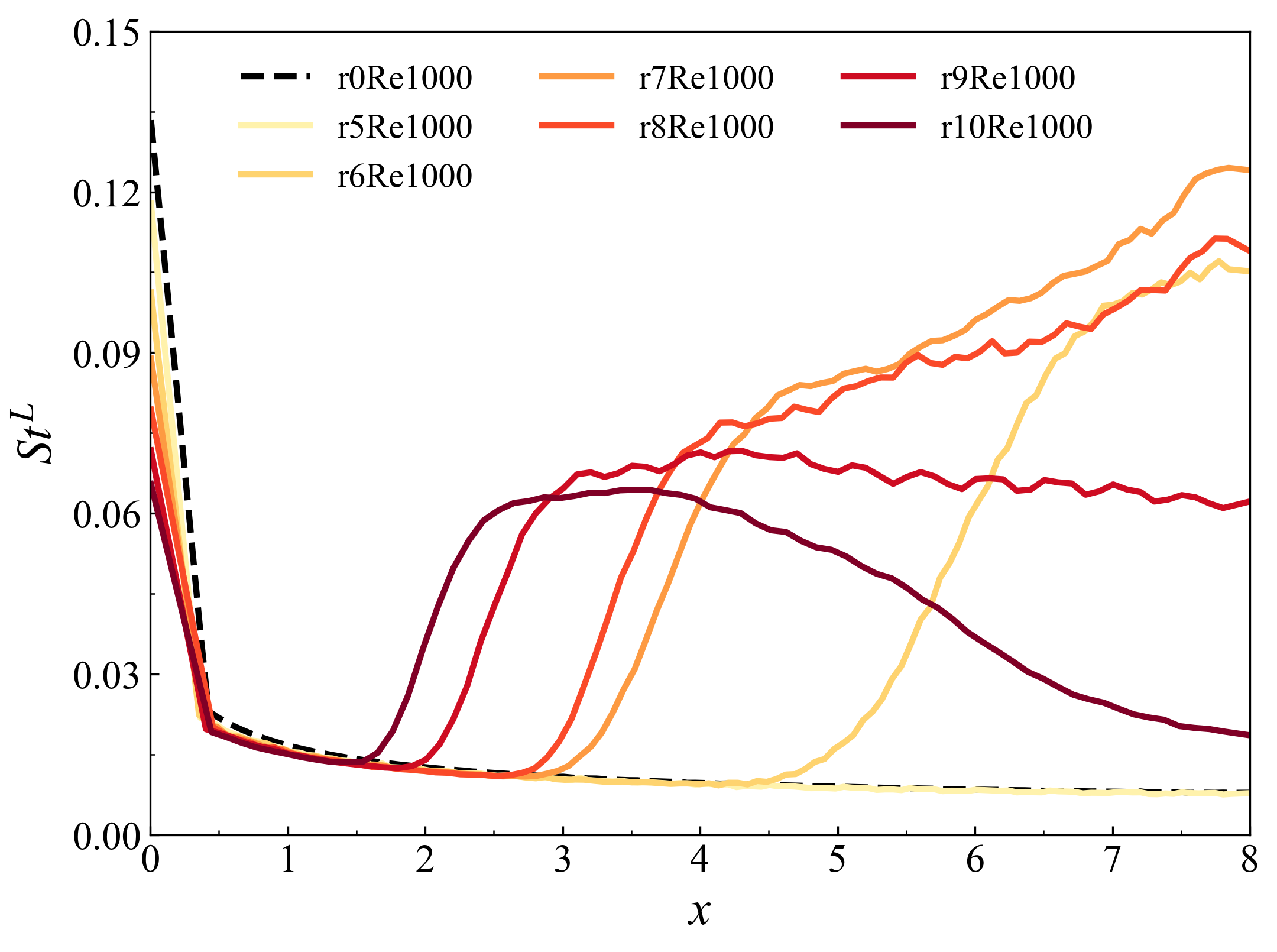}
\end{subfigure}
\hfill
\begin{subfigure}[t]{0.48\linewidth}
    \centering
    \caption{}
    \vspace{2mm}
    \includegraphics[width=\linewidth]{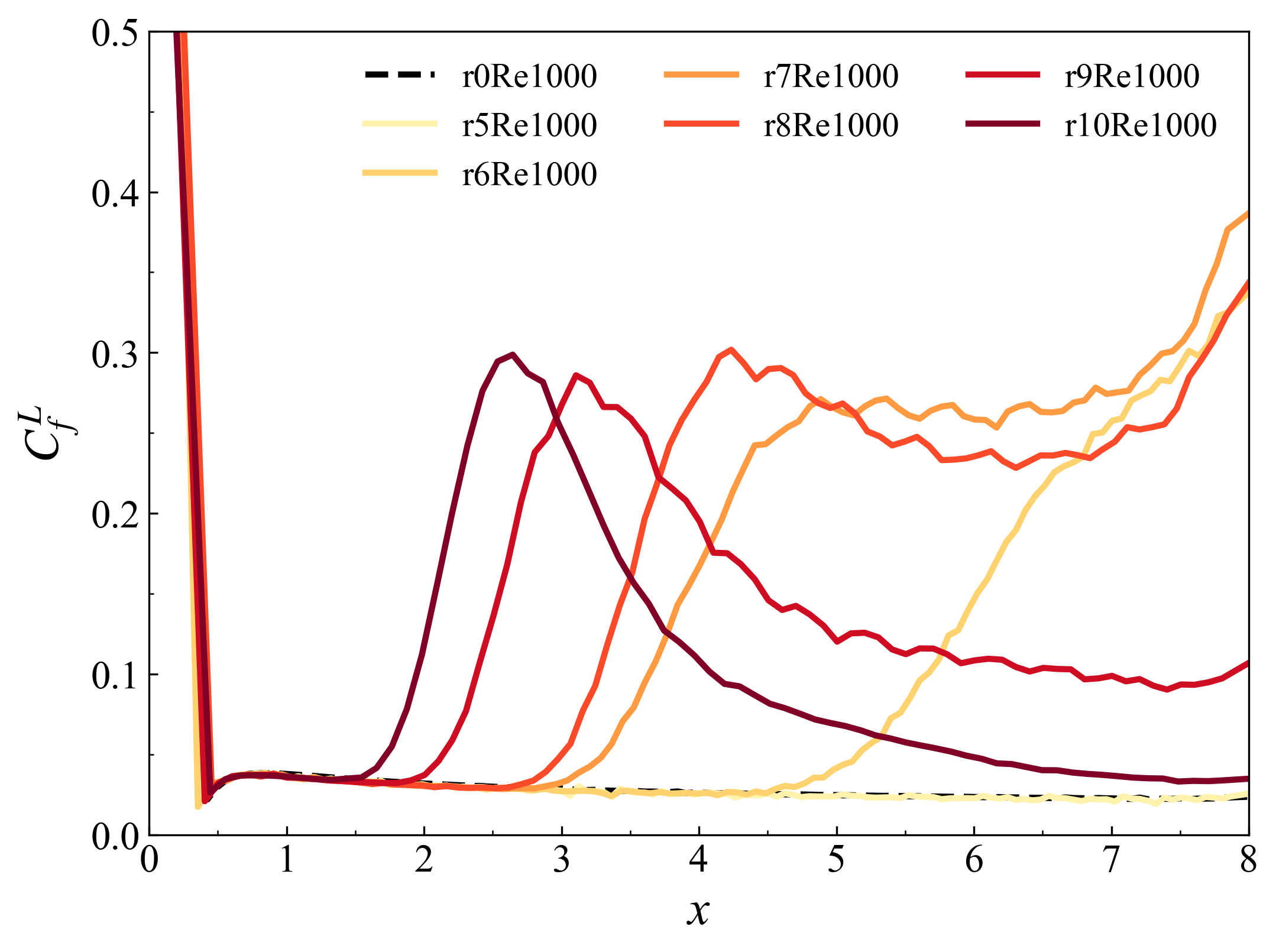}
\end{subfigure}

\caption{Streamwise distribution of (a) local Stanton number $St^L$ and (b) local friction factor $C_f^L$ along the perforated plates for Cases r5Re1000 to r10Re1000.}
\label{fig5}
\end{figure*}

\begin{figure*}[htbp]
\centering

\begin{subfigure}[t]{0.48\linewidth}
    \centering
    \caption{}
    \vspace{2mm}
    \includegraphics[width=\linewidth]{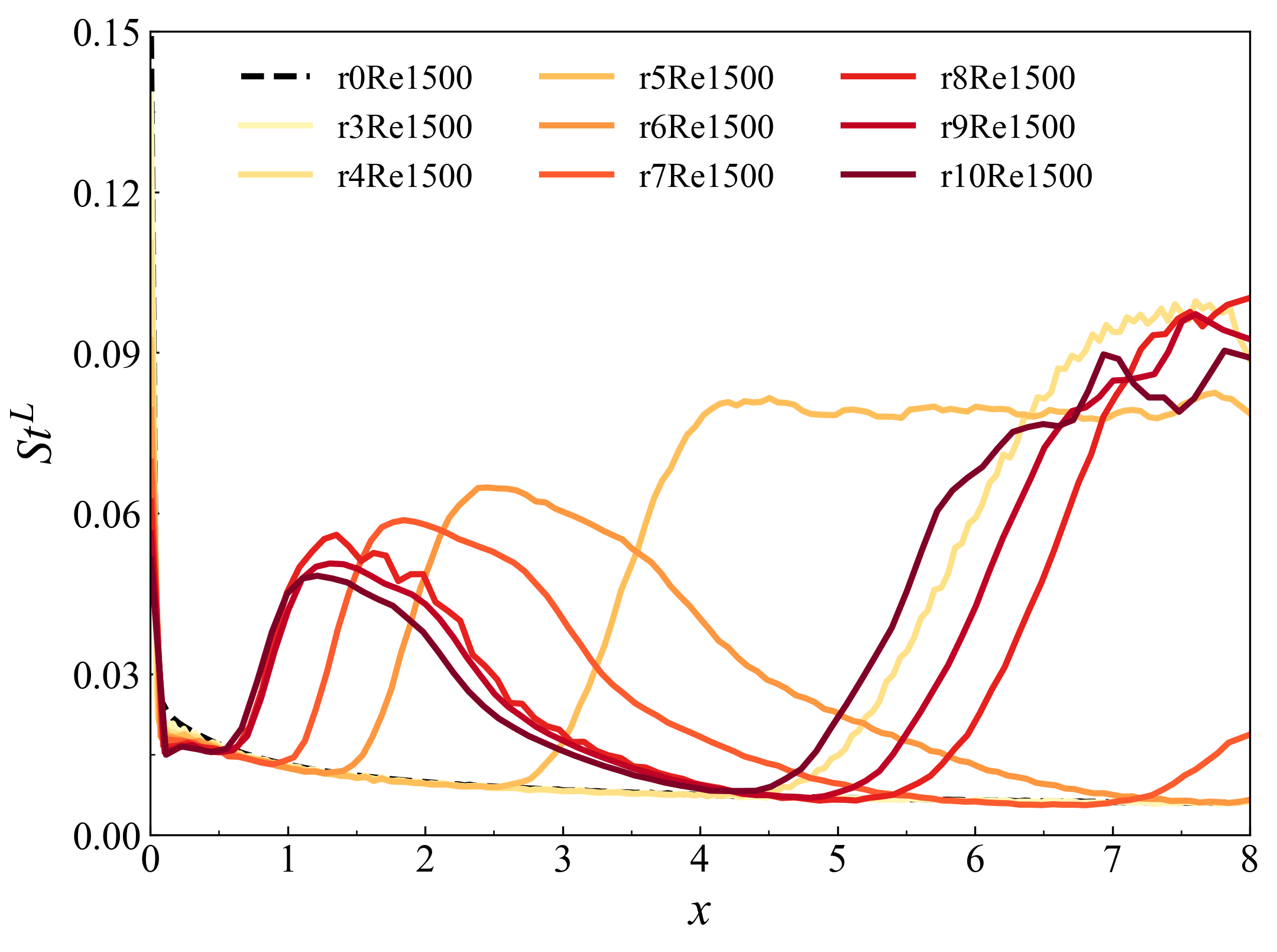}
\end{subfigure}
\hfill
\begin{subfigure}[t]{0.48\linewidth}
    \centering
    \caption{}
    \vspace{2mm}
    \includegraphics[width=\linewidth]{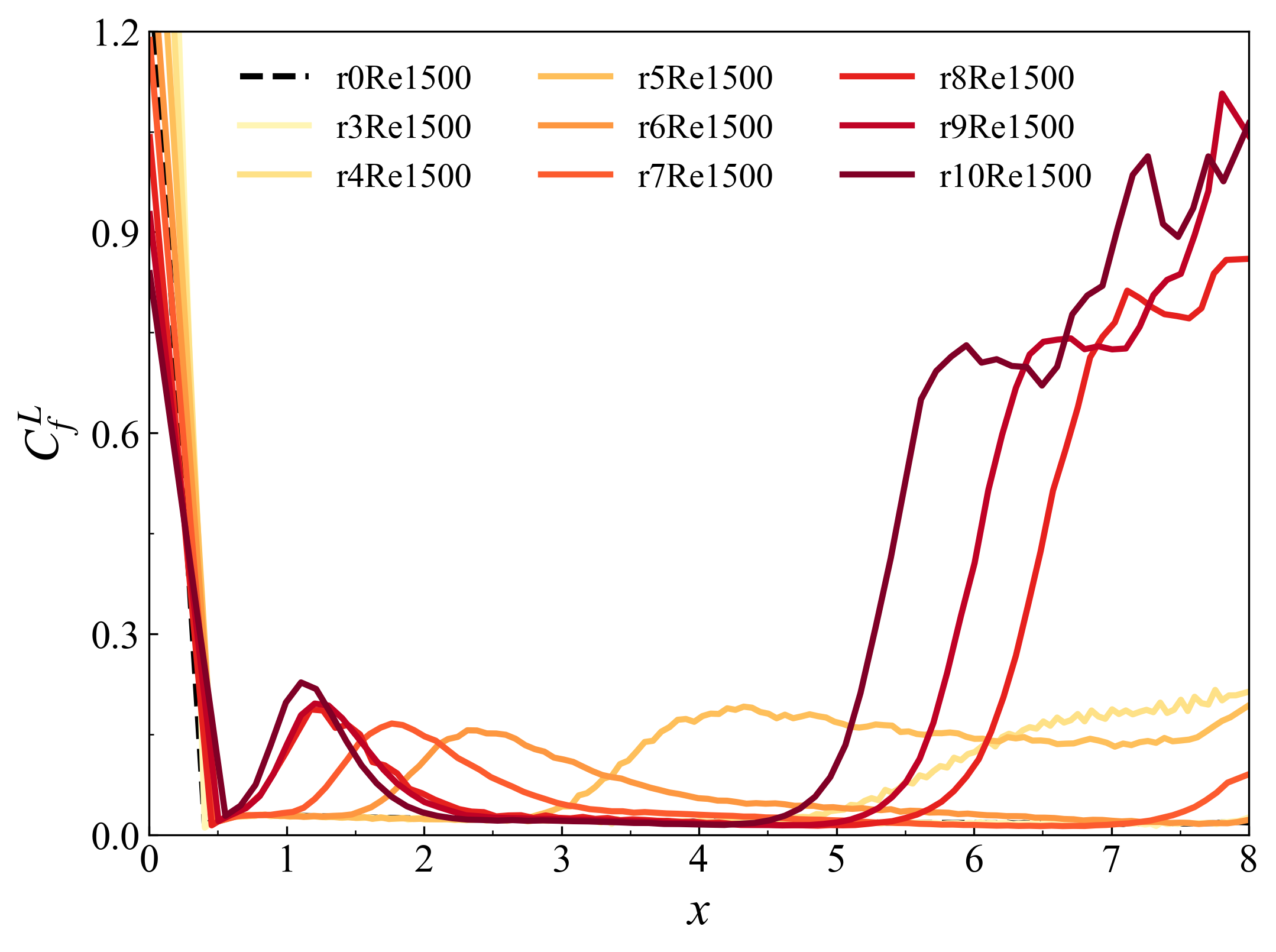}
\end{subfigure}

\caption{Streamwise distribution of (a) local Stanton number $St^L$ and (b) local friction factor $C_f^L$ along the perforated plates for Cases r3Re1500 to r10Re1500.}
\label{fig6}
\end{figure*}

To clarify the global performance presented in \S\ref{subsec:global_performance}, the local distributions of $St^L$, $C_f^L$, and $A^L$ are further plotted. Figures~\ref{fig5} and \ref{fig6} show the distributions of $St^L$ and $C_f^L$ along the streamwise direction at $Re = 1000$ and 1500, respectively. Here, we focus on these two Reynolds numbers because significant dissimilar heat transfer enhancement, i.e., $A^G/A_0^G > 1$, is achieved, and Cases r0Re1000 and r0Re1500 for zero-permeability solid plates are also shown with dashed black lines for comparison.

At $Re = 1000$, as $L_\mathrm{p}/L_\mathrm{s}$ increases from 5 to 10, the enhancement in both $St^L$ and $C_f^L$ moves upstream, while their values near the trailing edge gradually decrease. However, after reaching its peak, $St^L$ remains relatively high for a certain distance downstream, while $C_f^L$ declines rapidly. This explains why $A^G/A_0^G$ is larger than unity for large $L_\mathrm{p}/L_\mathrm{s}$ values of 9 and 10 at $Re = 1000$. The upstream movement of the enhancement and the decline near the trailing edge also appear at $Re = 1500$, while they shift further upstream. This explains the early enhancement of $A^G/A_0^G$ for $L_\mathrm{p}/L_\mathrm{s} \geq 4$, as shown in Table~\ref{tab5} and Fig.~\ref{fig4}. For $L_\mathrm{p}/L_\mathrm{s} > 8$ at $Re = 1500$, a second peak appears near the trailing edge ($x > 5$) in both $St^L$ and $C_f^L$, and $C_f^L$ increases more strongly than $St^L$ there, resulting in a reduction of $A^G/A_0^G$ in these cases.

\begin{figure*}[htbp]
\centering

\begin{subfigure}[t]{0.48\linewidth}
    \centering
    \caption{}
    \vspace{2mm}
    \includegraphics[width=\linewidth]{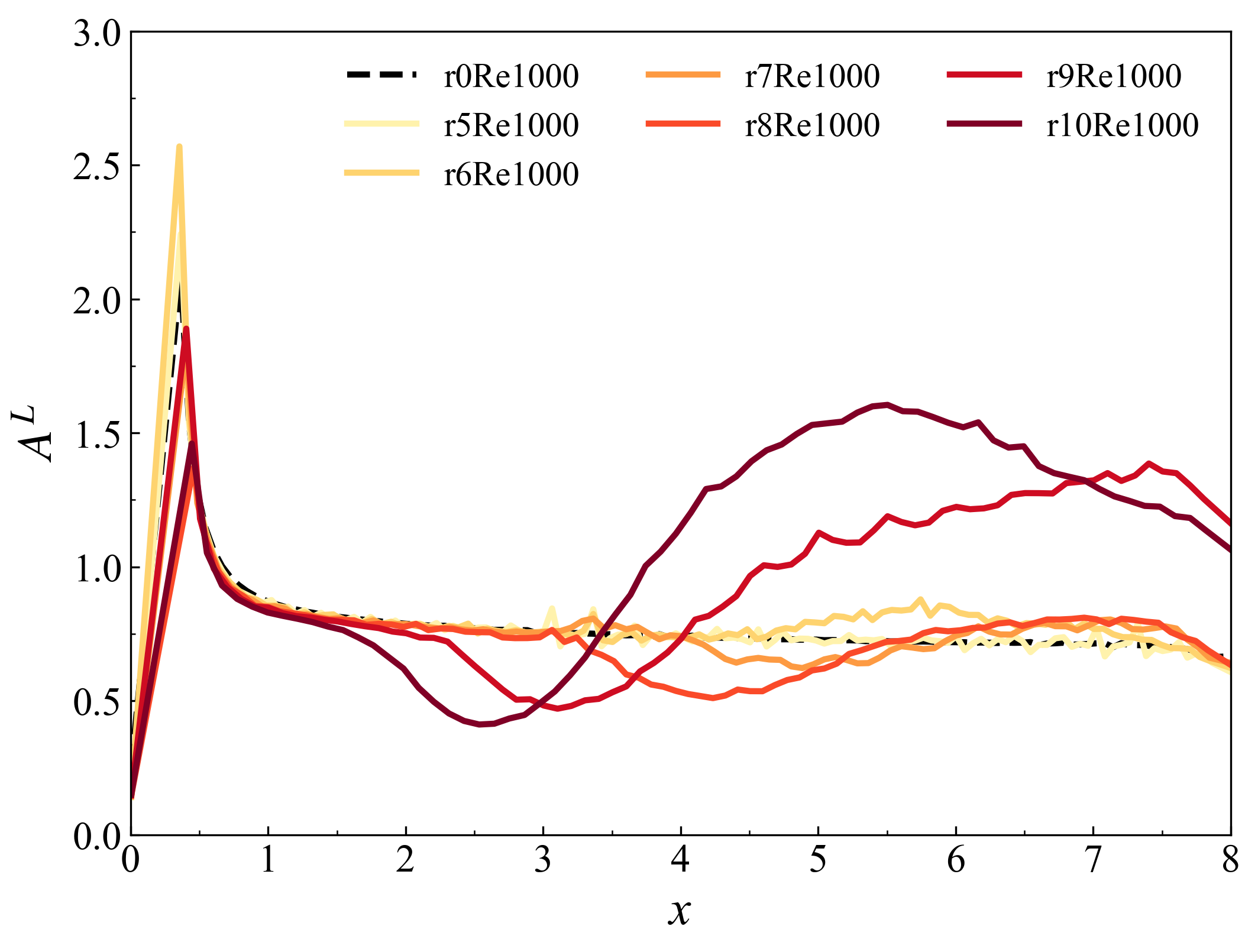}
\end{subfigure}
\hfill
\begin{subfigure}[t]{0.48\linewidth}
    \centering
    \caption{}
    \vspace{2mm}
    \includegraphics[width=\linewidth]{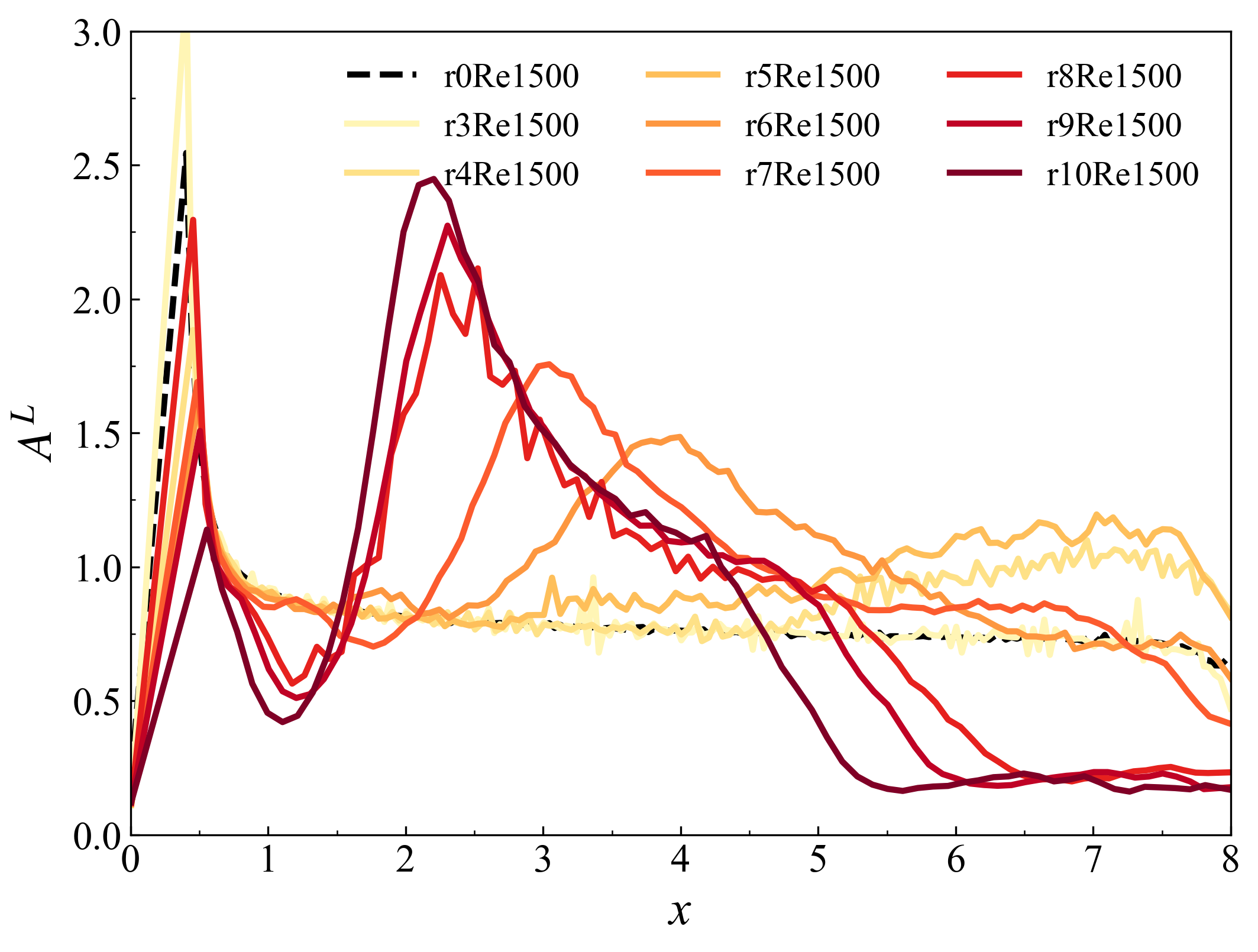}
\end{subfigure}

\caption{Streamwise distribution of local analogy factor $A^L$ along the perforated plates at (a) $Re = 1000$ and (b) $Re = 1500$.}
\label{fig7}
\end{figure*}

Figure~\ref{fig7} shows the streamwise distribution of the local analogy factor $A^L$ at $Re = 1000$ and 1500. As $L_\mathrm{p}/L_\mathrm{s}$ increases, the enhancement of $A^L$ not only moves upstream but also intensifies in magnitude. At $Re = 1000$, this enhancement appears from the middle of the perforated plates, while it shifts further upstream at $Re = 1500$. This results in the global performance enhancement of $A^G/A_0^G > 1.0$ for $L_\mathrm{p}/L_\mathrm{s} \geq 9$ at $Re = 1000$ and $4 \leq L_\mathrm{p}/L_\mathrm{s} \leq 7$ at $Re = 1500$.
Regardless of the Reynolds number, the region where $A^L > A^L_0$ occurs downstream of the peak locations of $St^L$ and $C_f^L$ (see, Figs.~\ref{fig5}--\ref{fig7}). This is because $C_f^L$ decays relatively quickly after reaching its peak, whereas $St^L$ decreases more gradually.

At $Re = 1500$, increasing $L_\mathrm{p}/L_\mathrm{s}$ beyond 7 remarkably reduces $A^L$, driven by the sharp rise in $C_f^L$ at the second peak near the trailing edge as mentioned before (see Fig.~\ref{fig6}(b)). 

\section{Flow characteristics}
\label{sec:flow_characteristics}
\subsection{Flow regimes}
\label{subsec:flow_regimes}

\begin{figure*}[htbp]
\centering

\begin{subfigure}[t]{0.98\linewidth}
    \centering
    \caption{}
    \vspace{2mm}
    \includegraphics[width=\linewidth]{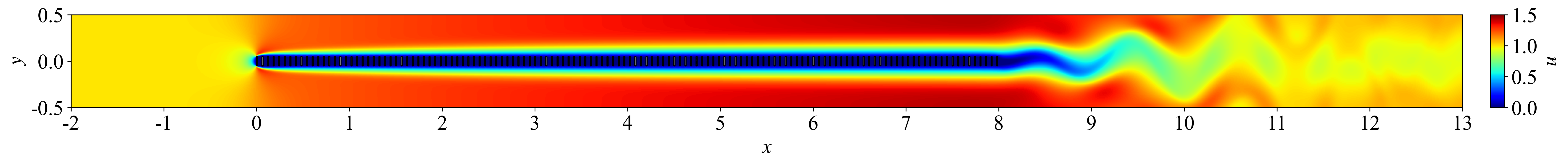}
    \includegraphics[width=\linewidth]{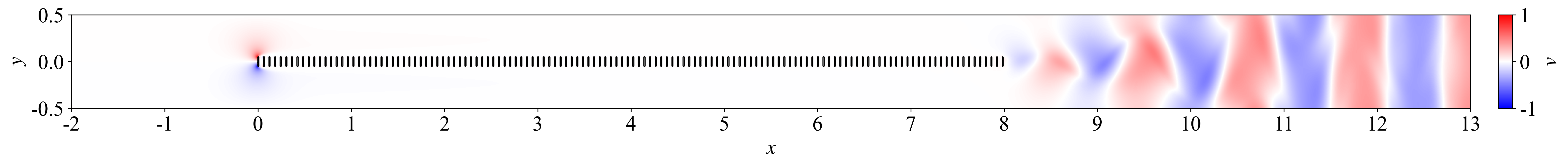}
    \includegraphics[width=\linewidth]{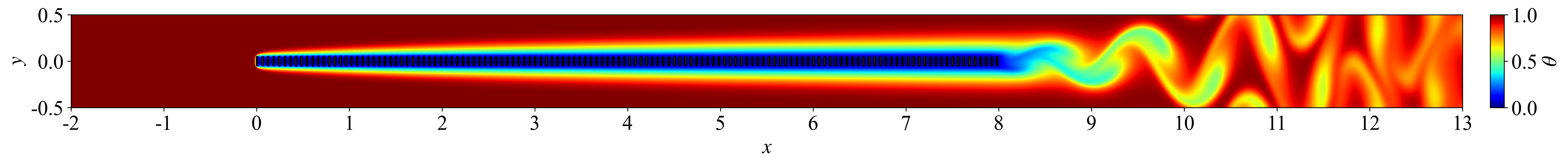}
\end{subfigure}

\begin{subfigure}[t]{0.98\linewidth}
    \centering
    \caption{}
    \vspace{2mm}
    \includegraphics[width=\linewidth]{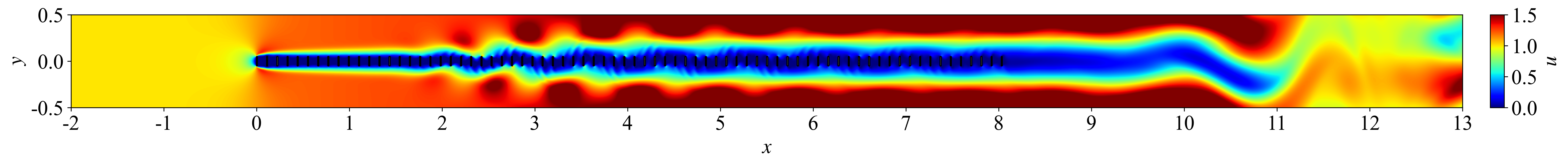}
    \includegraphics[width=\linewidth]{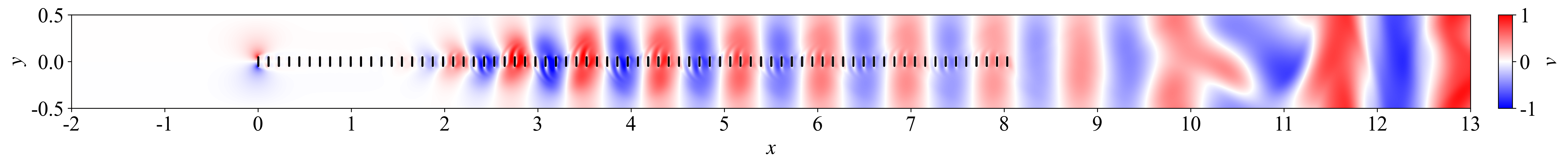}
    \includegraphics[width=\linewidth]{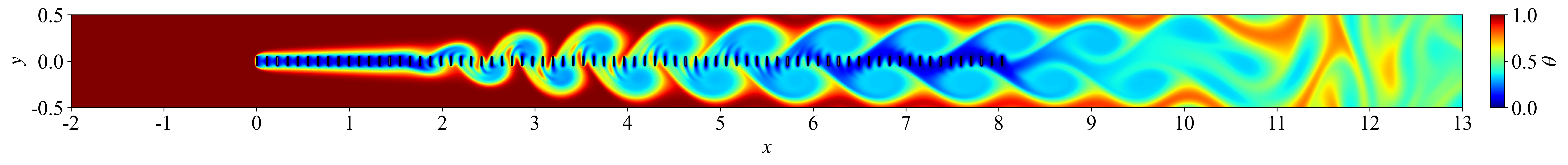}
\end{subfigure}

\begin{subfigure}[t]{0.98\linewidth}
    \centering
    \caption{}
    \vspace{2mm}
    \begin{minipage}[t]{0.48\linewidth}
        \includegraphics[width=\linewidth]{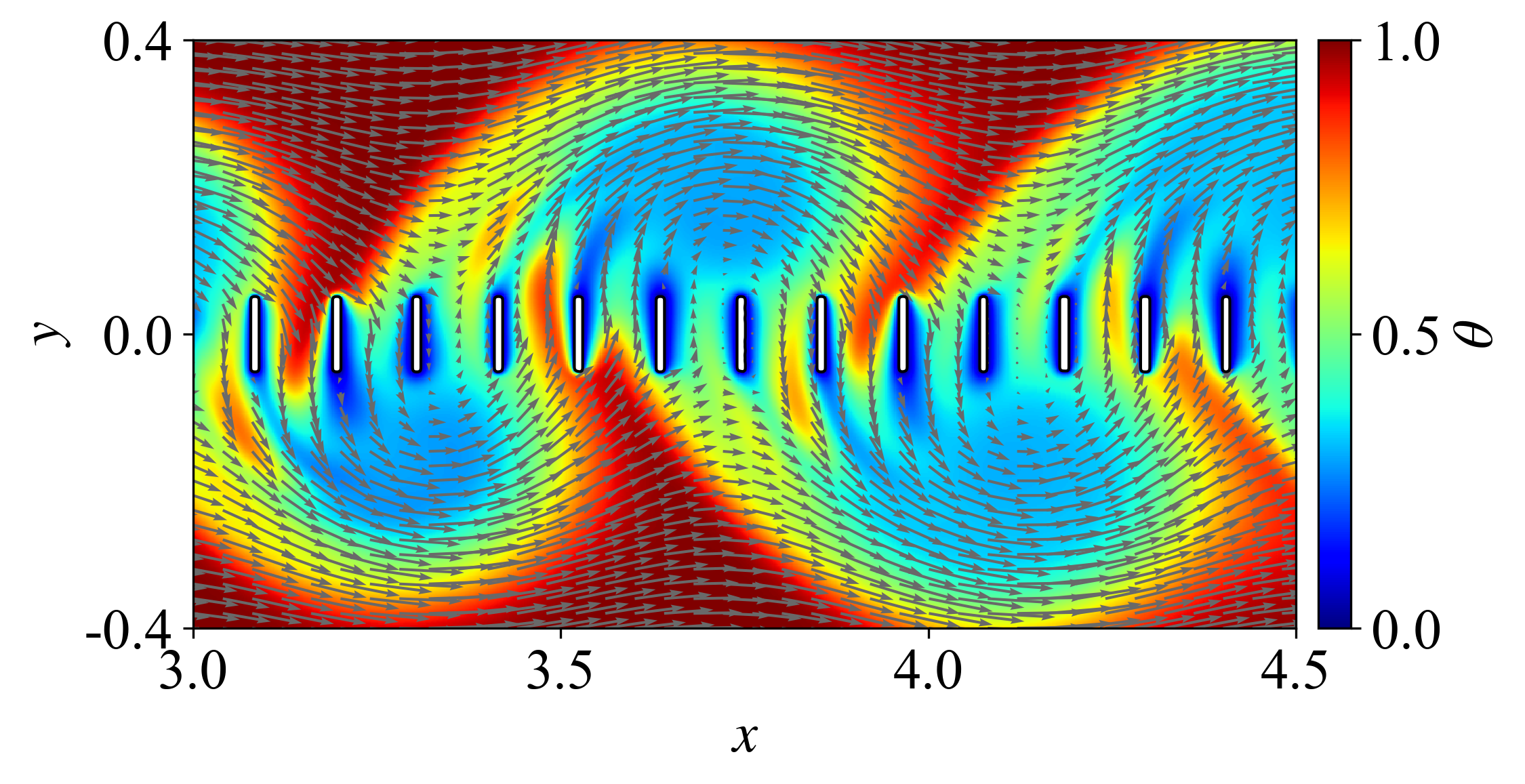}
    \end{minipage}
    \hfill
    \begin{minipage}[t]{0.48\linewidth}
        \includegraphics[width=\linewidth]{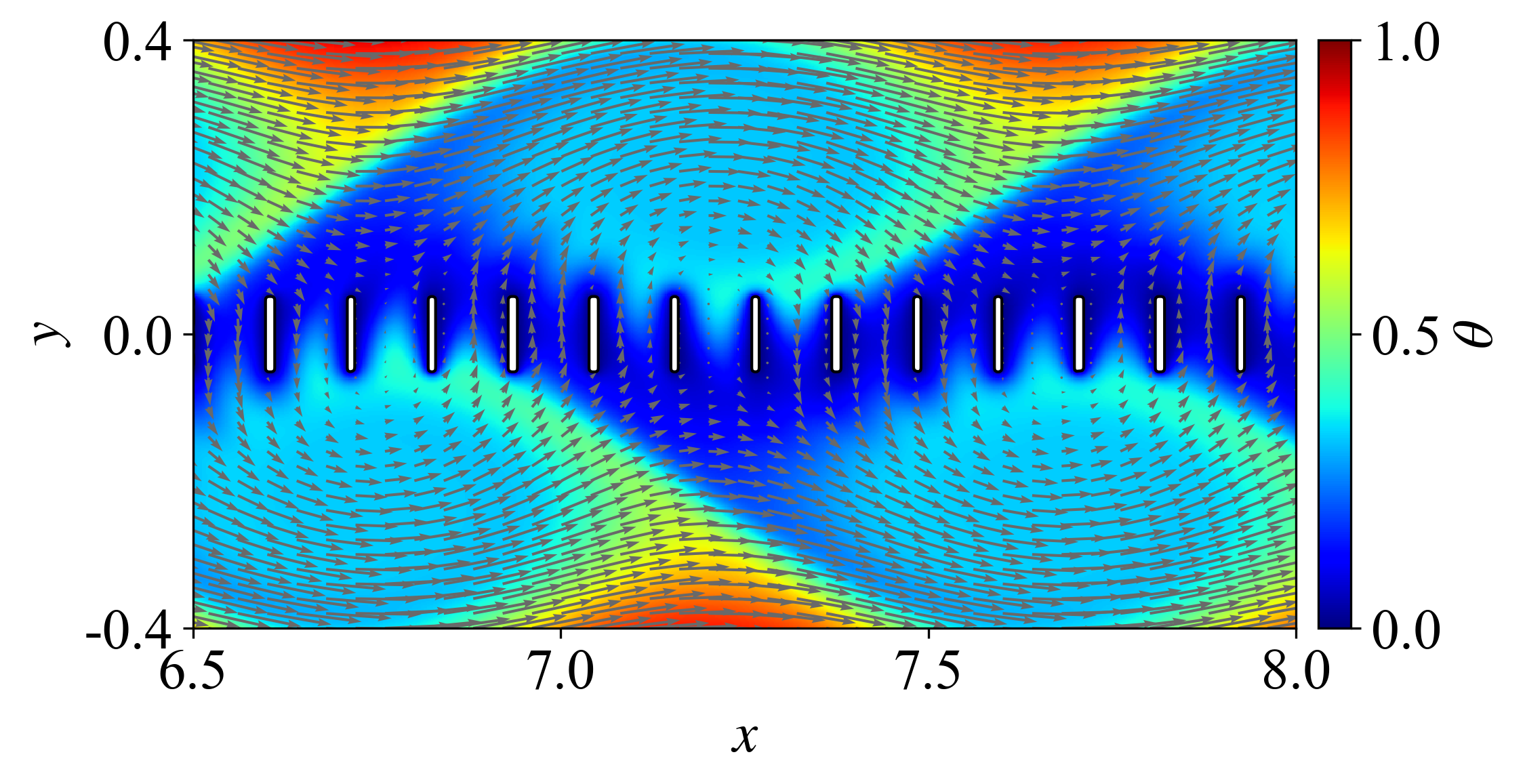}
    \end{minipage}
\end{subfigure}

\caption{Instantaneous streamwise velocity, wall-normal velocity, and temperature field (from top to bottom) of two typical flow regimes: (a) steady flow regime (Case r5Re1000) and (b) travelling-wave regime (Case r10Re1000), together with (c) zoomed temperature fields from $x = 3.0$ to 4.5 (left) and from $x = 6.5$ to 8.0 (right). The arrows represent the velocity, and the white solid line shows the solid interface.}
\label{fig8}
\end{figure*}

To eclucidate the underlying mechanisms govenring the global and local performance indices and their dependencies on the Reynolds number and the porous geometry shown in 
\S\ref{sec:global_local_performances}, this section further analyzes the velocity and temperature fields. In general, depending on the Reynolds number and the porous structure, we identified two distinct flow regimes: a steady flow regime, typified by Case r5Re1000 shown in Fig.~\ref{fig8}(a), and a travelling-wave regime, exemplified by Case r10Re1000 in Fig.~\ref{fig8}(b).
For the steady flow regime, the laminar boundary layer is attached to the upper and lower surfaces of the perforated plates from the leading edge to the trailing edge. Because the flow regime is quite similar to that over a zero-permeability solid plate, the corresponding global and local performances also exhibit the similar features as shown in Table~\ref{tab4}, Fig.~\ref{fig4}(b), and Figs.~\ref{fig5}, \ref{fig6}, and \ref{fig7}(a). For the travelling-wave regime, the laminar boundary layer is periodically disrupted by a travelling-wave from the middle of the plate (see Fig.~\ref{fig8}(c)). As shown in Fig.~\ref{fig8}(b), the travelling-wave appears from $x \approx 2$, which corresponds to the location where the enhancement of $St^L$ and $C_f^L$ starts to develop as shown in Figs.~\ref{fig5} and \ref{fig6}. Therefore, the increases in $St^L$ and $C_f^L$ from those of a zero-permeability solid plate can be attributed to the emergence of the travelling-wave-like disturbance. We also note that the streamwise wavelength of the travelling-wave gradually increases as it is convected downstream, as shown in Fig.~\ref{fig8}(c). The detailed wave properties and their development in the streamwise direction will be discussed later.

\begin{figure*}[htbp]
\centering

\begin{subfigure}[t]{0.98\linewidth}
    \centering
    \caption{}
    \vspace{2mm}
    \includegraphics[width=\linewidth]{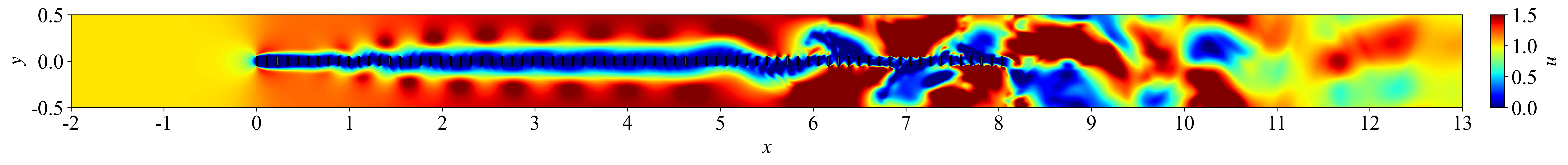}
\end{subfigure}

\begin{subfigure}[t]{0.98\linewidth}
    \centering
    \caption{}
    \vspace{2mm}
    \includegraphics[width=\linewidth]{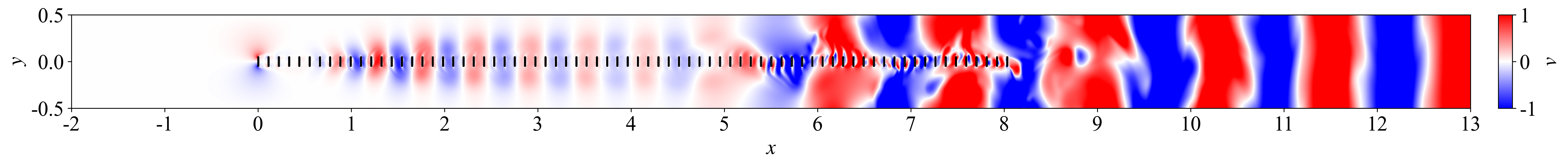}
\end{subfigure}

\begin{subfigure}[t]{0.98\linewidth}
    \centering
    \caption{}
    \vspace{2mm}
    \includegraphics[width=\linewidth]{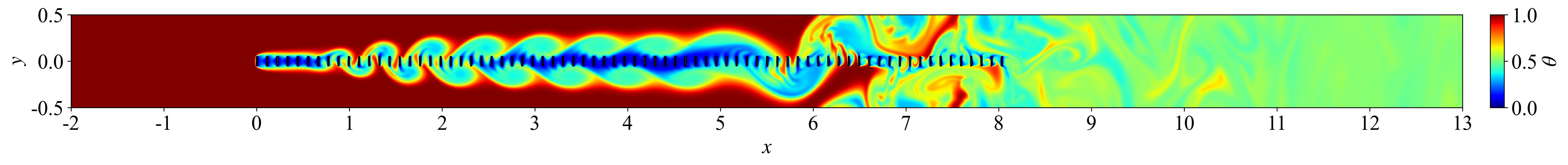}
\end{subfigure}

\begin{subfigure}[t]{0.98\linewidth}
    \centering
    \caption{}
    \vspace{2mm}
    \begin{minipage}[t]{0.48\linewidth}
        \includegraphics[width=\linewidth]{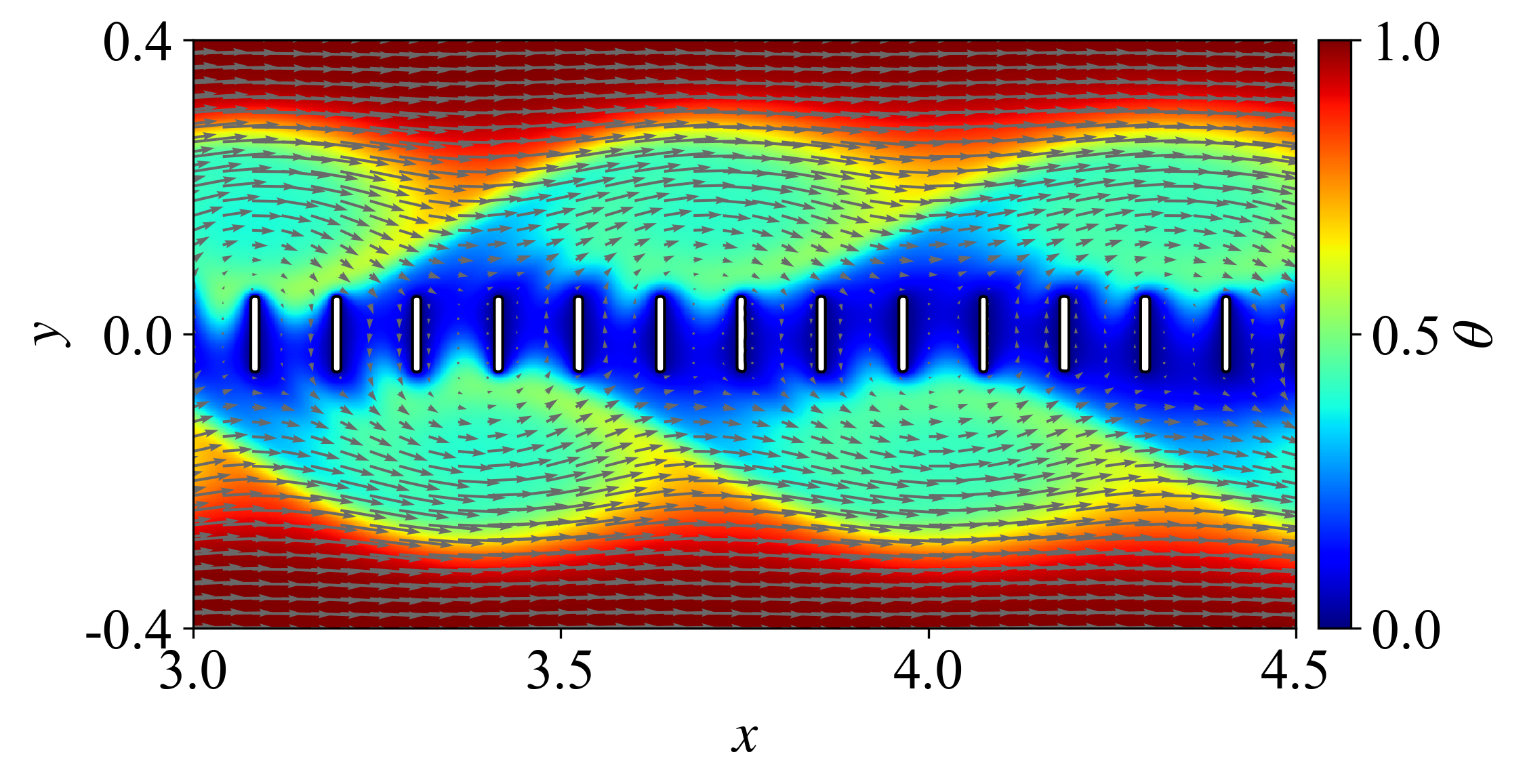}
    \end{minipage}
    \hfill
    \begin{minipage}[t]{0.48\linewidth}
        \includegraphics[width=\linewidth]{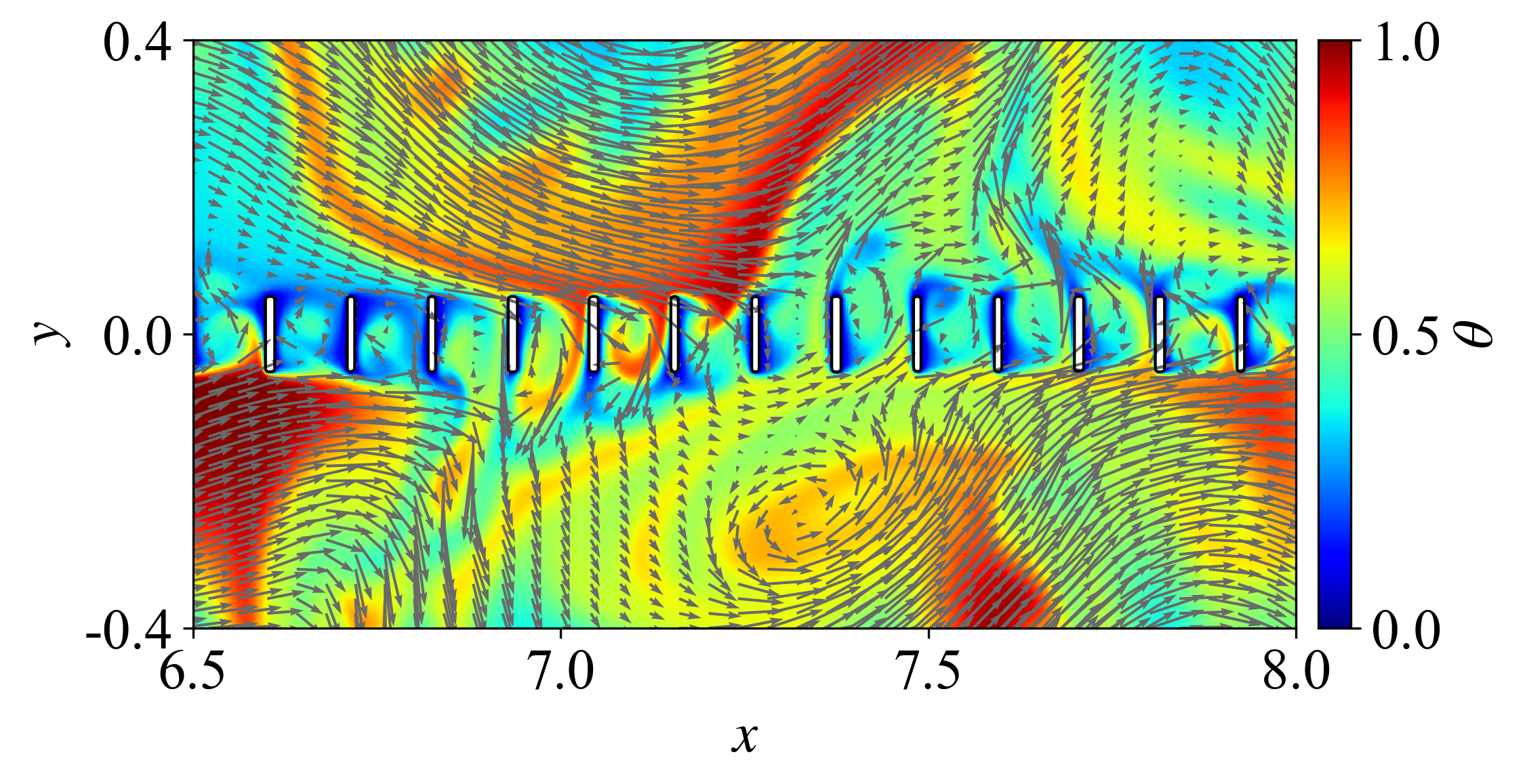}
    \end{minipage}
\end{subfigure}

\caption{Instantaneous (a) streamwise velocity, (b) wall-normal velocity, and (c) temperature field for Case r10Re1500 with a rise in $C_f^L$ near the trailing edge, together with (d) zoomed temperature fields from $x = 3.0$ to 4.5 (left) and from $x = 6.5$ to 8.0 (right).}
\label{fig9}
\end{figure*}

As shown in Figs.~\ref{fig6} and \ref{fig7}(b), further increase in $L_\mathrm{p}/L_\mathrm{s}$ beyond 7 deteriorates the dissimilarity between heat and momentum transfer at $Re = 1500$, owing to the more rapid increase in $C_f^L$ than in $St^L$ near the trailing edge. To understand its mechanisms, the instantaneous streamwise velocity, wall-normal velocity, and temperature field for Case r10Re1500 are also plotted in Fig.~\ref{fig9}. Although the travelling-wave-like disturbance is induced at $x \approx 1$, which corresponds to the peaks of $St^L$, $C_f^L$, and $A^L$, as shown in Figs.~\ref{fig6} and \ref{fig7}(b), it becomes chaotic as it develops downstream. Specifically, beyond $x \approx 5.5$ (see the right panel in Fig.~\ref{fig9}(d)), the induced travelling-wave breaks down into multiple small vortices, resulting in a remarkable rise in $C_f^L$ near the trailing edge as shown in Figs.~\ref{fig6} and \ref{fig7}(b), which in turn suppresses $A^L$ for $L_\mathrm{p}/L_\mathrm{s} > 7$ at $Re = 1500$.

\begin{figure*}[htbp]
\centering

\begin{subfigure}[t]{0.48\linewidth}
    \centering
    \caption{}
    \vspace{2mm}
    \includegraphics[width=\linewidth]{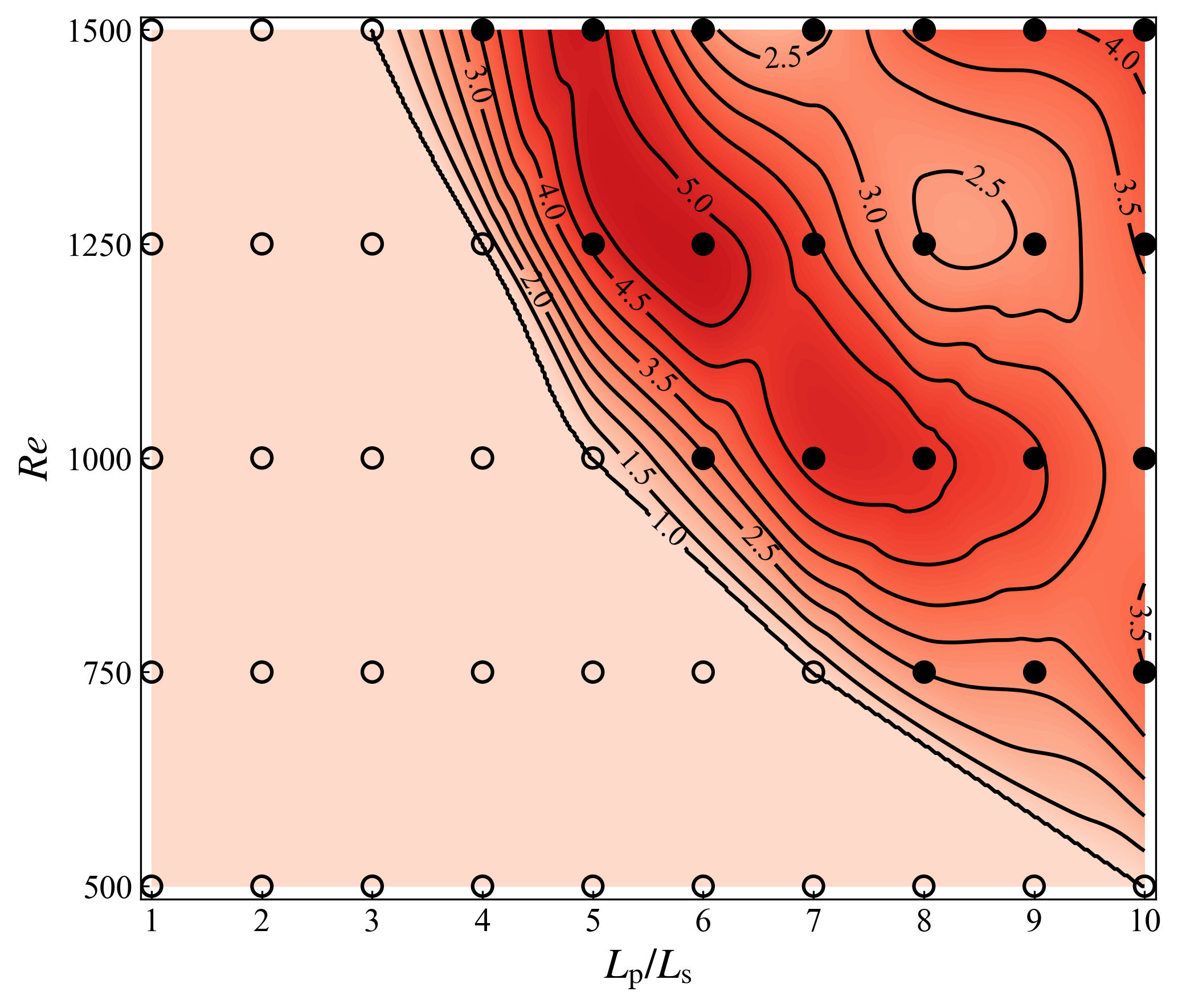}
\end{subfigure}
\hfill
\begin{subfigure}[t]{0.48\linewidth}
    \centering
    \caption{}
    \vspace{2mm}
    \includegraphics[width=\linewidth]{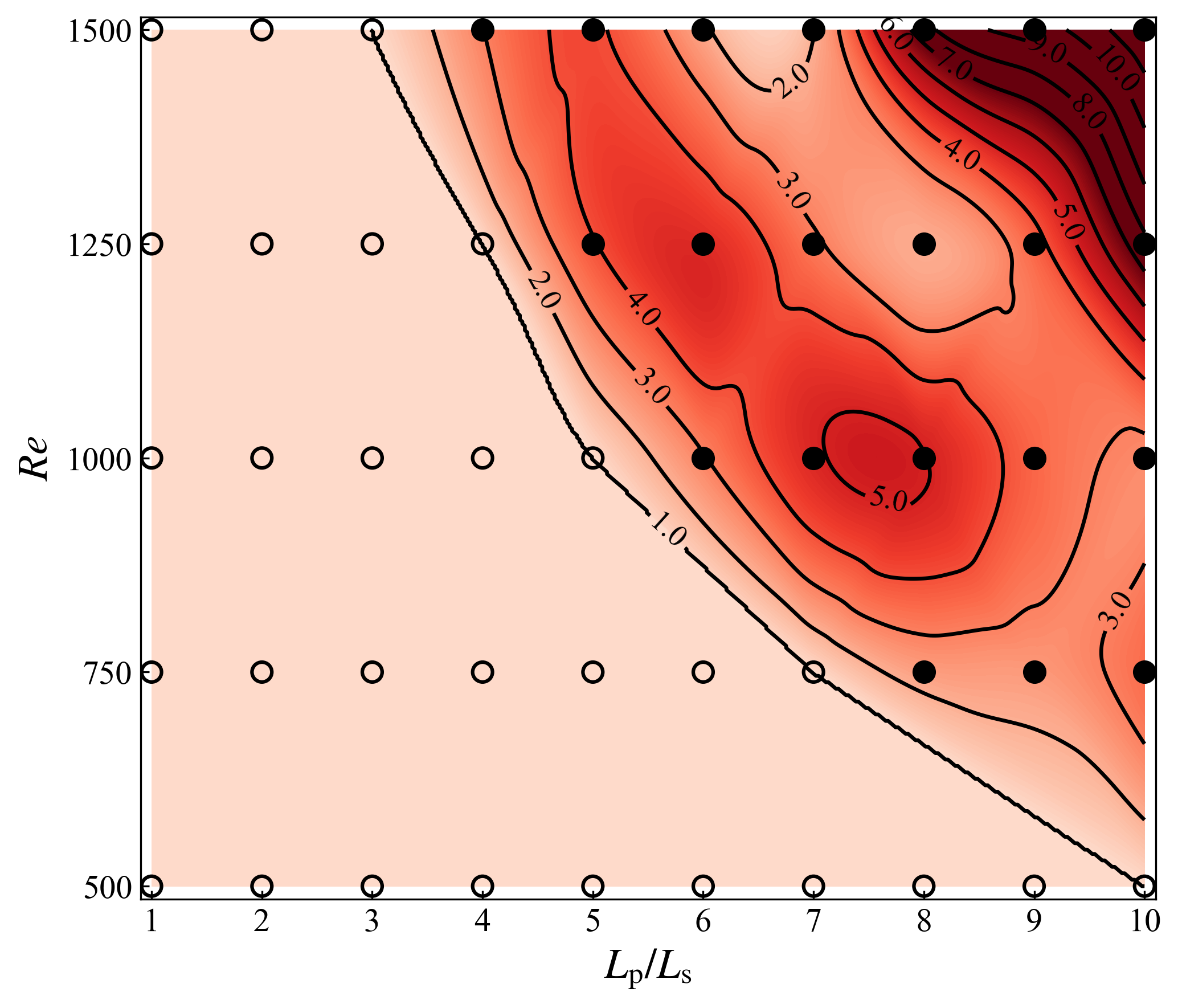}
\end{subfigure}

\vspace{4mm}

\begin{subfigure}[t]{0.48\linewidth}
    \centering
    \caption{}
    \vspace{2mm}
    \includegraphics[width=\linewidth]{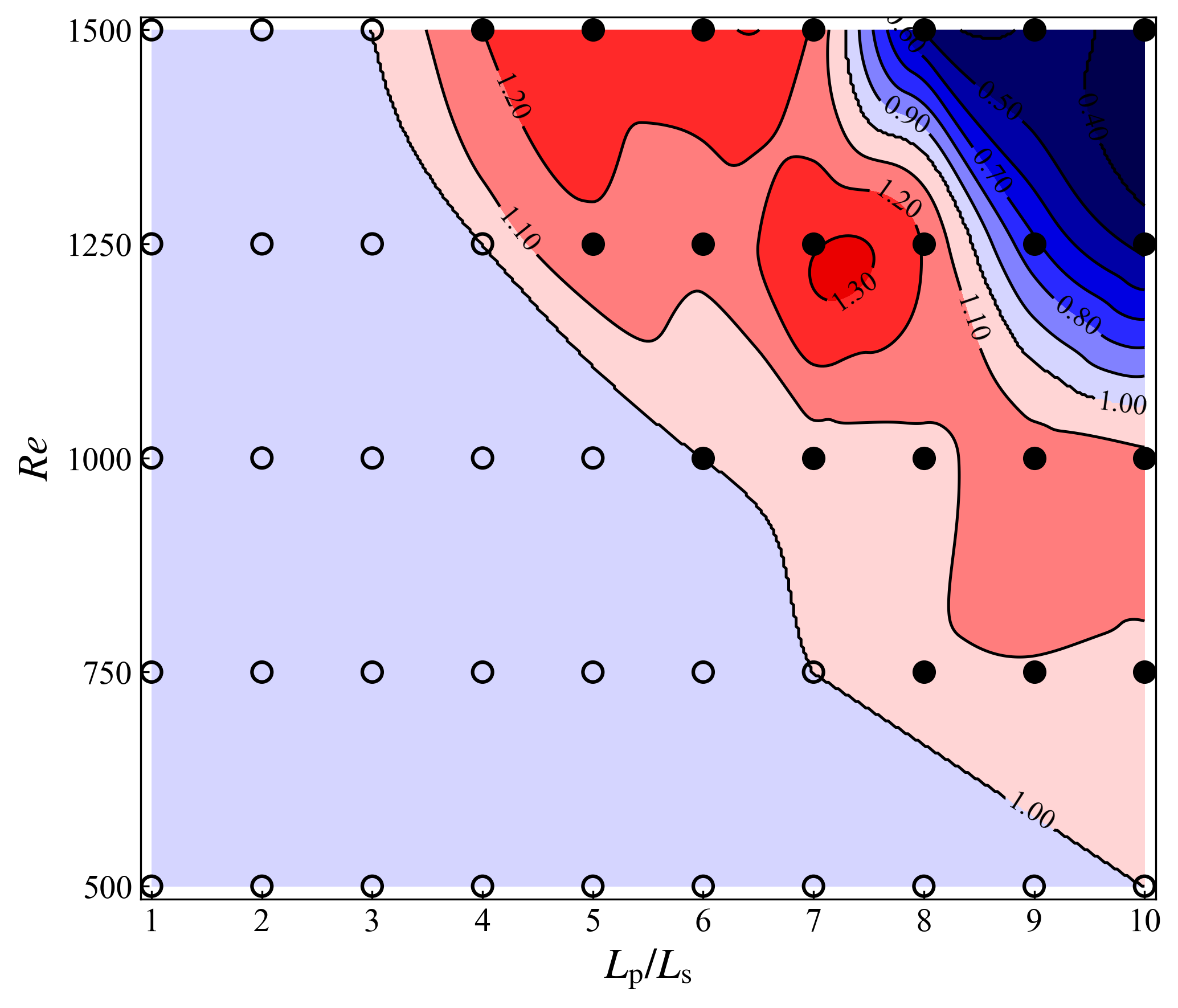}
\end{subfigure}

\caption{Performance indices of (a) $St^G/St_0^G$, (b) $C_f^G/C_{f,0}^G$, and (c) $A^G/A_0^G$ as a function of $L_\mathrm{p}/L_\mathrm{s}$ and $Re$. Here, the open and solid circles represent cases with steady and travelling-wave regimes, respectively.}
\label{fig10}
\end{figure*}

Performance indices normalized by their reference values such as $St_0^G$, $C_{f,0}^G$, and $A_0^G$ are plotted as a function of $L_\mathrm{p}/L_\mathrm{s}$ and $Re$ in Fig.~\ref{fig10}. Here, open and solid circles represent the cases of steady and travelling wave flow regimes, respectively. Simulations at two additional Reynolds numbers of $Re = 750$ and $1250$ are also conducted to clarify the dependencies of the global indices on $Re$ and $L_\mathrm{p}/L_\mathrm{s}$. Overall, the flow regime switches from the steady to travelling-wave regime as $L_\mathrm{p}/L_\mathrm{s}$ and $Re$ increase. Once the travelling wave is induced, both $C_f^G/C_{f,0}^G$ and $St^G/St_0^G$ are enhanced; however, the enhancement of $St^G/St_0^G$ is more prominent, thereby leading to $A^G/A_0^G > 1$ (see Fig.~\ref{fig10}(c)). Nonetheless, further increases in $L_\mathrm{p}/L_\mathrm{s}$ and $Re$ lead to a sharp rise in $C_f^G/C_{f,0}^G$. Consequently, dissimilar heat transfer enhancement is no longer achieved for $L_\mathrm{p}/L_\mathrm{s} > 7$ at $Re = 1500$ and $L_\mathrm{p}/L_\mathrm{s} > 9$ at $Re = 1250$.

\subsection{Characteristics of travelling-waves}
\label{subsec:travelling_wave_characteristics}
\begin{figure*}[htbp]
\centering

\begin{subfigure}[t]{0.98\linewidth}
    \centering
    \caption{}
    \begin{minipage}[t]{0.4\linewidth}
        \includegraphics[width=\linewidth]{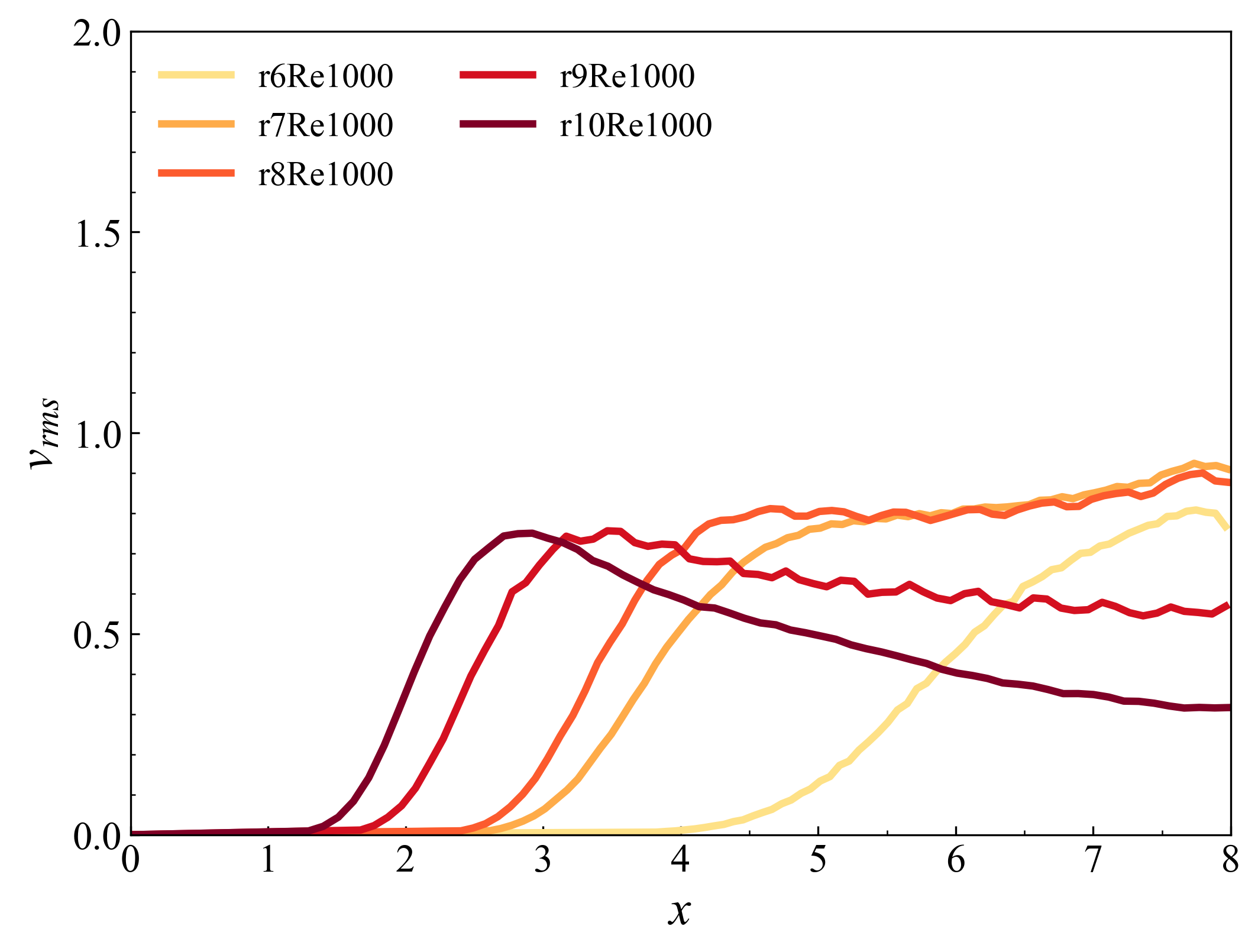}
    \end{minipage}
    \hfill
    \begin{minipage}[t]{0.4\linewidth}
        \includegraphics[width=\linewidth]{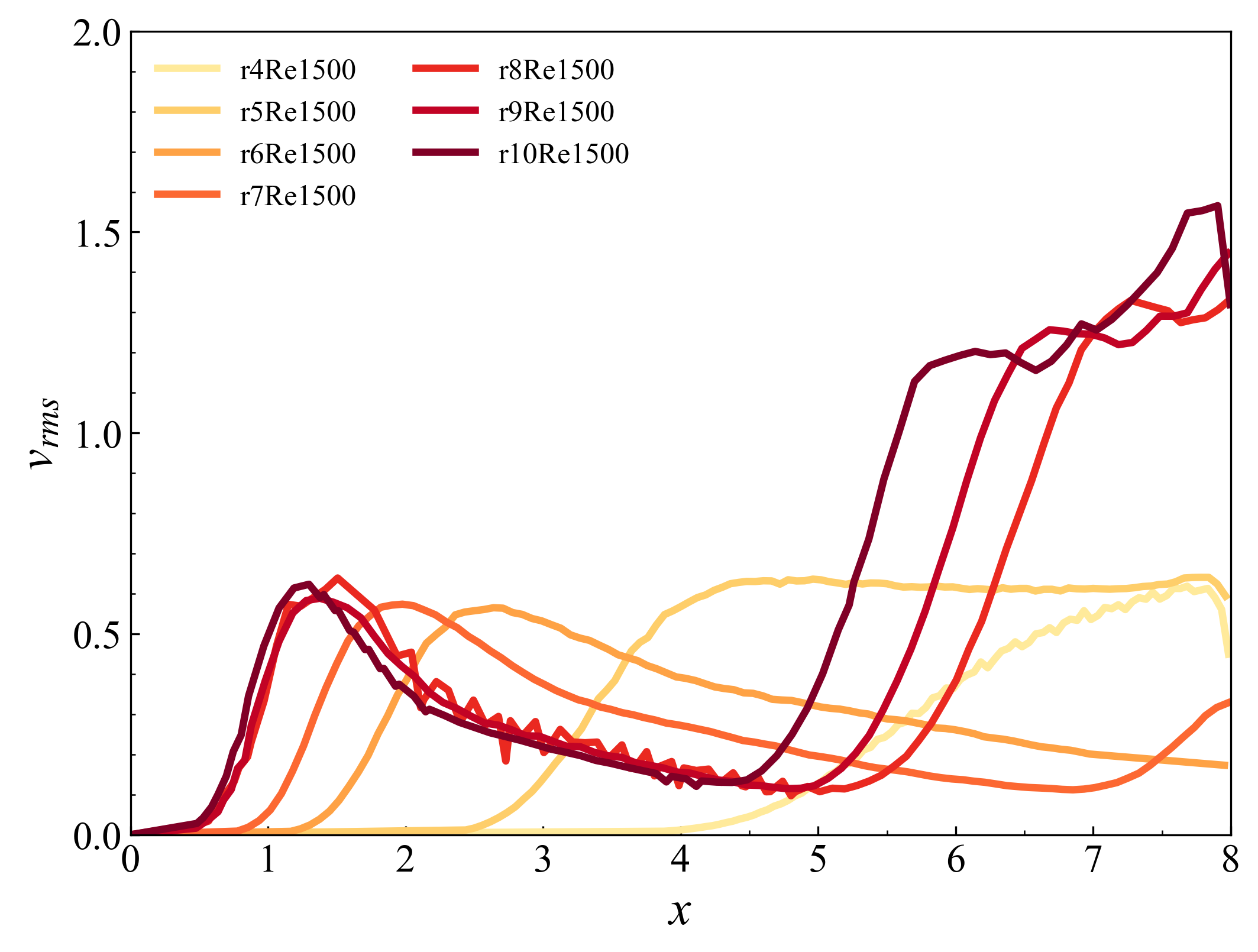}
    \end{minipage}
\end{subfigure}

\vspace{2mm}

\begin{subfigure}[t]{0.98\linewidth}
    \centering
    \caption{}
    \begin{minipage}[t]{0.4\linewidth}
        \includegraphics[width=\linewidth]{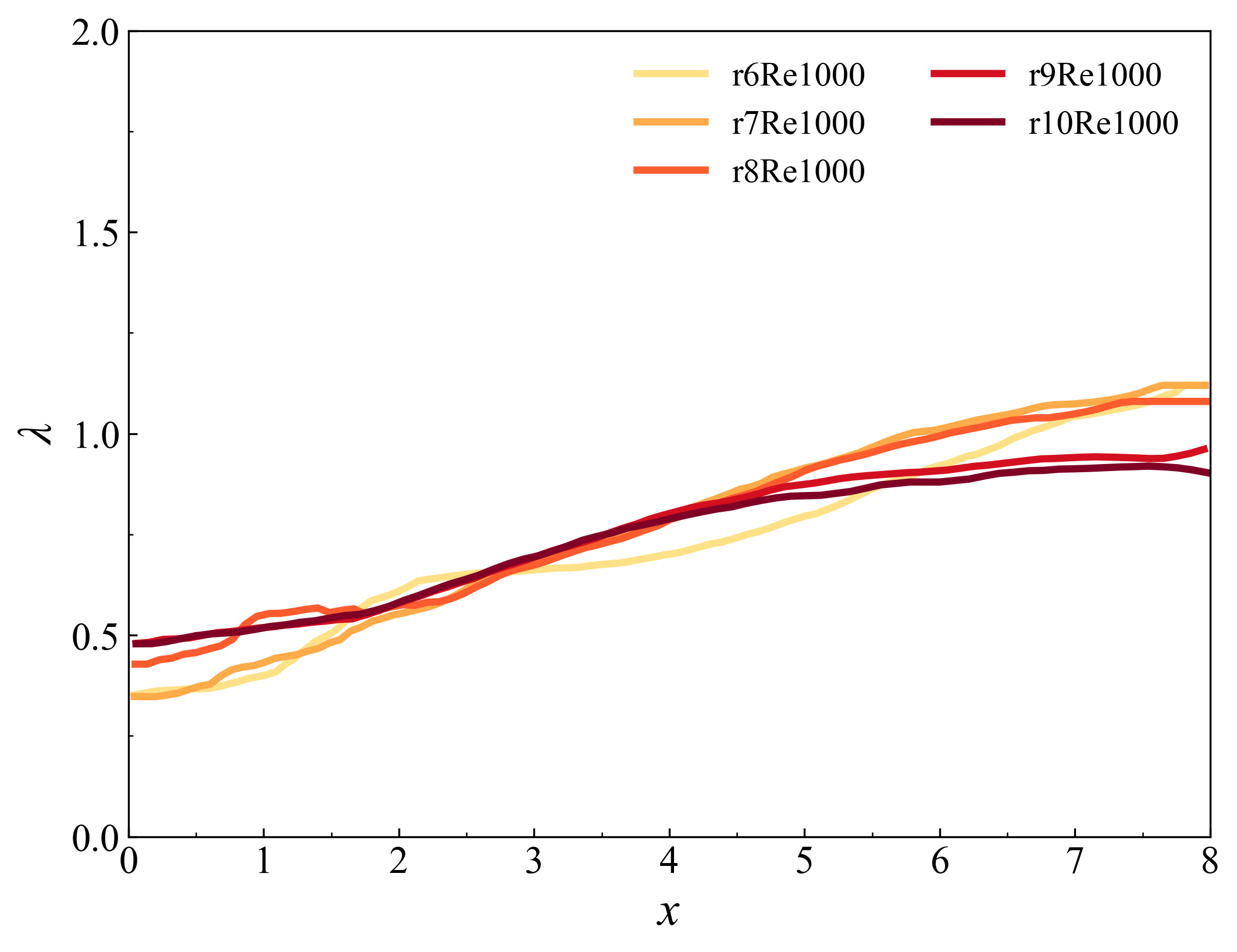}
    \end{minipage}
    \hfill
    \begin{minipage}[t]{0.4\linewidth}
        \includegraphics[width=\linewidth]{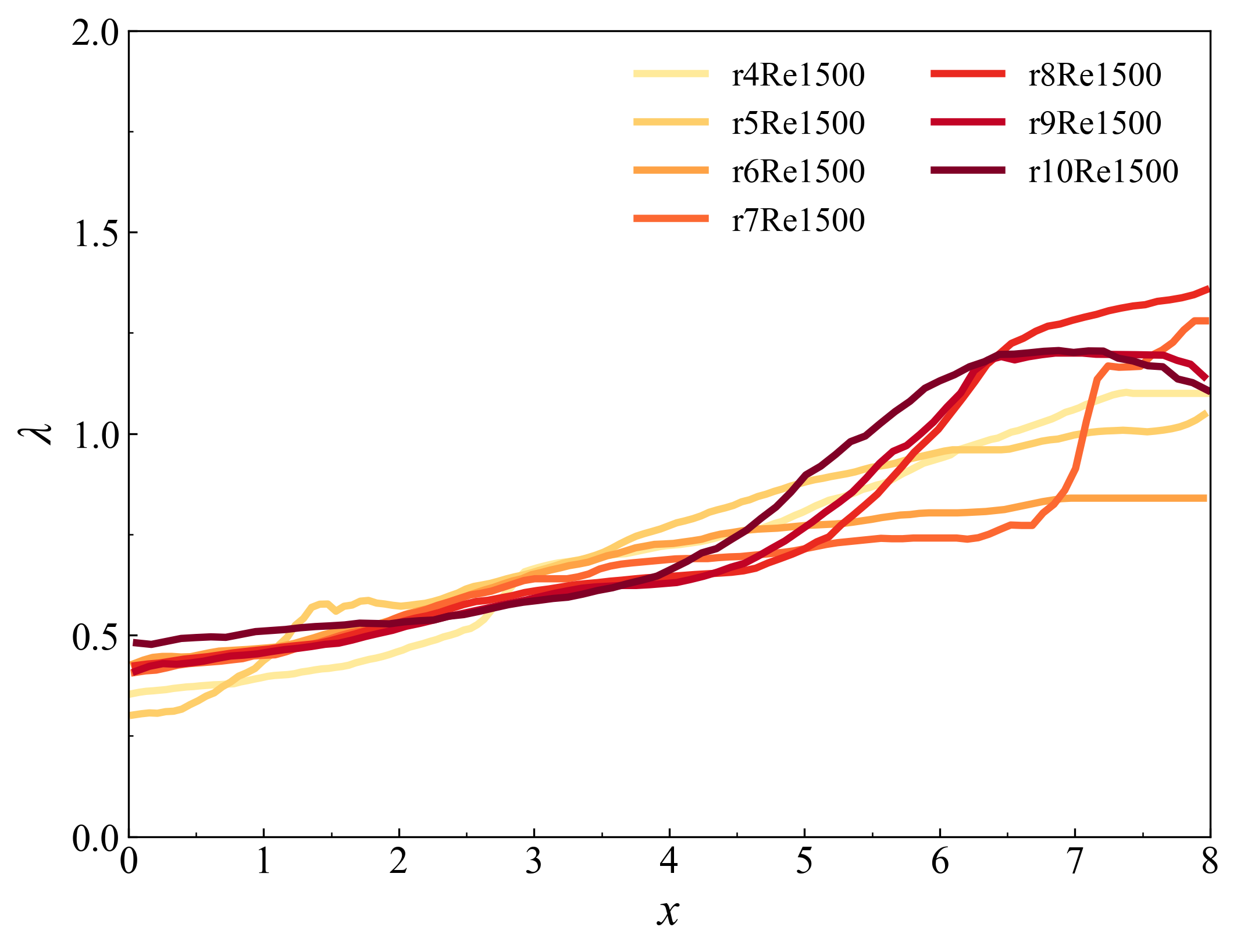}
    \end{minipage}
\end{subfigure}

\vspace{2mm}

\begin{subfigure}[t]{0.98\linewidth}
    \centering
    \caption{}
    \begin{minipage}[t]{0.4\linewidth}
        \includegraphics[width=\linewidth]{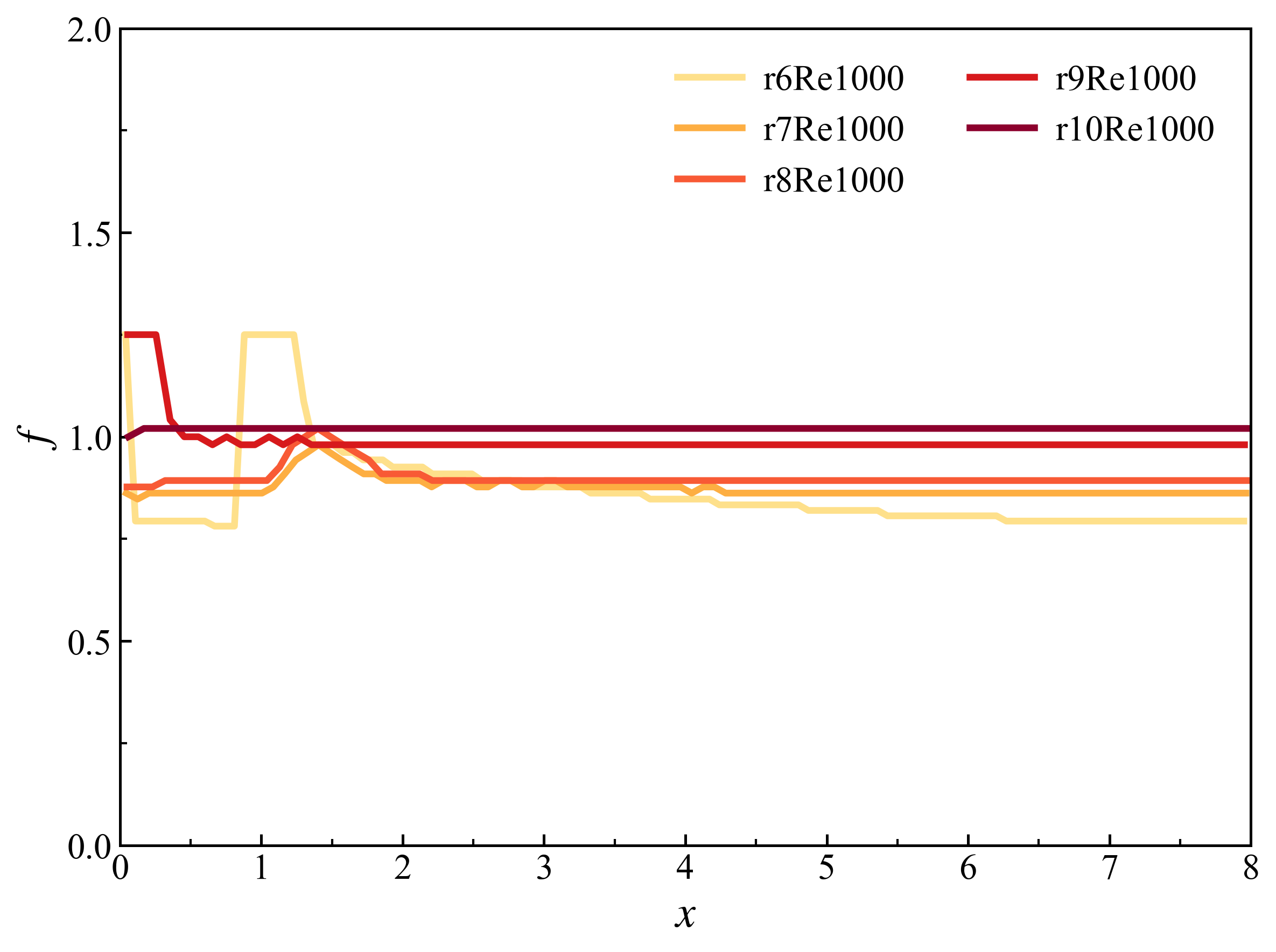}
    \end{minipage}
    \hfill
    \begin{minipage}[t]{0.4\linewidth}
        \includegraphics[width=\linewidth]{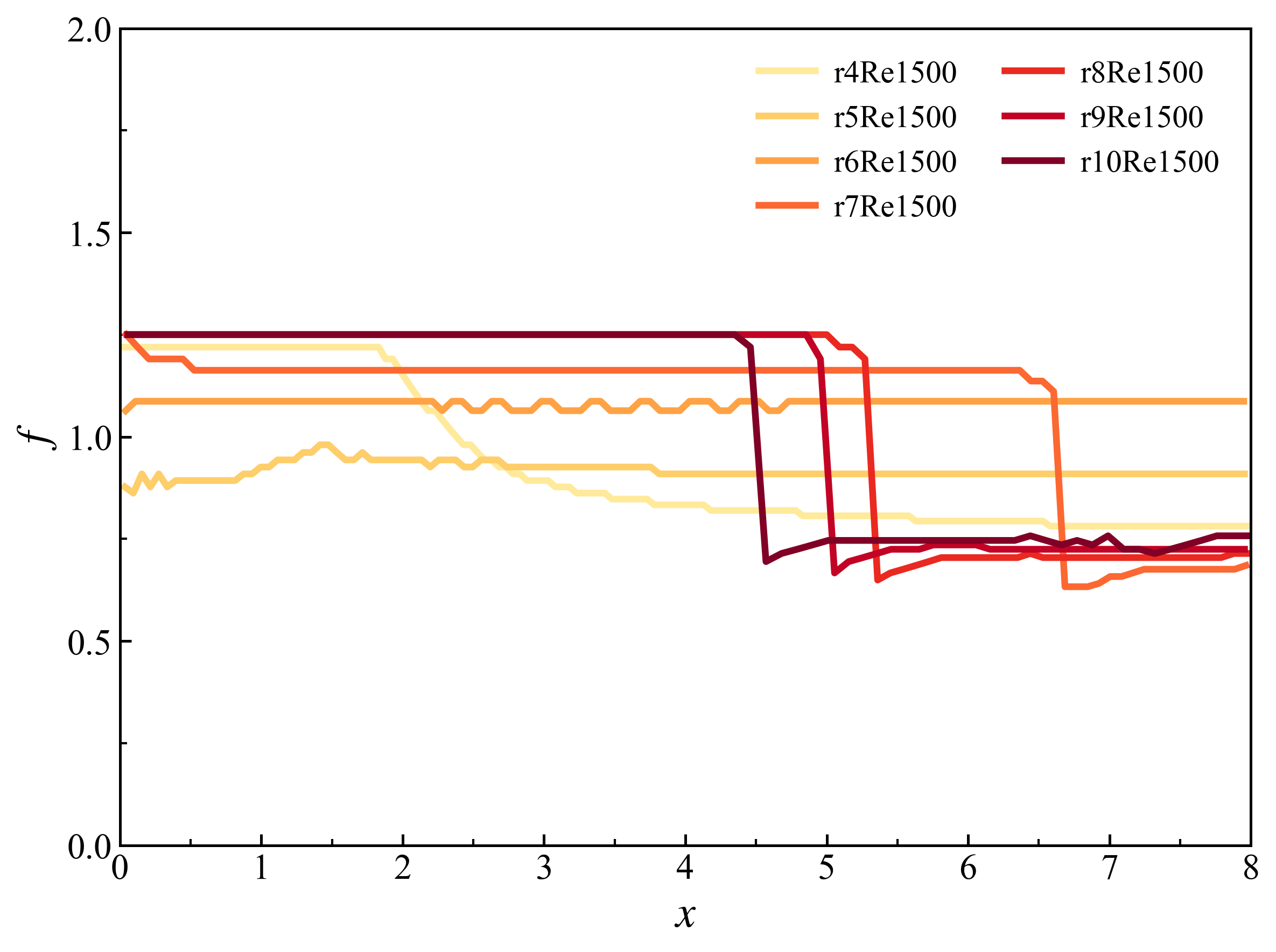}
    \end{minipage}
\end{subfigure}

\vspace{2mm}

\begin{subfigure}[t]{0.98\linewidth}
    \centering
    \caption{}
    \begin{minipage}[t]{0.4\linewidth}
        \includegraphics[width=\linewidth]{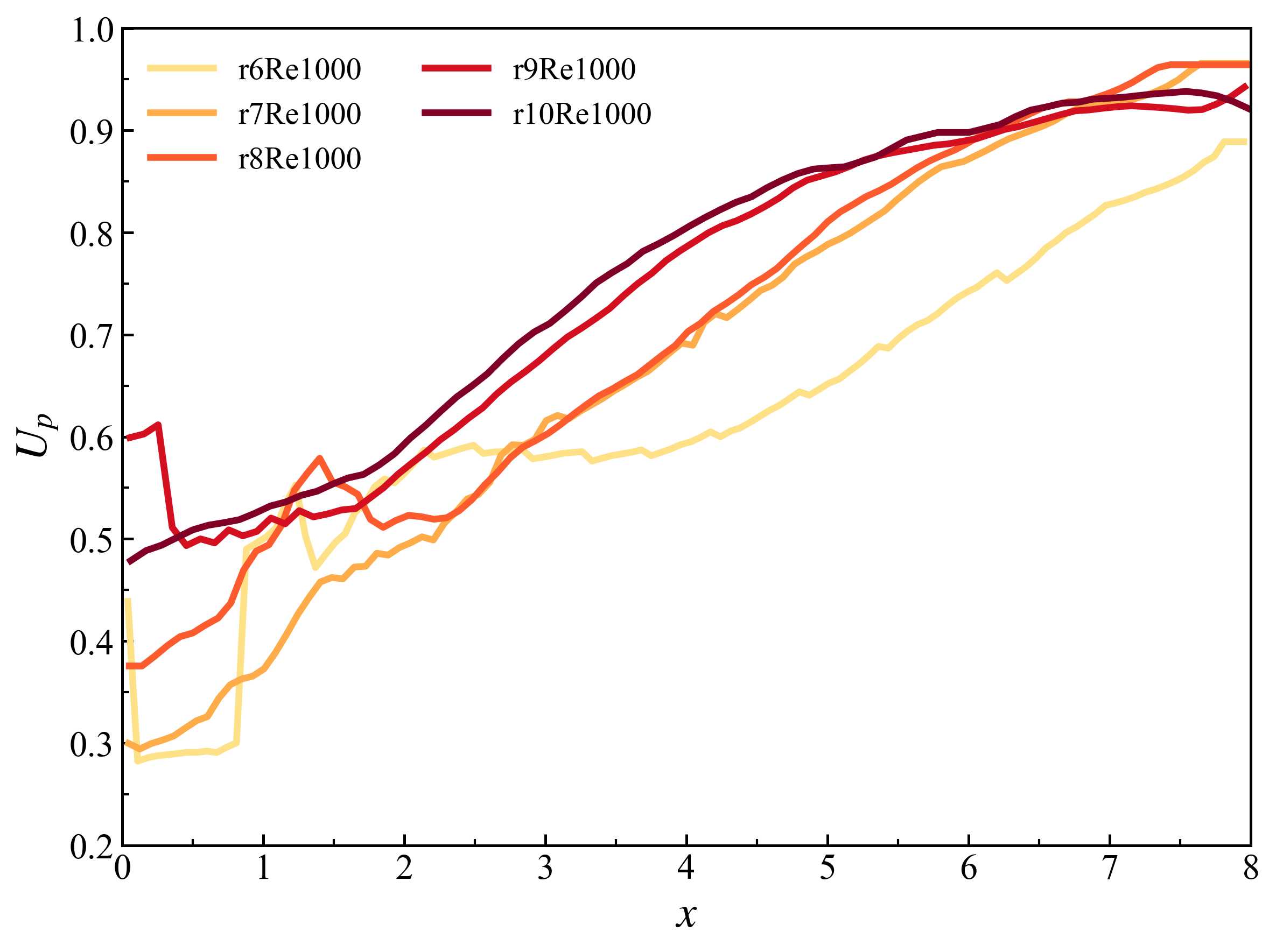}
    \end{minipage}
    \hfill
    \begin{minipage}[t]{0.4\linewidth}
        \includegraphics[width=\linewidth]{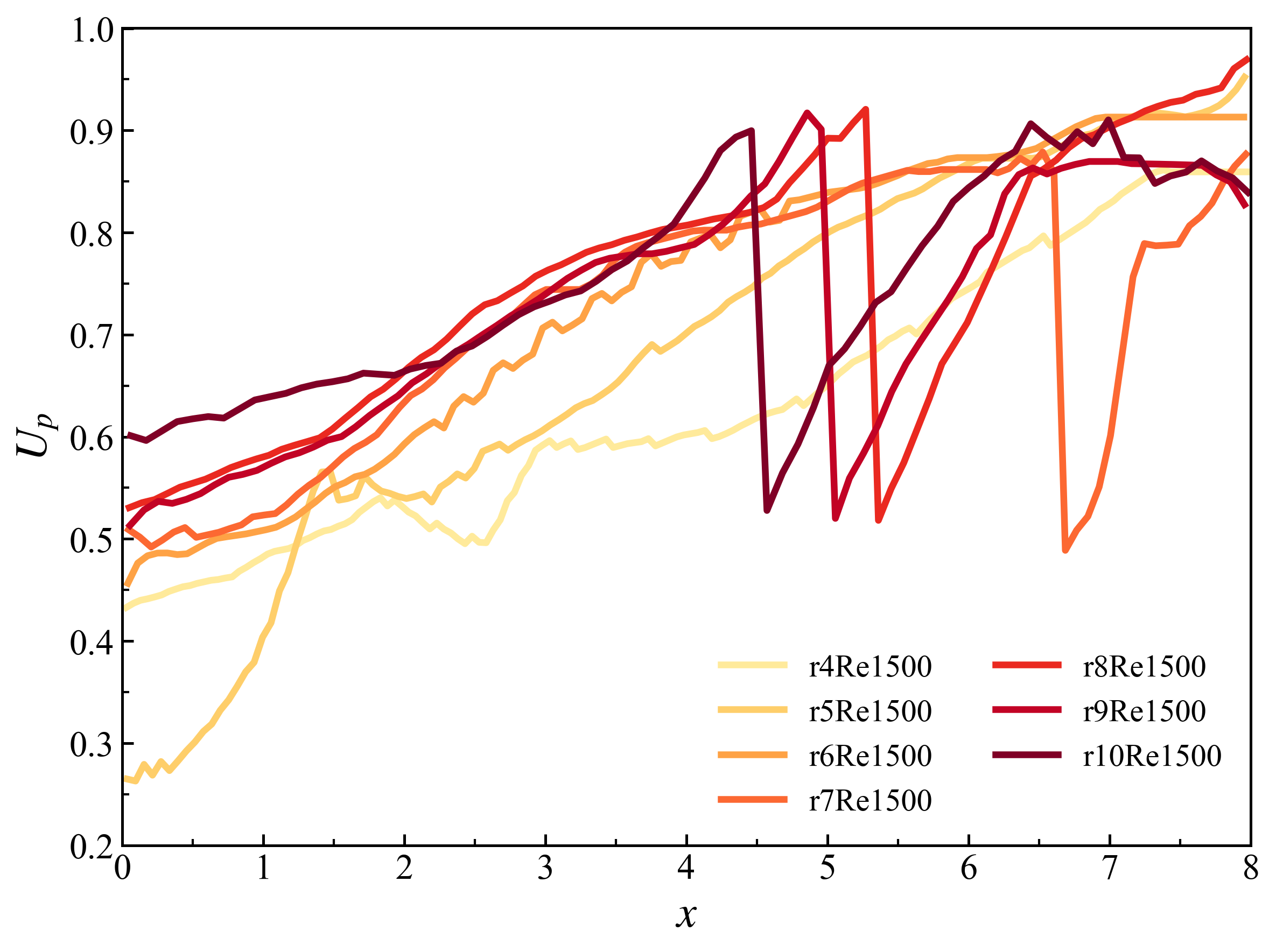}
    \end{minipage}
\end{subfigure}

\vspace{2mm}

\caption{Distributions of (a) wall-normal velocity fluctuation $v_{rms}$, (b) wavelength $\lambda$, (c) wave frequency $f$, and (d) phase speed $U_p$ along the streamwise direction for the cases in the travelling-wave regime at $Re = 1000$ (left) and 1500 (right).}
\label{fig11}
\end{figure*}

As presented in \citet{KaithakkalKametaniHasegawa2020, KaithakkalKametaniHasegawa2021}, the dissimilarity between flow and heat transfer induced by travelling-wave-like wall blowing and suction strongly depends on the wave properties, such as the wave magnitude, wavelength, wave frequency and phase speed. The root-mean-square (RMS) of the fluctuating wall-normal velocity, $v_{rms}$, at the center of each pore is calculated to represent the wave magnitude. Two-point correlation analysis is also performed for the wall-normal velocity along the center line of the porous plate, i.e., $y = 0$, at each streamwise location to determine the spatial correlation coefficient, and the local wavelength $\lambda(x)$ is then determined from the streamwise distance between two neighboring positive peaks. The wave frequency $f$ is calculated similarly based on the temporal two-point correlation. The phase speed $U_p$ is then calculated as $U_p = \lambda f$. Figure~\ref{fig11} plots the distributions of wall-normal velocity fluctuation $v_{rms}$, wavelength $\lambda$, wave frequency $f$, and phase speed $U_p$ along the streamwise direction for the cases in the travelling-wave regime at $Re = 1000$ and 1500.

The distribution of $v_{rms}$ along the streamwise direction exhibits a feature similar to that of $St^L$ (see Figs.~\ref{fig5}(a) and \ref{fig6}(a)), where the enhancement moves upstream as $L_\mathrm{p}/L_\mathrm{s}$ increases. At $Re = 1000$, as $L_\mathrm{p}/L_\mathrm{s}$ increases, $v_{rms}$ near the trailing edge rises until $L_\mathrm{p}/L_\mathrm{s} = 8$ and then decreases. As shown in Fig.~\ref{fig5}(b), a larger $v_{rms}$ near the trailing edge leads to a remarkable rise of $C_f^L$, resulting in a slight decline in $A^G/A_0^G$ (see Case r8Re1000). In contrast, a moderate $v_{rms}$, i.e., $v_{rms} < 1$, near the trailing edge (see Cases r9Re1000 and r10Re1000) contributes to the global dissimilarity $A^G/A_0^G$ larger than unity. At $Re = 1500$, the enhancement of $v_{rms}$ moves further upstream. $v_{rms}$ near the trailing edge gradually decreases for $4 \leq L_\mathrm{p}/L_\mathrm{s} \leq 6$, resulting in dissimilar heat transfer enhancement with $A^G/A_0^G > 1$. With further increasing $L_\mathrm{p}/L_\mathrm{s}$, a significant increase in $v_{rms}$ significantly enhances $C_f^L$, resulting in a decrease in $A^G/A_0^G$. These features suggest a strong relationship between the performance indices and wave magnitude. Specifically, a moderate wave magnitude of $v_{rms} < 1$ leads to dissimilar heat transfer enhancement, while an excessively high $v_rms$ shifts the analogy in an unfavorable direction by increasing a pressure drop without substantial enhancement of heat transfer.

The peak of $A^L$ is primarily attributed to the increase in $v_{rms}$, indicating that strong wall-normal velocity fluctuations promote the enhancement of $A^L$ in the upstream and middle regions. However, at $Re = 1500$, $A^L$ is suppressed despite the presence of even larger $v_{rms} > 1$. 
In particular, from Cases r7Re1500 to r10Re1500, the normalized wavelength exceeds unity downstream of $x \approx 5$, which coincides with the region where $A^L$ decreases below 1 at $Re = 1500$. 
These findings imply that whether $A^L$ remains above or below unity is governed not merely by the presence of a traveling wave, but heavily by its wave properties.
Specifically, an increase in $A^L$ in the upstream part of the perforated plate is associated with a traveling wave of relatively high fluctuation intensity and short wavelength, whereas a decrease in $A^L$ is related to the induction of a traveling wave with a longer wavelength and even stronger amplitude of $v_{rms}$. At $Re = 1000$, the wavelength exhibits a similar increasing trend and gradually saturates toward unity, i.e., the channel height, near the trailing edge in all the cases as shown in the left panel of Fig.~\ref{fig11} (b). This may suggest that as the laminar boundary layer develops, the wavelength of the traveling wave increases and eventually saturates once it becomes comparable to the channel height. This is also consistent with the visualization of the travelling-wave in Fig.~\ref{fig8}(c).


At $Re = 1000$, the frequency $f$ exhibits an almost constant value, i.e., $f\approx 1$, along the perforated plate, slightly increasing as $L_\mathrm{p}/L_\mathrm{s}$ increases (see, the left panel of Fig.~\ref{fig11}(c)). Similar behavior is also found at $Re = 1500$ (see, the right panel of Fig.~\ref{fig11}(c)), while $f$ abruptly turns to decrease in the downstream region for $L_\mathrm{p}/L_\mathrm{s} > 7$. 
This sharp drop in the frequency $f$ at $Re = 1500$ occurs simultaneously with the rapid increase in wavelength as shown in the right panel of Fig.~\ref{fig11}(b), which is potentially associated with the breakdown of the traveling wave and the subsequent onset of chaotic flow observed in this region.

\subsection{Reynolds stress and turbulent heat flux}
\label{subsec:reynolds_stress_heat_flux}


While the previous sections focused on the local distributions of drag and heat transfer on the perforated surface, and their relationship with the properties of induced traveling waves, this section considers the wall-normal Reynolds stress $-\overline{u'v'}$ and convective heat flux $-\overline{\theta'v'}$ to investigate the underlying momentum and heat transport mechanisms within the fluid domain. Here, the over-bar $\overline{\left( \cdot \right)}$ represents a time averaged quantity, while a dashed quantity $\left( \cdot \right)'$ is a deviation from the time average, namely, $\xi^{\prime}\left(x, y, t\right) = \xi\left(x, y, t\right) - \overline{\xi}\left(x, y\right)$, where $\xi$ is an arbitrary physical quantity. 

\begin{figure*}[htbp]
\centering

\begin{subfigure}[t]{0.70\linewidth}
    \centering
    \caption{}
    \includegraphics[width=\linewidth]{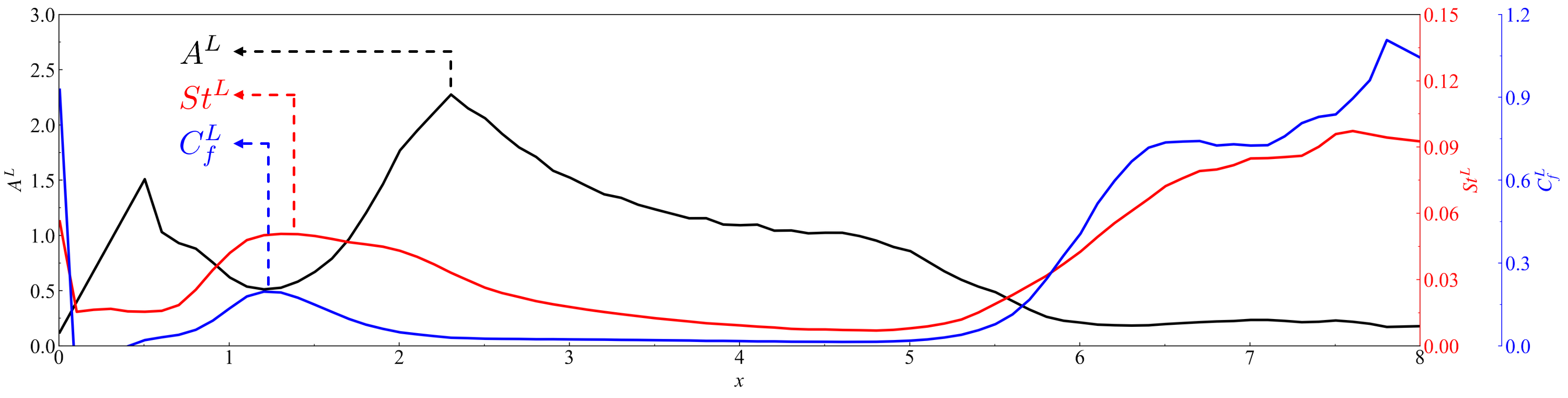}
\end{subfigure}

\begin{subfigure}[t]{0.70\linewidth}
    \centering
    \caption{}
    \includegraphics[width=\linewidth]{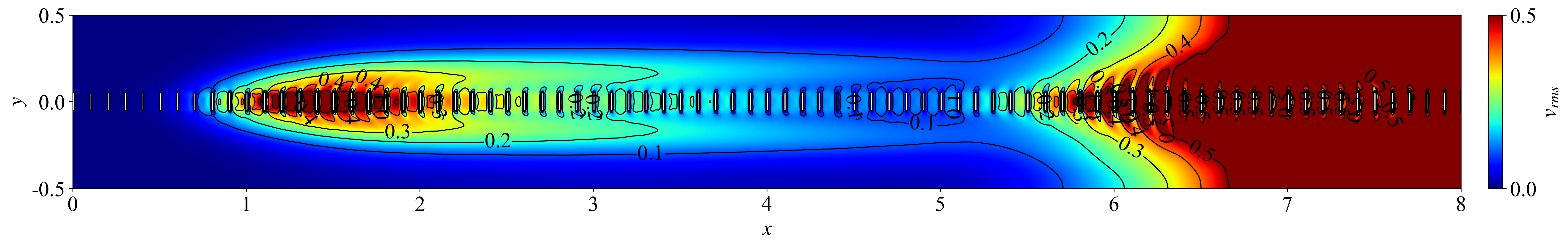}
\end{subfigure}

\begin{subfigure}[t]{0.70\linewidth}
    \centering
    \caption{}
    \includegraphics[width=\linewidth]{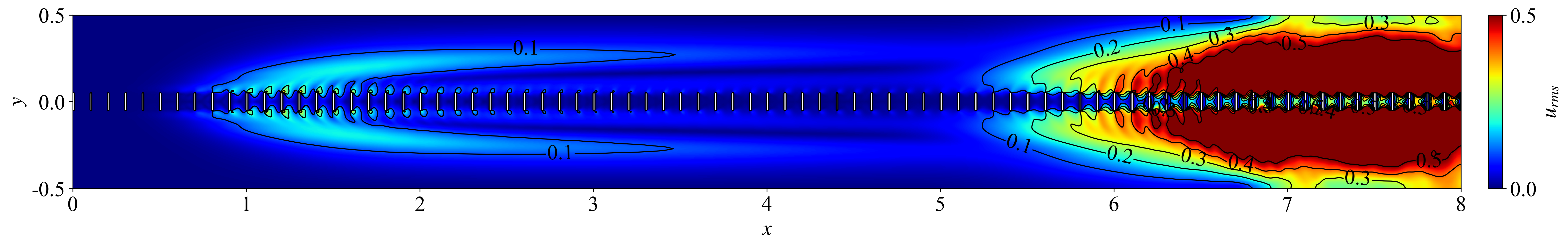}
\end{subfigure}

\begin{subfigure}[t]{0.70\linewidth}
    \centering
    \caption{}
    \includegraphics[width=\linewidth]{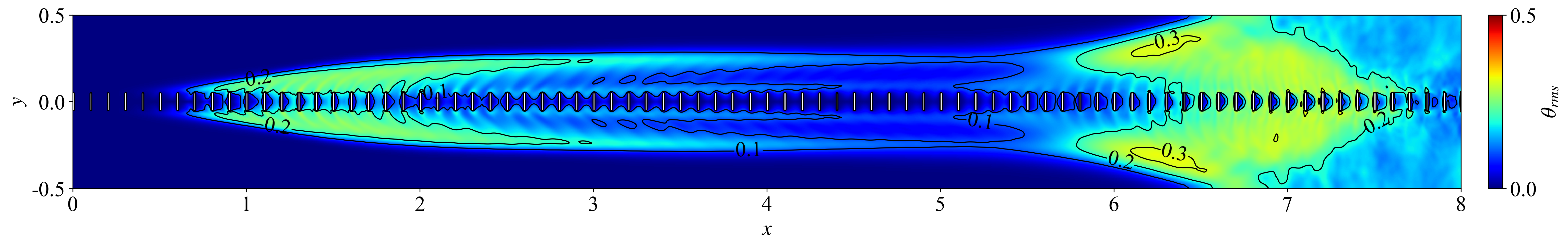}
\end{subfigure}

\begin{subfigure}[t]{0.70\linewidth}
    \centering
    \caption{}
    \includegraphics[width=\linewidth]{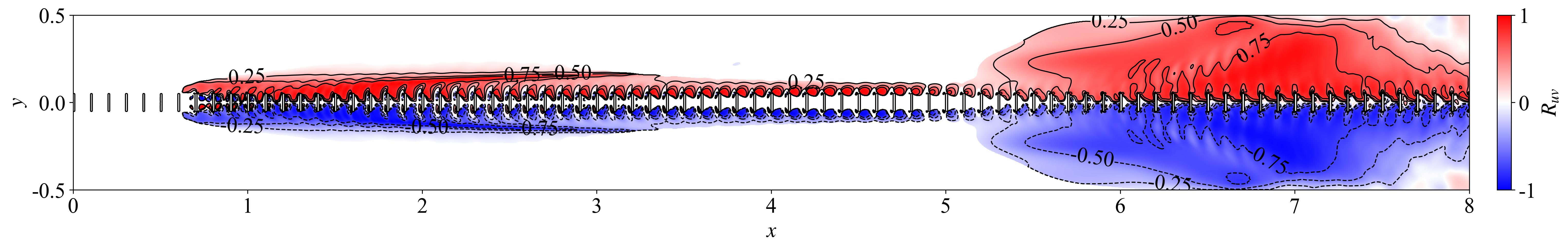}
\end{subfigure}

\begin{subfigure}[t]{0.70\linewidth}
    \centering
    \caption{}
    \includegraphics[width=\linewidth]{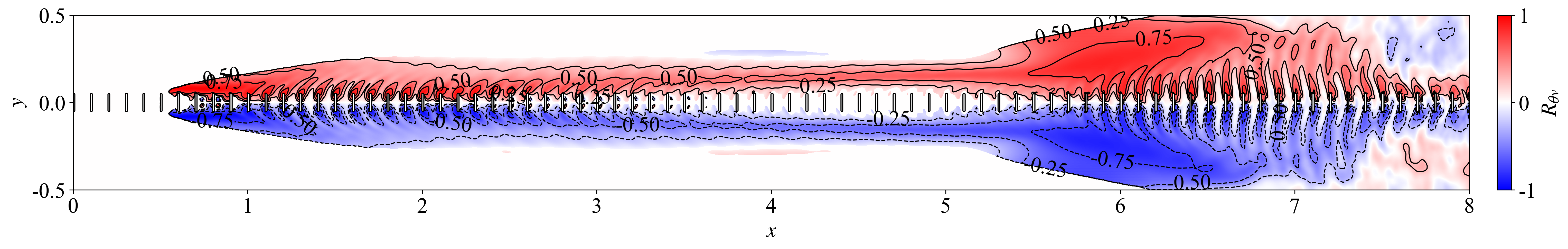}
\end{subfigure}

\begin{subfigure}[t]{0.70\linewidth}
    \centering
    \caption{}
    \includegraphics[width=\linewidth]{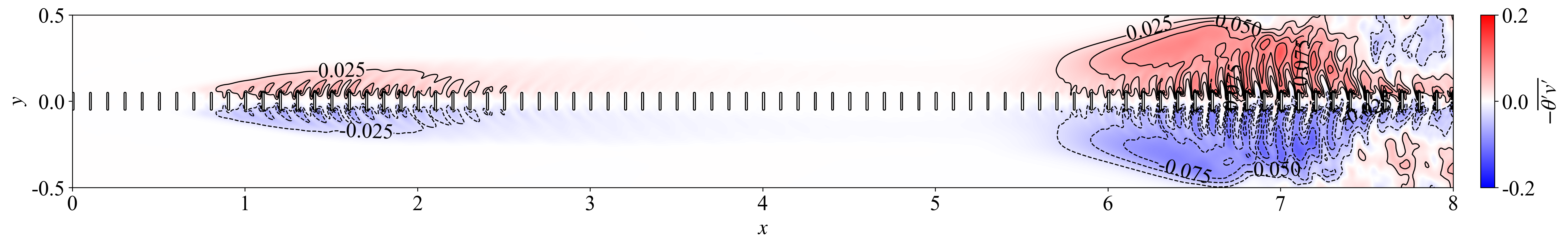}
\end{subfigure}

\begin{subfigure}[t]{0.70\linewidth}
    \centering
    \caption{}
    \includegraphics[width=\linewidth]{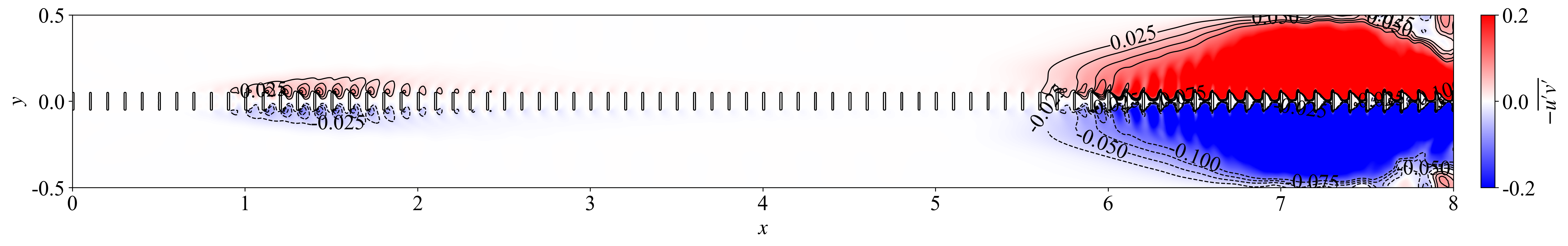}
\end{subfigure}

\caption{(a) Local performance indices $St^L$, $C_f^L$, and $A^L$ along the streamwise direction; distributions of (b) wall-normal velocity fluctuation, (c) streamwise velocity fluctuation, and (d) temperature fluctuation; correlation coefficients between (e) wall-normal and streamwise velocity $R_{uv}$ and (f) wall-normal velocity and temperature $R_{\theta v}$; and time-averaged (g) Reynolds stress $-\overline{u^\prime v^\prime}$ and (h) turbulent heat flux $-\overline{\theta^\prime v^\prime}$ for Case r9Re1500. The solid and dashed lines indicate contours with positive and negative values, respectively.}
\label{fig12}
\end{figure*}

We particularly focus on Case r9Re1500, where local dissimilar heat transfer enhancement is first achieved in the upstream traveling wave region and then breaks down after the induced traveling wave becomes chaotic near the trailing edge. The local performance indices $St^L$, $C_f^L$, and $A^L$ in this case are plotted in Fig.~\ref{fig12}(a). The distributions of the wall-normal velocity fluctuation $v_{rms}$, streamwise velocity fluctuation $u_{rms}$, and temperature fluctuation $\theta_{rms}$ are shown in Figs.~\ref{fig12}(b)--\ref{fig12}(d), respectively. Significant rises in $v_{rms}$, and the associated $u_{rms}$ and $\theta_{rms}$ appear around the perforated plates near $x \approx 1$. The streamwise development of $v_{rms}$ up to $x \approx 4.5$ is consistent with the traveling wave shown in Fig.~\ref{fig9}(c). Compared with $u_{rms}$, the high-$\theta_{rms}$ region extends further downstream, reaching $x \approx 4.5$. Note that the governing equations and wall boundary conditions for the streamwise velocity component and temperature are similar in the present configuration. Thus, their difference is attributed to the continuity constraint on the velocity field, which manifests through the pressure gradient term in the governing equation.

The correlation coefficients between wall-normal and streamwise velocity, $R_{uv}$, and between wall-normal velocity and temperature, $R_{\theta v}$, are shown in Figs.~\ref{fig12}(e) and \ref{fig12}(f), respectively. Although they generally show similar distributions, a closer look reveals that $R_{\theta v}$ has a wider wall-normal distribution and extends farther downstream than $R_{uv}$. Consequently, in the travelling-wave region, the turbulent heat flux $-\overline{\theta^\prime v^\prime}$ remains enhanced from $x \approx 1$ to $x \approx 4$, whereas the Reynolds stress $-\overline{u^\prime v^\prime}$ decays rapidly and almost disappears around $x \approx 2$ near the perforated plates. This explains for larger enhancement of heat transfer than drag in the region of $2 < x < 4$, resulting in local dissimilar heat transfer enhancement, i.e., $A^L > 1$, as shown in Fig.~\ref{fig12}(a). In contrast, further downstream beyond $x \approx 6$, where the induced traveling wave breaks down into a chaotic flow regime, $v_{rms}$, $u_{rms}$, and $\theta_{rms}$ are intensified again (see Fig.~\ref{fig9}(d)), with $\theta_{rms}$ remaining much weaker than $u_{rms}$. This leads to a larger enhancement of $-\overline{u^\prime v^\prime}$ than of $-\overline{\theta^\prime v^\prime}$ around the trailing edge. Accordingly, $C_f^L$ exhibits a more significant rise than $St^L$, reducing $A^L$ in this region. 

\begin{figure*}[htbp]
\centering

\begin{subfigure}[t]{0.48\linewidth}
    \centering
    \caption{}
    \vspace{2mm}
    \includegraphics[width=\linewidth]{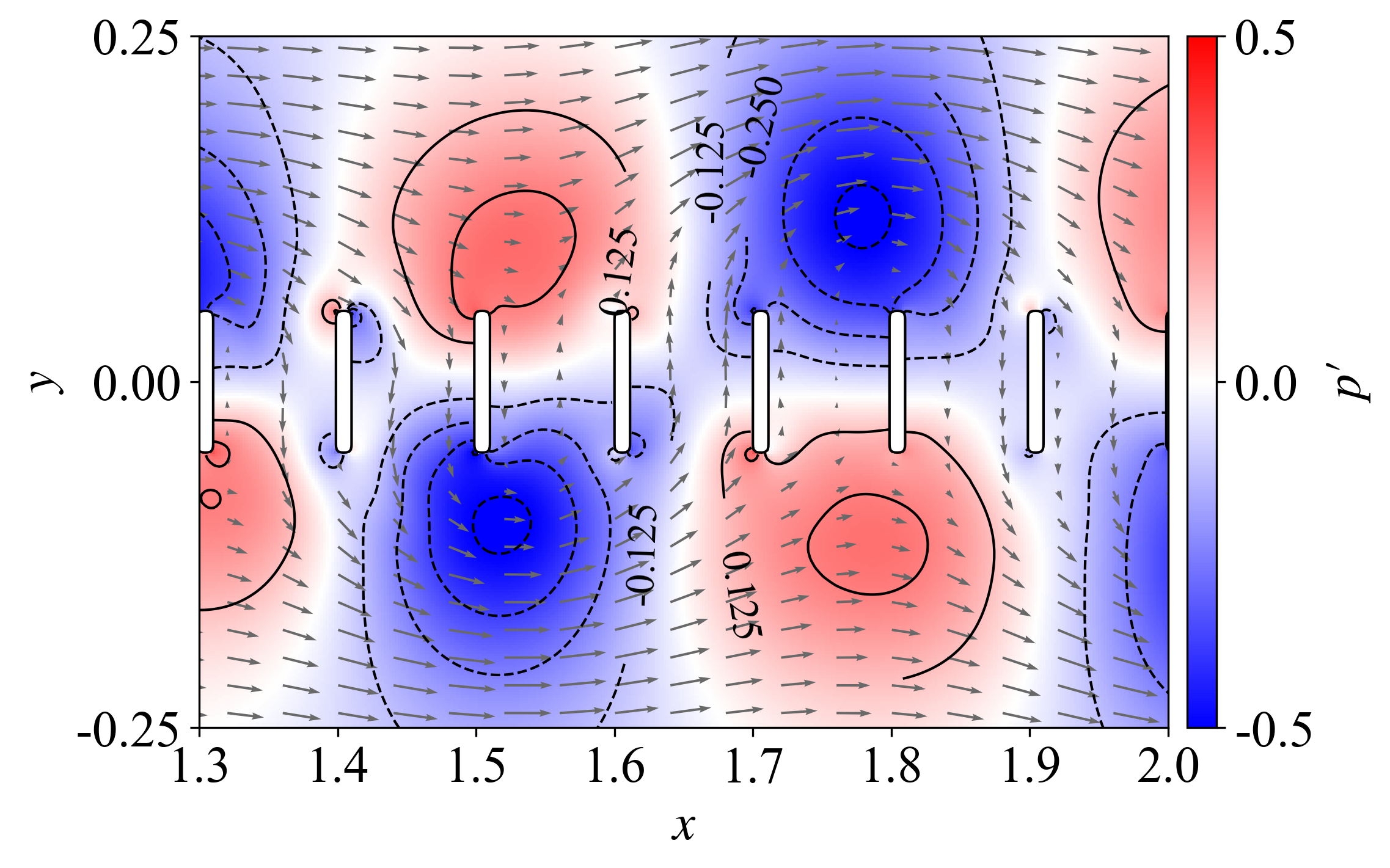}
\end{subfigure}
\hfill
\begin{subfigure}[t]{0.48\linewidth}
    \centering
    \caption{}
    \vspace{2mm}
    \includegraphics[width=\linewidth]{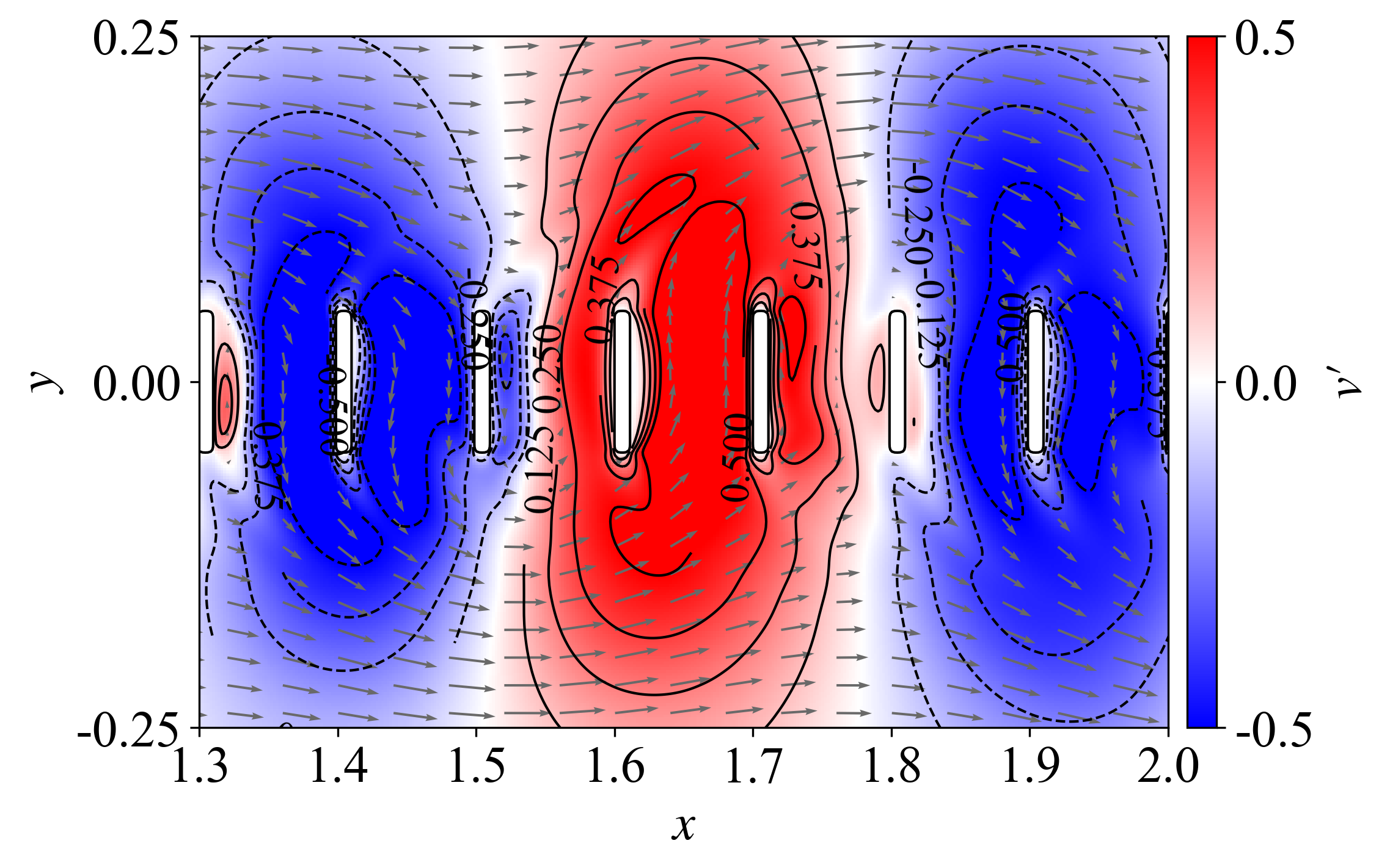}
\end{subfigure}

\vspace{4mm}

\begin{subfigure}[t]{0.48\linewidth}
    \centering
    \caption{}
    \vspace{2mm}
    \includegraphics[width=\linewidth]{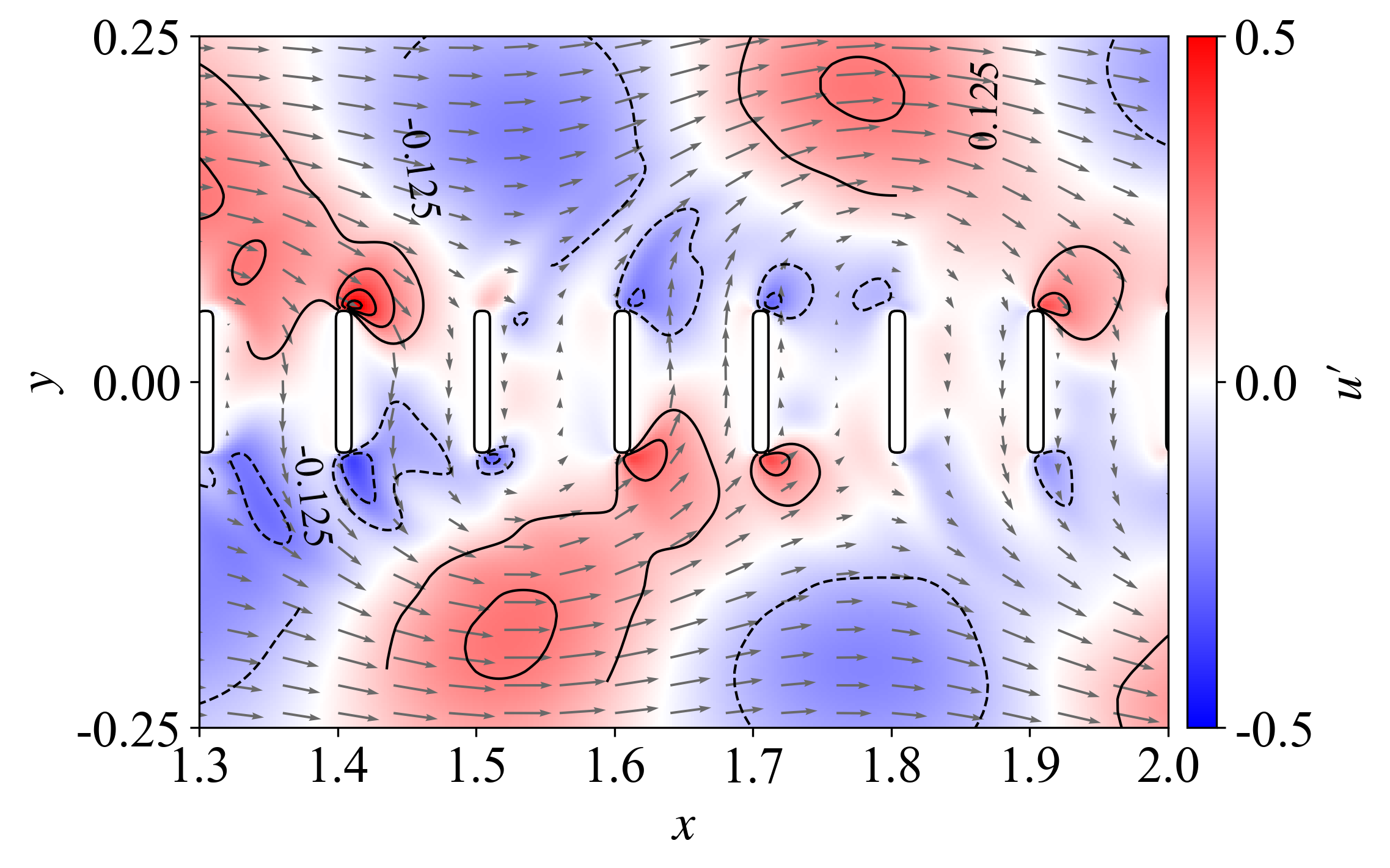}
\end{subfigure}
\hfill
\begin{subfigure}[t]{0.48\linewidth}
    \centering
    \caption{}
    \vspace{2mm}
    \includegraphics[width=\linewidth]{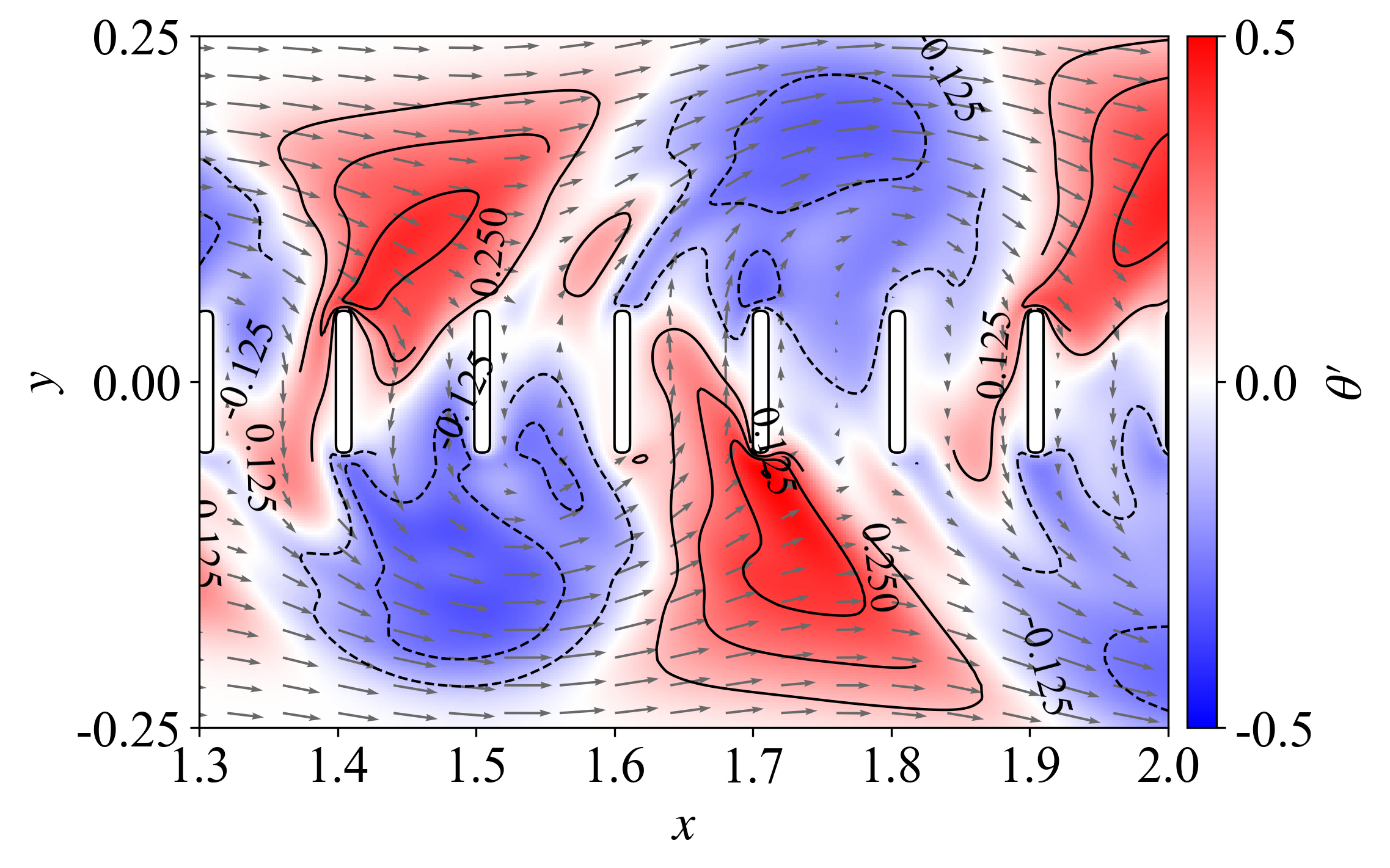}
\end{subfigure}

\vspace{4mm}

\begin{subfigure}[t]{0.48\linewidth}
    \centering
    \caption{}
    \vspace{2mm}
    \includegraphics[width=\linewidth]{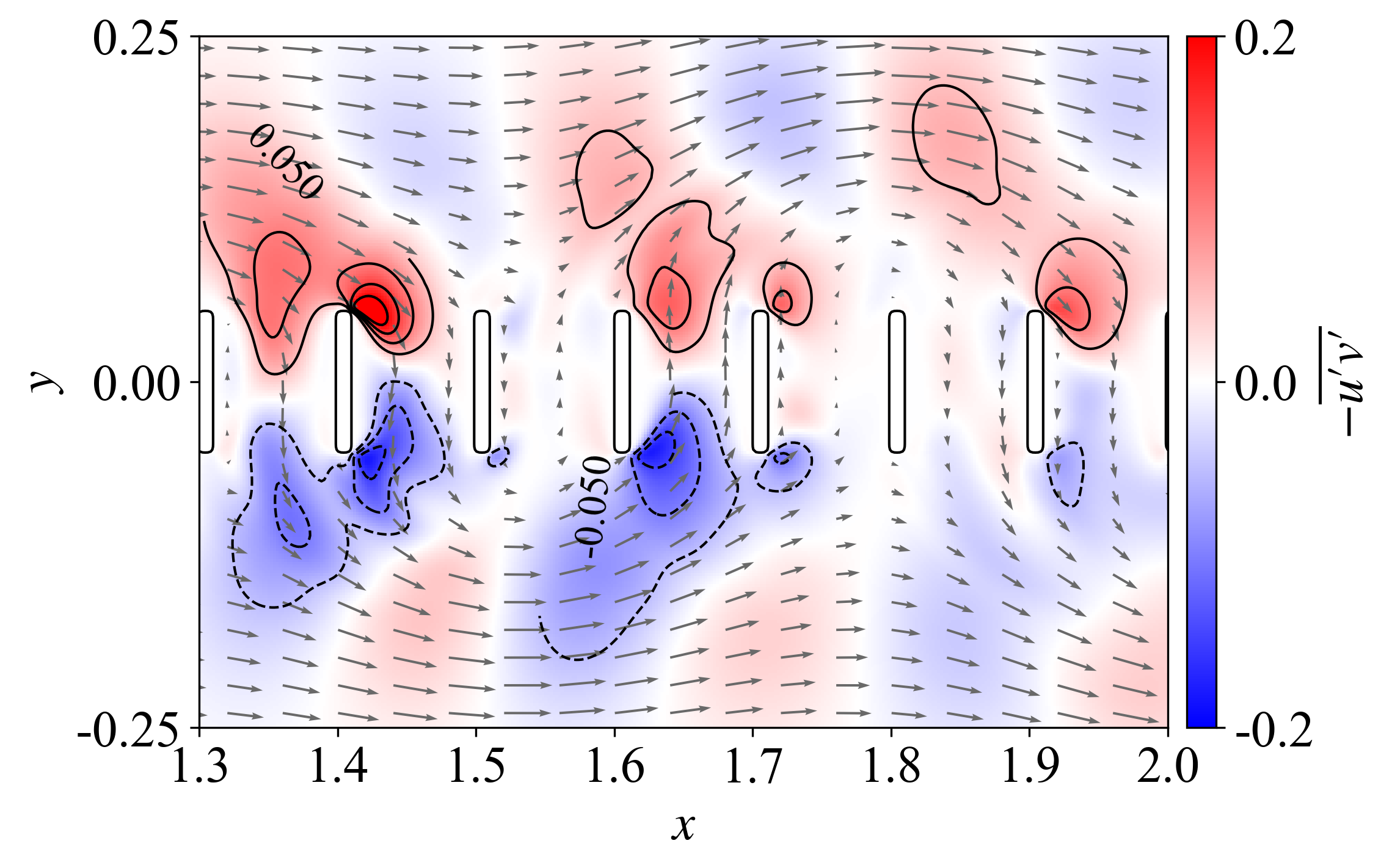}
\end{subfigure}
\hfill
\begin{subfigure}[t]{0.48\linewidth}
    \centering
    \caption{}
    \vspace{2mm}
    \includegraphics[width=\linewidth]{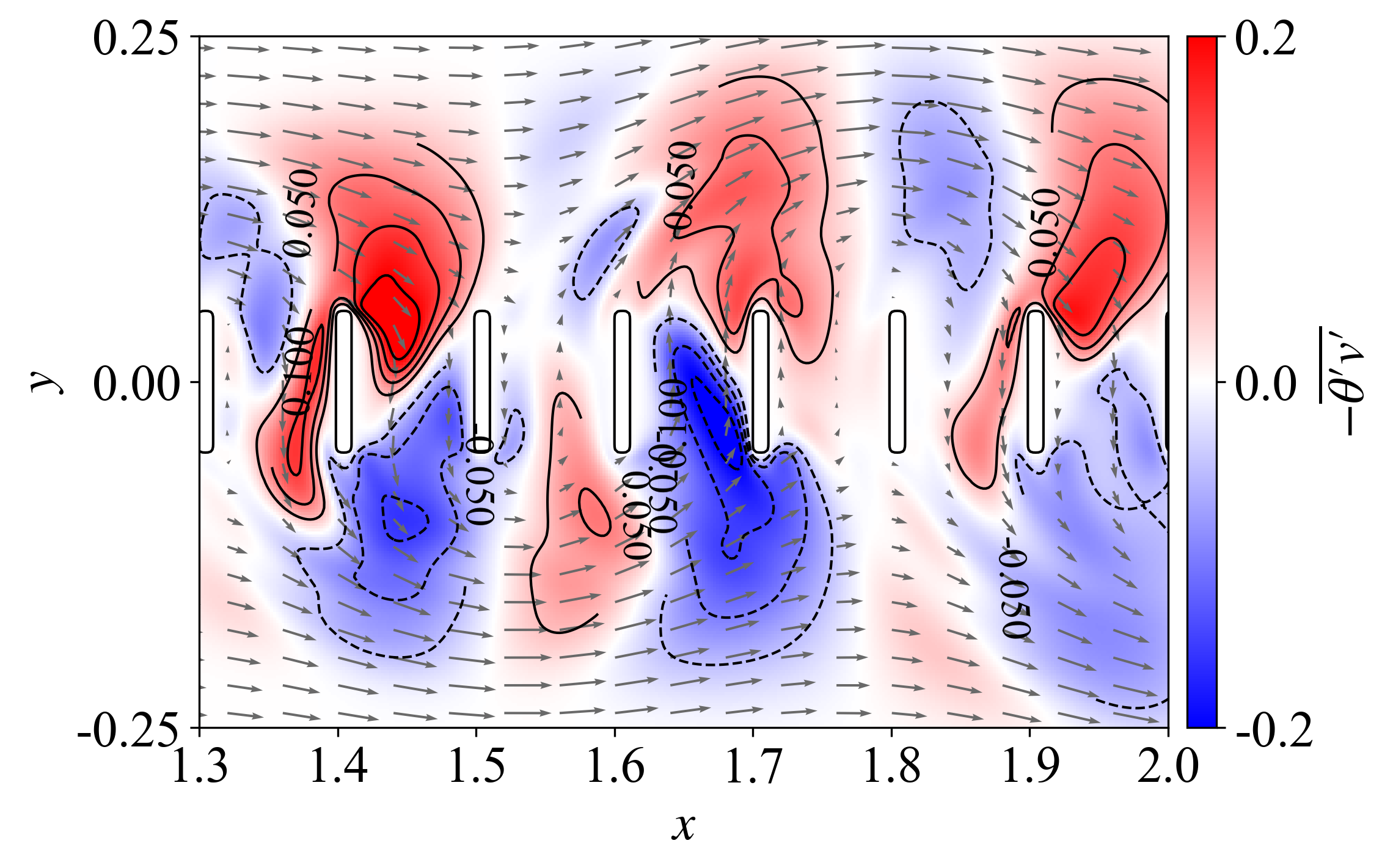}
\end{subfigure}

\vspace{4mm}

\caption{Distributions of instantaneous (a) $p^\prime$, (b) $v^\prime$, (c) $u^\prime$, (d) $\theta^\prime$, (e) $-u^\prime v^\prime$, and (f) $-\theta^\prime v^\prime$ from $x = 1.3$ to 2.0 for Case r9Re1500. The solid and dashed lines indicate the corresponding contours with positive and negative values, respectively. The grey arrows show the instantaneous velocity vectors.}
\label{fig13}
\end{figure*}

\begin{figure*}[htbp]
\centering

\begin{subfigure}[t]{0.48\linewidth}
    \centering
    \caption{}
    \vspace{2mm}
    \includegraphics[width=\linewidth]{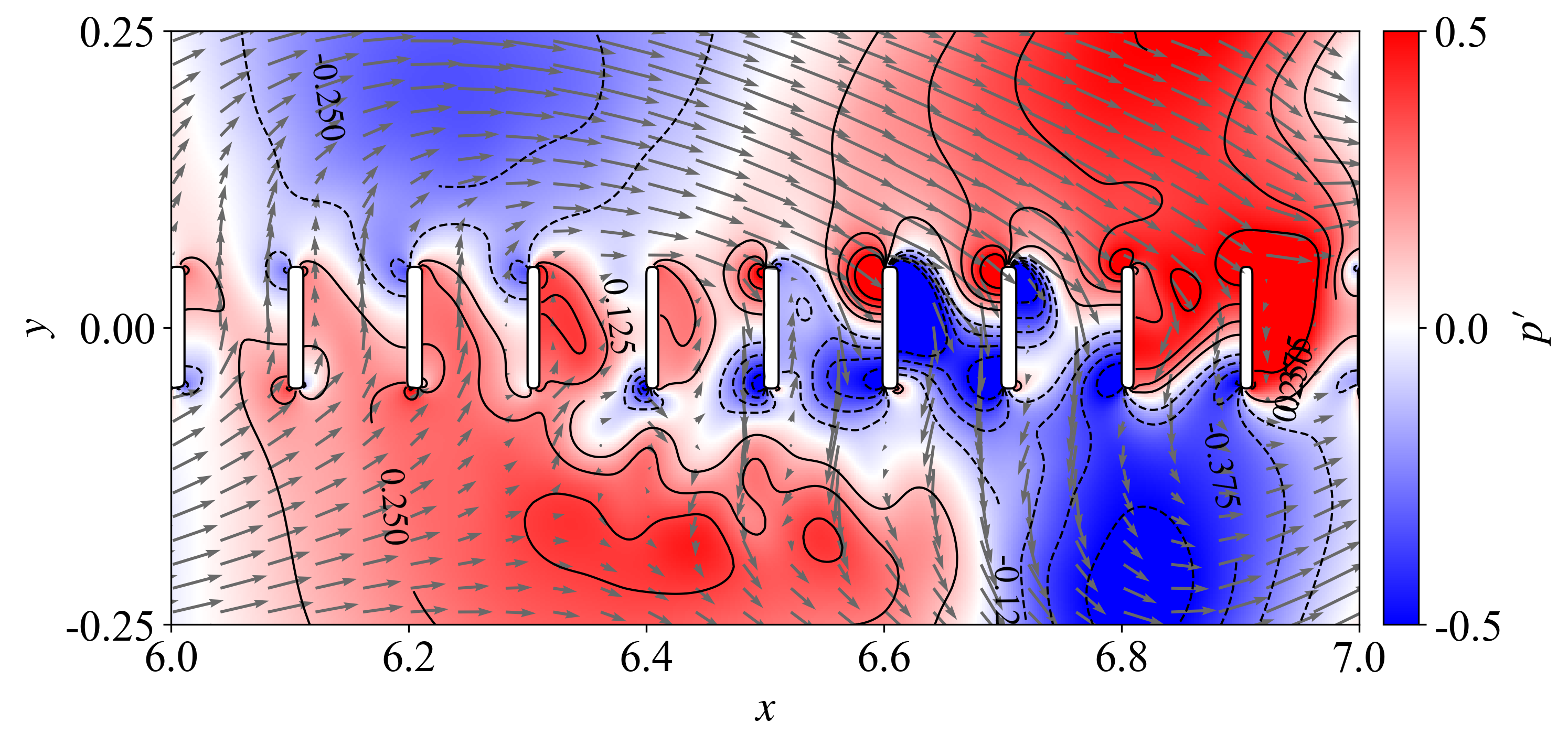}
\end{subfigure}
\hfill
\begin{subfigure}[t]{0.48\linewidth}
    \centering
    \caption{}
    \vspace{2mm}
    \includegraphics[width=\linewidth]{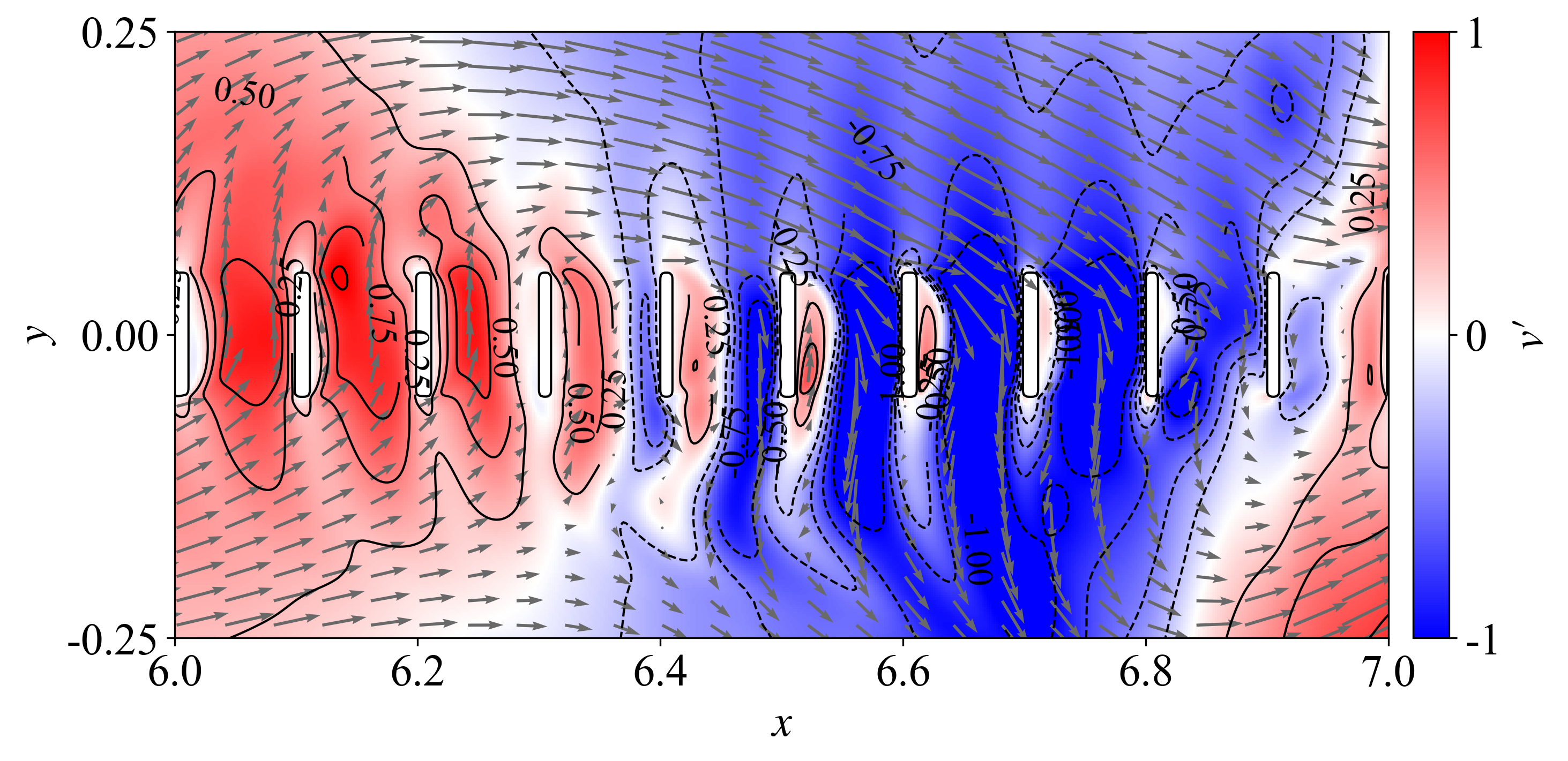}
\end{subfigure}

\vspace{4mm}

\begin{subfigure}[t]{0.48\linewidth}
    \centering
    \caption{}
    \vspace{2mm}
    \includegraphics[width=\linewidth]{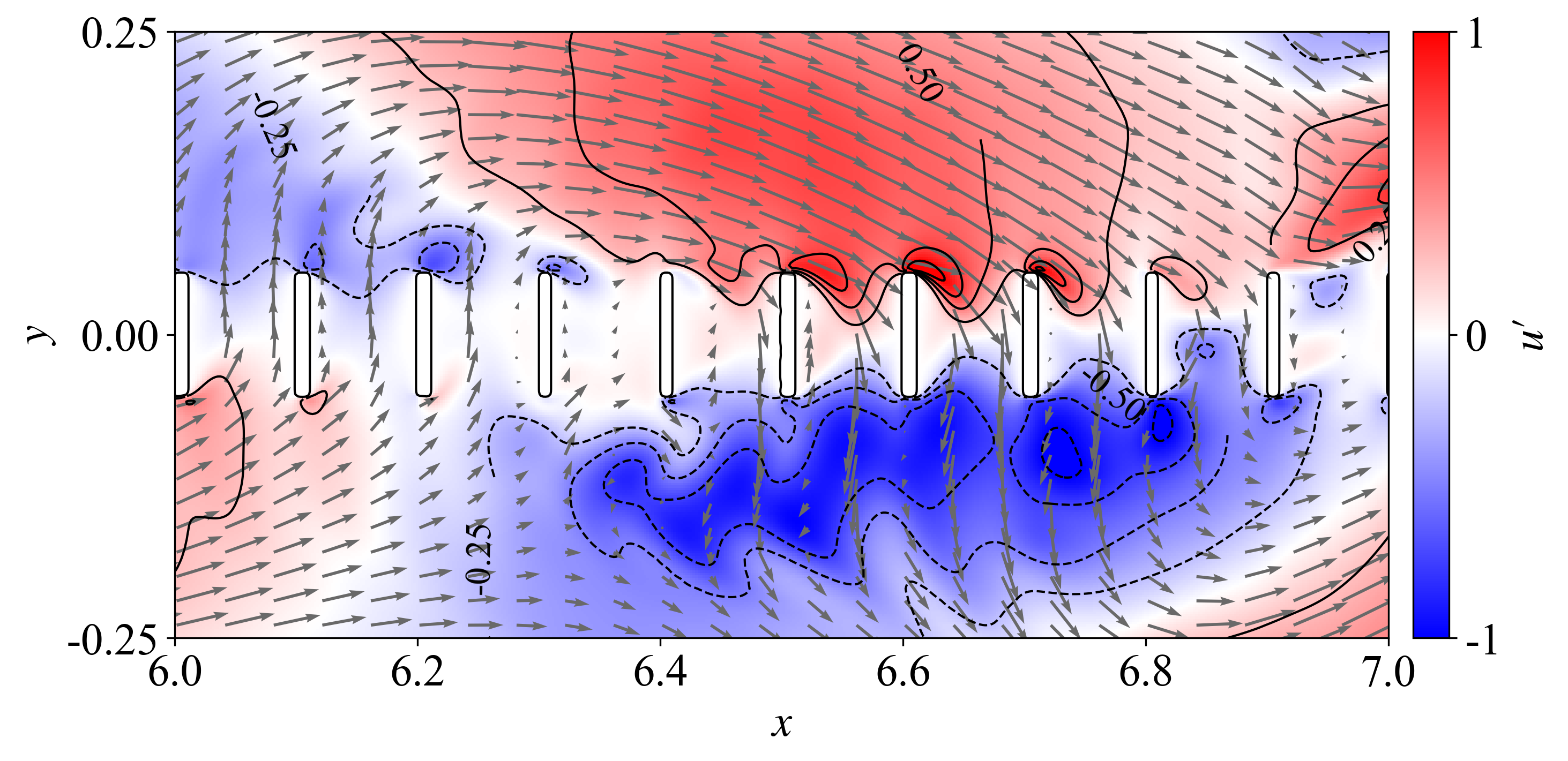}
\end{subfigure}
\hfill
\begin{subfigure}[t]{0.48\linewidth}
    \centering
    \caption{}
    \vspace{2mm}
    \includegraphics[width=\linewidth]{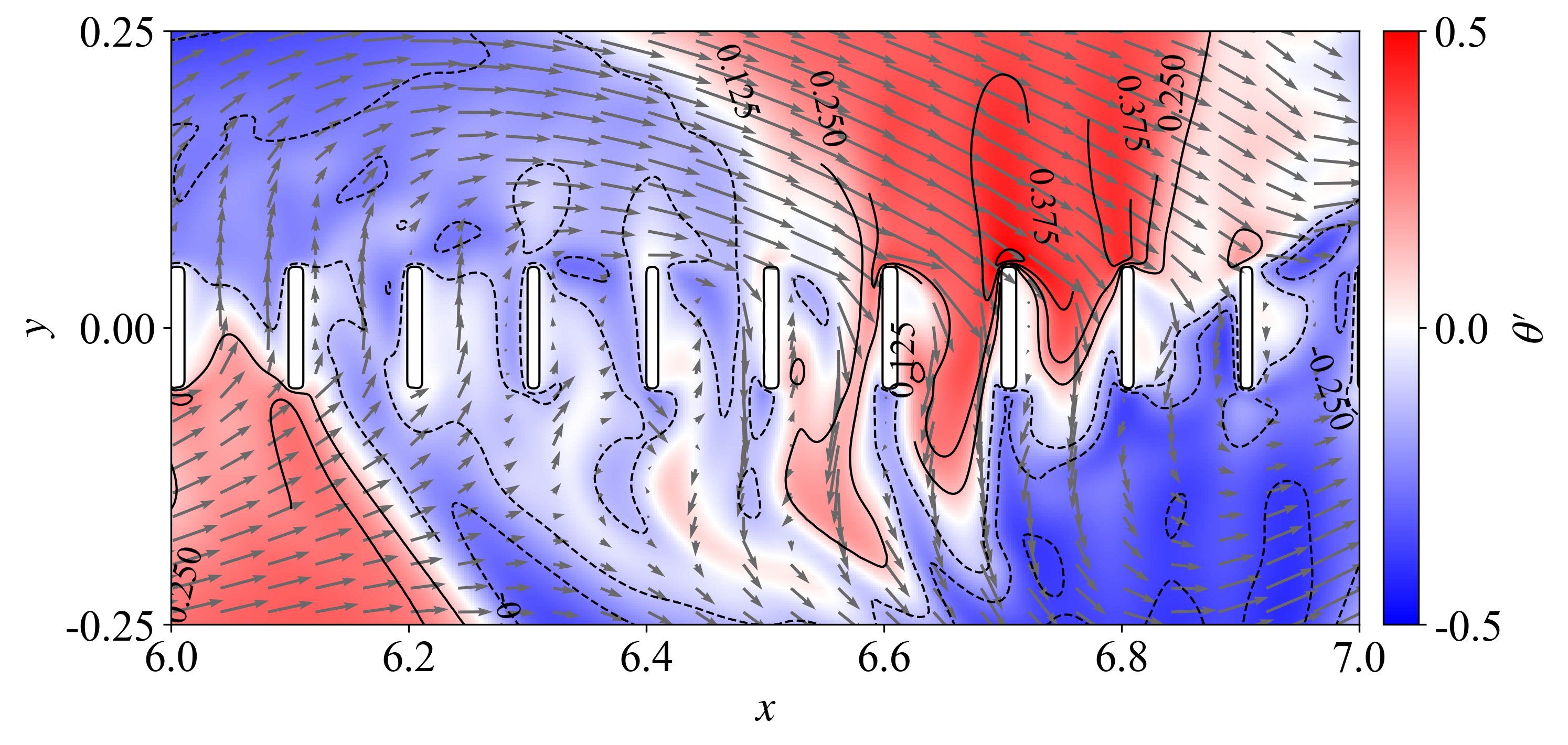}
\end{subfigure}

\vspace{4mm}

\begin{subfigure}[t]{0.48\linewidth}
    \centering
    \caption{}
    \vspace{2mm}
    \includegraphics[width=\linewidth]{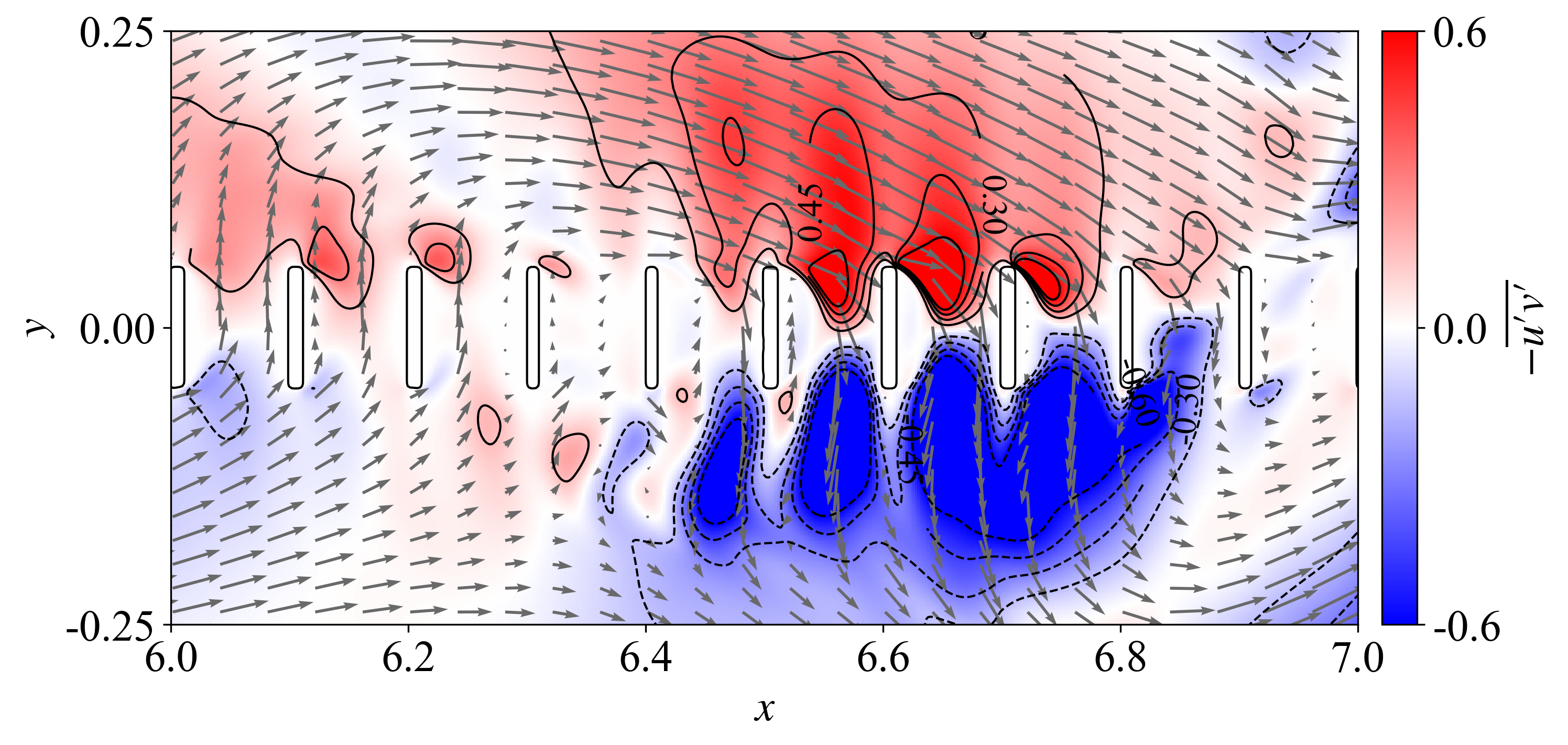}
\end{subfigure}
\hfill
\begin{subfigure}[t]{0.48\linewidth}
    \centering
    \caption{}
    \vspace{2mm}
    \includegraphics[width=\linewidth]{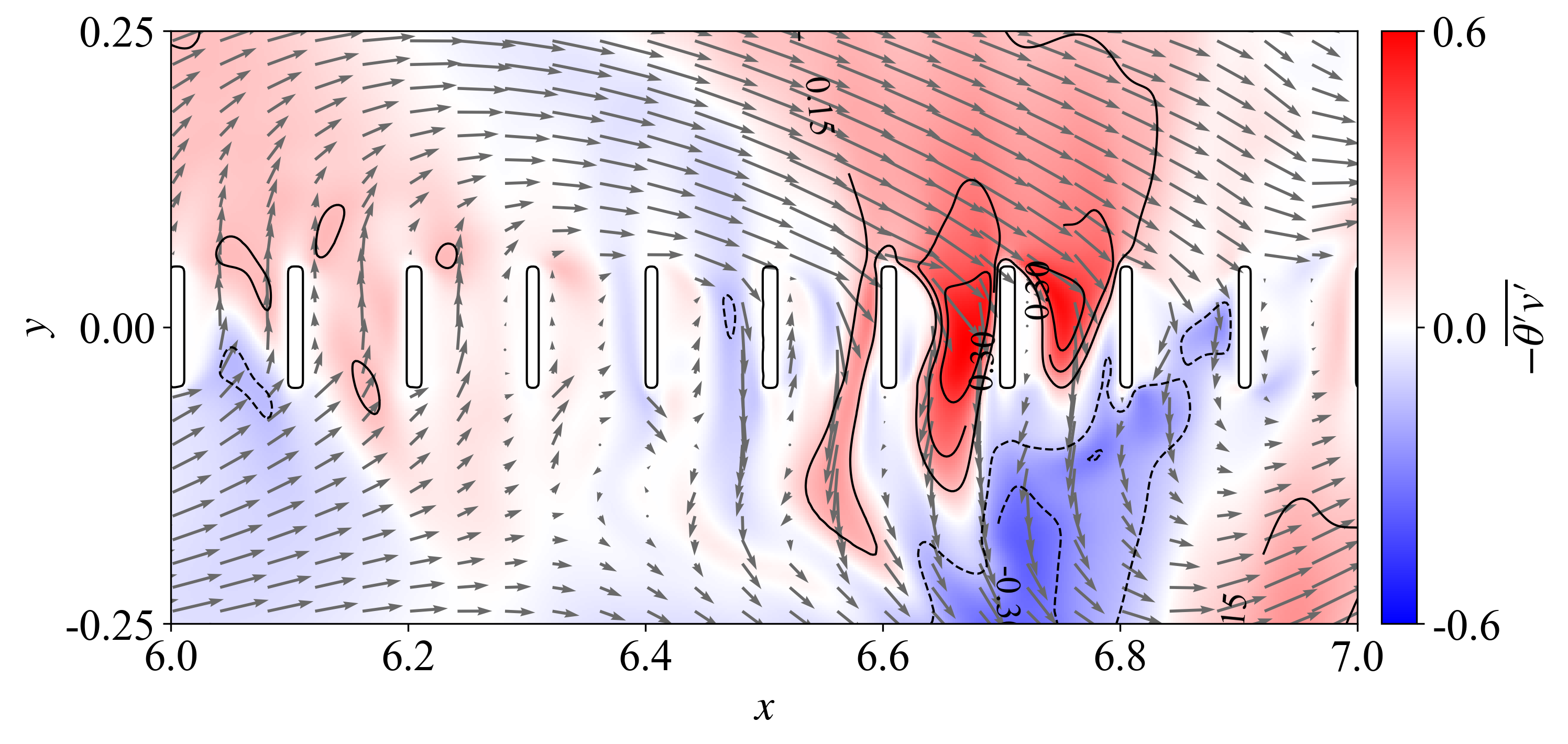}
\end{subfigure}

\vspace{4mm}

\caption{Distributions of instantaneous (a) $p^\prime$, (b) $v^\prime$, (c) $u^\prime$, (d) $\theta^\prime$, (e) $-u^\prime v^\prime$, and (f) $-\theta^\prime v^\prime$ from $x = 6.0$ to 7.0 for Case r9Re1500. The solid and dashed lines indicate the corresponding contours with positive and negative values, respectively. The grey arrows show the instantaneous velocity vectors.}
\label{fig14}
\end{figure*}

To elucidate the underlying momentum and heat transport mechanisms around the perforated plate, the zoomed-in instantaneous fields of $p^\prime$, $v^\prime$, $u^\prime$, $\theta^\prime$, $-u^\prime v^\prime$, and $-\theta^\prime v^\prime$ are shown in Figs.~\ref{fig13} and \ref{fig14} for the upstream region of $1.3 < x < 2.0$, where the traveling wave is induced, and also the downstream region of $6.0 < x < 7.0$, where the induced traveling wave breaks down into a chaotic flow, respectively. Figures~\ref{fig13}(a) and \ref{fig13}(b) show that, at $x \approx 1.55$, where $v^\prime$ around the perforated plate switches from negative to positive, a positive pressure gradient in the $y$ direction is generated. This induces an acceleration in the downward fluid motion, sustaining a streamwise traveling wave of the wall-normal velocity fluctuation. Figures~\ref{fig13}(c) and \ref{fig13}(d) show that, in the immediate vicinity above the perforated plate ($0 < y < 0.1$), negative $u^\prime$ and $\theta^\prime$ are observed for $1.55 < x < 1.75$ where the wall-normal velocity $v^\prime$ is positive. The upward flow across the perforated plate carries low-velocity, low-temperature fluid from the perforated plate. Similarly, positive $u^\prime$ and $\theta^\prime$ are found where $v^\prime$ is negative for $1.35 < x < 1.55$ or $1.8 < x < 2.0$. In the region further away from the wall ($y > 0.1$), however, $\theta^\prime$ maintains its sign and the positive or negative structures expand their distributions, whereas the sign of $u^\prime$ reverses. 
The dissimilarity between $u^\prime$ and $\theta^\prime$ can be attributed to the streamwise gradient of the pressure field shown in Fig.~\ref{fig13}(a). While the velocity field is constrained by the continuity equation and thus influenced by the pressure gradient, the temperature field is not subject to such a constraint. Indeed, it can be confirmed that the streamwise velocity fluctuation away from the wall correlates negatively with the pressure fluctuation, consistent with Bernoulli's principle assuming an inviscid fluid.   
The inherent difference between $u^\prime$ and $\theta^\prime$ leads to significant dissimilarity between the instantaneous distributions of $-u^\prime v^\prime$ and $-\theta^\prime v^\prime$ especially away from the wall, as shown in Figs.~\ref{fig13}(e) and \ref{fig13}(f). This explains the dissimilarity between the turbulent heat flux $-\theta^\prime v^\prime$ and the Reynolds shear stress $-u^\prime v^\prime$ shown in Fig.~\ref{fig12}, thereby enhancing $St$ over $C_f$.

In the downstream chaotic-flow region of $6.0 < x < 7.0$, Figs.~\ref{fig14}(a) and \ref{fig14}(b) show that pressure fluctuations still induce a vertical pressure gradient across the perforated plate and generate wall-normal velocity fluctuations. Unlike the upstream region of $ 1.3 < x < 2.0$ shown in Figs.~\ref{fig13}, however, Figs.~\ref{fig14}(c) and \ref{fig14}(d) show that the sign of $u^\prime$ is maintained even far from the wall. Furthermore, its magnitude exceeds that of $\theta^\prime$, and it exhibits a strong negative correlation with $v^\prime$. Consequently, in the chaotic-flow region, the instantaneous Reynolds stress becomes larger than the turbulent heat flux, as shown in Figs.~\ref{fig14}(e) and \ref{fig14}(f), leading to a reduction in the analogy factor.

It is confirmed that momentum and heat transport, as well as the emergence of their dissimilarity, differ significantly between the upstream region of $1.3 < x < 2.0$ where the traveling wave is induced and the downstream region of $6.0 < x < 7.0$ where the induced traveling wave develops and breaks down. Specifically, in the upstream region, the streamwise velocity fluctuation away from the wall is driven by the pressure gradient, making it dissimilar to the temperature fluctuation. Consequently, while the Reynolds shear stress decreases, the turbulent heat flux increases, resulting in a high dissimilar heat transfer enhancement. In contrast, the induced traveling wave is first suppressed and then strongly amplifies again in the downstream region, eventually breaking down into a chaotic flow. Correspondingly, the correlation between the streamwise and wall-normal velocity fluctuations strengthens, and the Reynolds shear stress is significantly enhanced. As a result, momentum transport dominates over heat transport, causing the dissimilarity factor to drop below unity.

\section{Conclusions}
\label{sec:conclusions}

Travelling-wave-like wall blowing and suction has been widely investigated and confirmed as an effective approach to achieving dissimilar heat transfer enhancement. However, whether a similar effect can be realized passively, especially at moderate or low Reynolds numbers, has remained unclear. In this study, pore-resolving simulations were performed to assess whether perforated plates can passively induce travelling-wave-like disturbances and thereby enhance heat transfer relative to drag in a spatially developing laminar flow. The volume penalization method was used to resolve the flow and temperature fields around plates with the pore-to-solid ratio of $L_\mathrm{p}/L_\mathrm{s} = 0$--10 at $Re = 500$, 1000 and 1500.

It was revealed that dissimilar heat transfer enhancement is achieved only within a limited region of the parameter space. Specifically, at $Re = 500$, the perforated plates show little improvement over the zero-permeability plate because the flow remains nearly the steady regime due to the dominance of the viscosity. At $Re = 1000$, dissimilar heat transfer enhancement is obtained for $9 \leq L_\mathrm{p}/L_\mathrm{s} \leq 10$, while at $Re = 1500$ it occurs in a wider range of $4 \leq L_\mathrm{p}/L_\mathrm{s} \leq 7$. The analogy factor reaches its maximum at $Re = 1500$ and $L_\mathrm{p}/L_\mathrm{s} = 6$, exceeding that of the pure-solid plate by 31\%, i.e.,  $A^G/A_0^G \approx 1.31$.

For the heat transfer and drag, both reach their local peaks at $Re=1500$ and $L_\mathrm{p}/L_\mathrm{s} = 5$, exhibiting increases to 428\% and 311\% relative to the zero-permeability plate, respectively. As a result, the analogy factor improves by 28\%.

Furthermore, as the porosity increases, i.e., with larger $L_\mathrm{p}/L_\mathrm{s}$, the onset of the traveling wave occurs even upstream. However, its breakdown into a chaotic flow downstream causes drag to increase more than heat transfer, so that the overall analogy factor decays.
 

Instantaneous flow field analysis confirms that, in the upstream region where the traveling wave-like disturbance is induced, the wall-normal velocity fluctuation is driven by the pressure fluctuation across the perforated surface. Furthermore, because the streamwise velocity is also influenced by the gradient of these pressure fluctuations, a dissimilarity with the temperature field arises, thereby achieving a dissimilar heat transfer enhancement where heat transfer outweighs momentum transport. 
Since the pressure gradient appears solely in the governing equation for the velocity field and is determined such that the continuity is satisfied, this mechanism is essentially attributed to the fundamental difference between the divergence-free velocity field and the conservative temperature field.

On the other hand, as the Reynolds number and porosity, i.e., $L_\mathrm{p}/L_\mathrm{s}$, increase, the breakdown of the induced traveling wave generates strong wall-normal velocity fluctuation. This enhances its correlation with the streamwise velocity fluctuation, and thereby causing the Reynolds shear stress to dominate over the turbulent heat flux. Consequently, the dissimilar heat transfer enhancement is diminished.

In this study, it was demonstrated that by employing a perforated plate, a traveling wave-like disturbance can be induced to achieve significant dissimilar heat transfer enhancement even in the low-Reynolds-number regime, where the flow normally remains laminar. On the other hand, the ranges of the Reynolds number and the porosity over which the dissimilar heat transfer enhancement is obtained are limited, indicating that the geometry of the perforated plate should be appropriately configured to achieve desired heat transfer and pressure drop characteristics. While the present study considered a heat transfer surface with a uniform porosity in the streamwise direction, further performance improvements could be expected by varying the porosity along the streamwise direction, taking into account the downstream development of the flow. Furthermore, although the length of the heat transfer surface was kept constant in this study, shortening the length of the perforated plate before the location where the traveling wave breaks down under high-Reynolds-number and high-porosity conditions could achieve better heat transfer and pressure drop characteristics within a more compact total volume. Finally, although the present study performed two-dimensional simulations due to the high computational cost of resolving the perforated geometry, three-dimensional simulations would be necessary, as three-dimensionality is expected to play an important role after the breakdown of the traveling wave. These issues remain as future work.

\begin{bmhead}[Funding]
Y.H. gratefully acknowledges the supports from JSPS KAKENHI Grant Number JP23K26034, and also Adaptable and Seamless Technology Transfer Program through Target-driven R\&D (A-STEP) from Japan Science and Technology Agency (JST) Grant Number JPMJTR232D. 
\end{bmhead}

\begin{bmhead}[Declaration of interests]
The authors report no conflict of interest.
\end{bmhead}

\begin{bmhead}[Author ORCIDs]
M. Liu, \url{https://orcid.org/0000-0001-6794-5023};
Y. Hasegawa, \url{https://orcid.org/0000-0002-1878-972X}
\end{bmhead}

\begin{appen}
\renewcommand{\theHsection}{appendix.\Alph{section}}

\section{Verification of numerical method}
\label{app:verification}
The channel flow under an isothermal wall condition in a laminar flow regime is considered to verify the present VPM solver. The quantities are normalized in the same way as described in \S\ref{subsec::problem}. Here, two solid flat plates (500 $\times$ 0.1 in streamwise and vertical directions) are placed at the top and bottom of a computational domain of size $[0, 500]\times[-0.6, 0.6]$, as shown in Fig.~\ref{figA1}. The normalized streamwise length of the channel is set as 500, which is much longer than the hydrodynamic entry length of laminar channel flow (typically 300 at $Re = 1500$ as reported in \citet{Cengel2002}) and is sufficient to reach a fully developed state.

Uniform Cartesian grids of size 0.02 $\times$ 0.02 are employed. No-slip and isothermal conditions are imposed on the solid plate surfaces. A uniform unit velocity profile is given at the inlet for the fluid region (grey region in Fig.~\ref{figA1}), and zero velocity is imposed by VPM in the black solid region in Fig.~\ref{figA1}. The dimensionless inlet temperature is set as unity, and zero-gradient conditions are given for velocity and temperature at the outlet. As for pressure, the reference zero value is fixed at the outlet, and the zero-gradient condition is given at the inlet.

\begin{figure*}[htbp]
  \centerline{\includegraphics[width=5in]{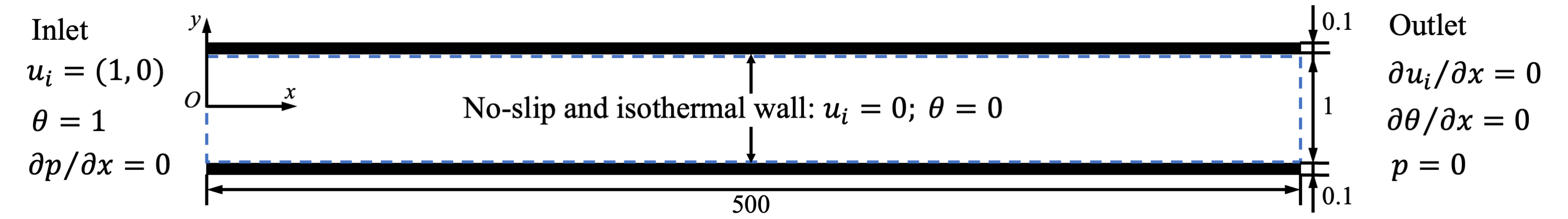}}%
  \caption{Computational domain and boundary condition for the verification case.}
\label{figA1}
\end{figure*}

The Reynolds number is defined based on the inlet mean velocity and the channel height. Six Reynolds numbers of $Re = 250$, 500, 750, 1000, 1250 and 1500 are considered for comparison with the analytical solutions of the Stanton number and the friction factor for fully developed laminar flow in a flat channel with constant wall temperature. The analytical solutions of the Stanton number and the friction factor for laminar flow in a flat channel with constant wall temperature \citep{Cengel2002} are used to verify the heat transfer and pressure-loss indices obtained from the present VPM solver.

\begin{equation}
St
=
\frac{3.77}{Re\,Pr}
\qquad
(Re \leq 2000)
\end{equation}

\begin{equation}
C_f
=
\frac{12}{Re}
\qquad
(Re \leq 2000)
\end{equation}
Figure~\ref{figA2} shows the simulated $St$ and $C_f$ by the VPM solver at various Reynolds numbers. It can be confirmed that the present simulation results agree well with the analytical solitons \citep{Cengel2002}, validating the accuracy of the present VPM solver.

\begin{figure*}[htbp]
\centering

\begin{subfigure}[t]{0.48\linewidth}
    \centering
    \caption{}
    \vspace{2mm}
    \includegraphics[width=\linewidth]{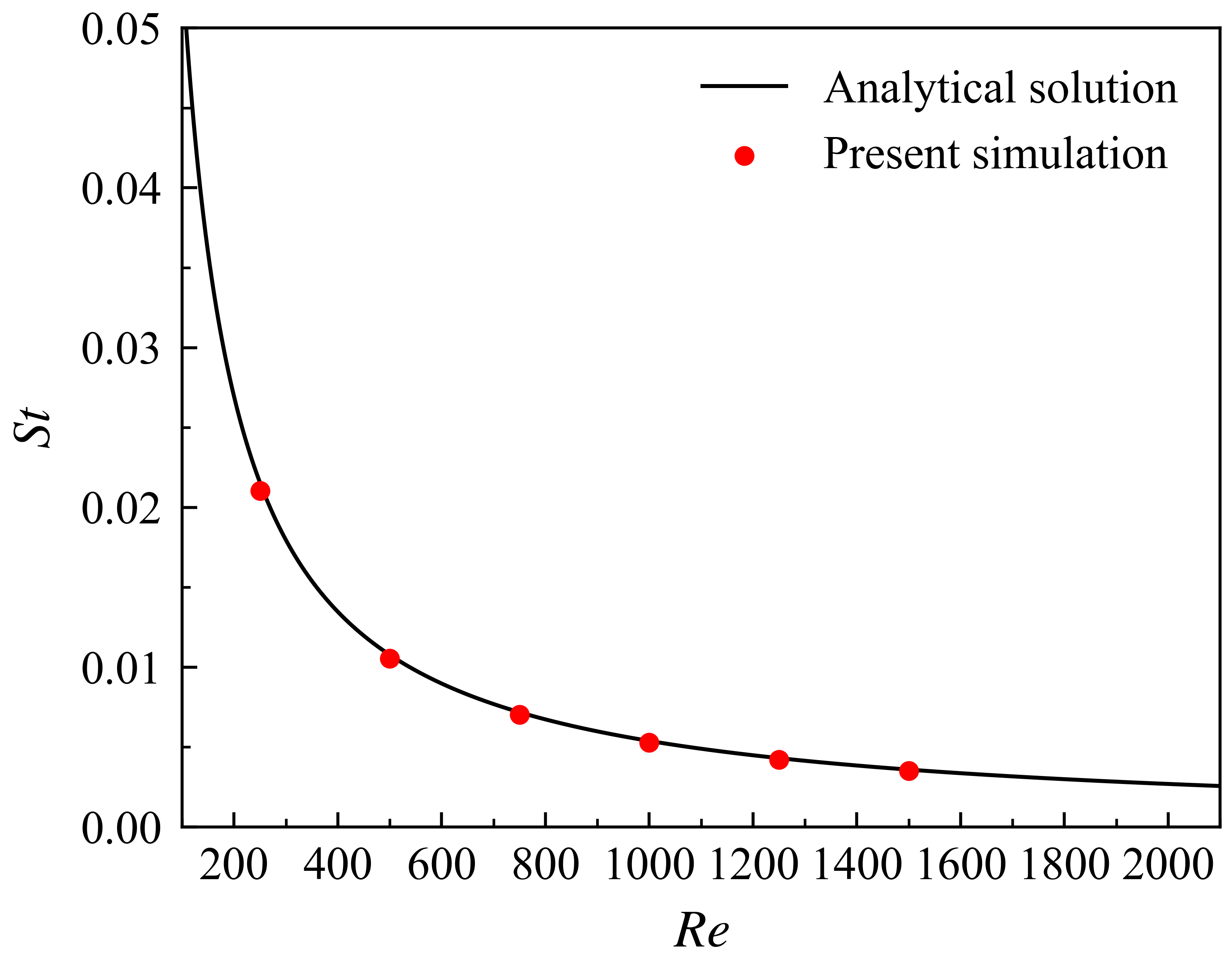}
\end{subfigure}
\hfill
\begin{subfigure}[t]{0.48\linewidth}
    \centering
    \caption{}
    \vspace{2mm}
    \includegraphics[width=\linewidth]{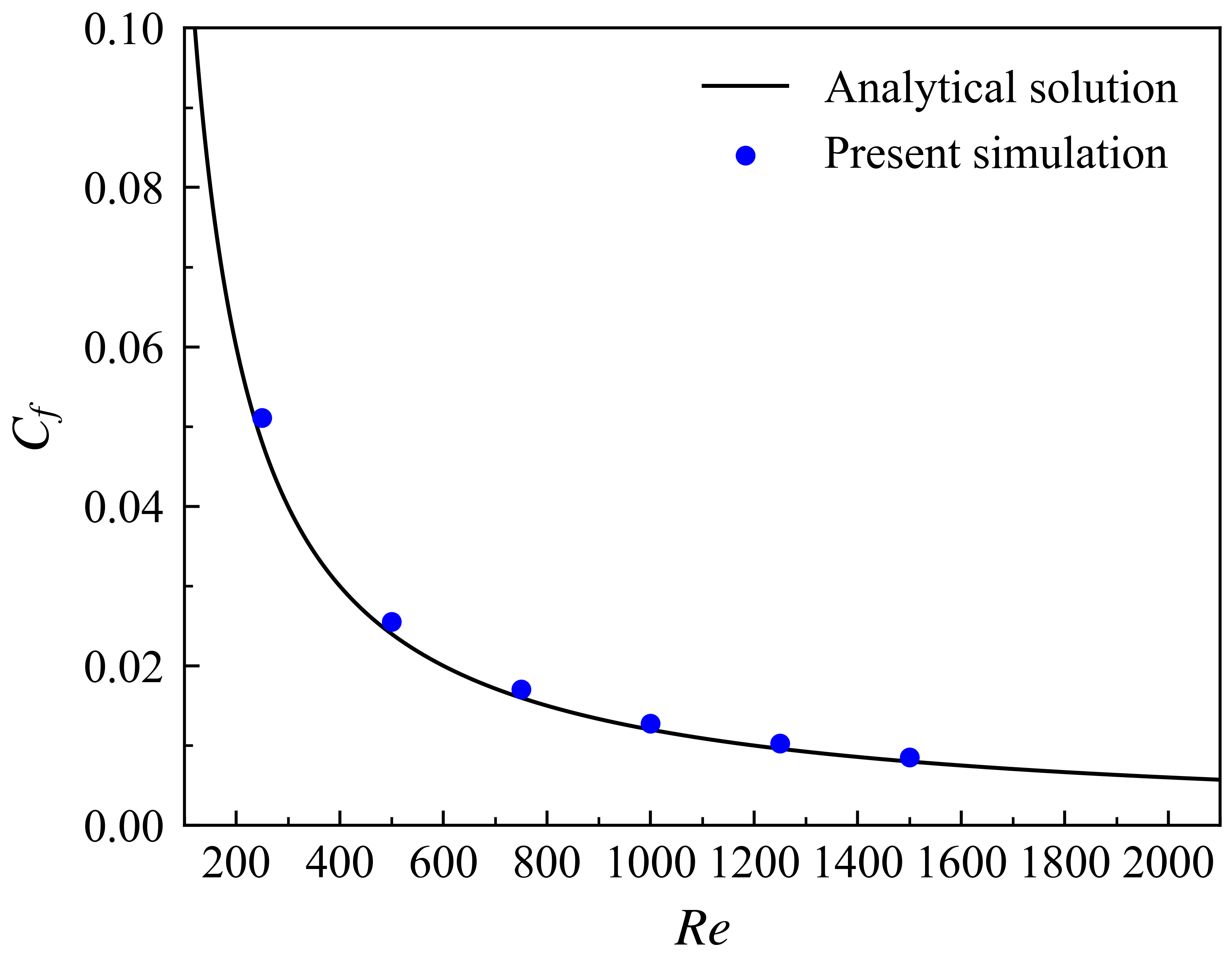}
\end{subfigure}

\vspace{4mm}

\caption{Comparison of $St$ (a) and $C_f$ (b) with their analytical solutions in a fully developed laminar channel flow.}
\label{figA2}
\end{figure*}

\end{appen}

\bibliographystyle{jfm}
\bibliography{jfm}

\end{document}